\documentclass[twocolumn]{aastex631}

\newcommand{\dens}{g cm$^{-3}$}

\begin{document}

\title{A Comparison of the Composition of Planets in Single- and Multi-Planet Systems Orbiting M dwarfs}

\correspondingauthor{Romy Rodr\'iguez Mart\'inez}
\email{rodriguezmartinez.2@osu.edu}

\author[0000-0003-1445-9923]{Romy Rodr\'iguez Mart\'inez}
\affiliation{Department of Astronomy, The Ohio State University, 140 W. 18th Avenue, Columbus OH 43210, USA}

\author[0000-0002-7595-6360]{David V. Martin}
\altaffiliation{NASA Sagan Fellow}
\affiliation{Department of Astronomy, The Ohio State University, 140 W. 18th Avenue, Columbus OH 43210, USA}

\author[0000-0003-0395-9869]{B. Scott Gaudi}
\affiliation{Department of Astronomy, The Ohio State University, 140 W. 18th Avenue, Columbus OH 43210, USA}

\author[0000-0003-3570-422X]{
Joseph G. Schulze}
\affiliation{School of Earth Sciences, The Ohio State University, 125 South Oval Mall, Columbus, OH 43210, USA}

\author[0000-0002-8823-8237
]{Anusha Pai Asnodkar}
\affiliation{Department of Astronomy, The Ohio State University, 140 W. 18th Avenue, Columbus OH 43210, USA}

\author[0000-0001-8153-639X]{Kiersten M. Boley}
\altaffiliation{NSF Graduate Research Fellow}
\affiliation{Department of Astronomy, The Ohio State University, 140 W. 18th Avenue, Columbus OH 43210, USA}

\author[0000-0002-3247-5081]{Sarah Ballard}
\affiliation{University of Florida Department of Astronomy, 1772 Stadium Rd, Gainesville, FL 32607, USA}

\begin{abstract}

We investigate and compare the composition of M-dwarf planets in systems with only one known planet (``singles") to those residing in multi-planet systems (``multis") and the fundamental properties of their host stars. We restrict our analysis to planets with directly measured masses and radii, which comprise a total of 70 planets: 30 singles and 40 multis in 19 systems. We compare the bulk densities for the full sample, which includes planets ranging in size from 0.52$R_{\oplus}$ to 12.8$R_\oplus$, and find that single planets have significantly lower densities on average than multis, which we cannot attribute to selection biases. We compare the bulk densities normalized by an Earth model for planets with $R_{p} < 6R_{\oplus}$, and find that multis are also denser with 99\% confidence. We calculate and compare the core/water mass fractions (CMF/WMF) of low-mass planets ($M_p <10 M_{\oplus}$), and find that the likely rocky multis (with $R_p <1.6 R_{\oplus}$) have lower CMFs than singles. We also compare the [Fe/H] metallicity and rotation period of all single versus multi-planet host stars with such measurements in the literature and find that multi-planet hosts are significantly more metal-poor than those hosting a single planet. Moreover, we find that host star metallicity decreases with increasing planet multiplicity. In contrast, we find only a modest difference in the rotation period. The significant differences in planetary composition and metallicity of the host stars point to different physical processes governing the formation of single- and multi-planet systems in M dwarfs.

\end{abstract}

\keywords{Exoplanet systems -- exoplanet structure -- exoplanet formation}

\section{Introduction} \label{sec:intro}

The NASA $Kepler$ \citep{Borucki:2010} and TESS missions \citep{Ricker:2015} have discovered hundreds of systems with multiple planets orbiting stars of almost every stellar type. At the time of writing, we know of $\sim$800 multi-planet systems\footnote{https://exoplanetarchive.ipac.caltech.edu/index.html} displaying a staggering diversity of architectures \citep{Winn:2015}. This large sample of multis enables statistically meaningful comparisons between systems with only one detected planet and those with 2 or more. Such a comparison can provide crucial insights into planet formation and evolution, as different formation mechanisms may be observably imprinted in the orbital and physical properties of the planets in these two different architectures. In addition, a comparison could help elucidate whether single- and multi- planet systems belong to the same underlying population (but perhaps have different dynamical histories), or to completely separate planetary populations.

Several authors have compared the properties of singles and multis\footnote{In this paper, we use the term `single' interchangeably to mean both 1) a planet in a system in which there is only one (detected) planet, and 2) a system in which a single planet is known.} and their host stars, although the majority have focused on FGK stars. \citet{Wright:2009} compared the distribution of multis and singles and found that the multis have lower masses and lower eccentricities. \citet{Morton:2014} found that singly transiting planets have higher obliquities than those in multiples. \citet{Limbach:2015} found that for radial velocity (RV) planets, eccentricity decreases with increasing planet multiplicity. Similarly, \citet{Xie:2016} and \citet{VanEylen:2019} found that singles have significantly higher eccentricities than multis. \citet{Latham:2011} compared a sample of multis and singles and found that small, sub-Neptunes are more frequent in multis (present in 86$^{+2}_{-5}$\% of multis) than in singles (69$^{+2}_{-3}$\%). \citet{Latham:2011} suggested that giant planets may disrupt the orbits of sub-Neptunes in flat systems, either ejecting them or altogether preventing their formation. All of the aforementioned lines of evidence seem to point to single planets having hotter or more `violent' dynamical histories than multis. More recently, \citet{Weiss:2018b} compared the planetary and stellar properties of singles and multis in FGK stars and found no substantial difference in the stellar mass ($M_{\star}$),  metallicity ([Fe/H]), and projected rotation speed ($v\sin i$), nor in the radii and orbital periods of the planets. \citet{Weiss:2018b} interpreted this lack of correlation as indicative of a common origin for multis and singles and concluded that host star properties are poor predictors of planet multiplicity, at least amongst FGK stars.

In this paper, we extend this line of research to planets orbiting M dwarfs. We examine how the fundamental stellar and planetary properties vary for planets in systems with only one planet versus those in multi-planet systems. \citet{Ballard:2016} was one of the first detailed studies of the $Kepler$ M-dwarf sample. They found an excess of single planetary systems, with roughly half of M-dwarf systems being singles and half multis. This implies that the so-called \textit{Kepler} dichotomy seemingly extends to M dwarfs. The \textit{Kepler} dichotomy is the discovery of an unexpectedly abundant population of singly-transiting systems, more than one would expect from transit geometry biases alone. It was originally discovered based on FGK stars only. Whilst the majority of planet discoveries and consequent studies have been for FGK stars, the most common stars ($\approx 70\%$) in the galaxy are the smallest ones: M dwarfs ($M_{\star} < 0.6 M_{\odot}$). M dwarfs actually have a higher occurrence rate of small planets \citep{Howard:2012,Mulders:2015,Dressing:2015,Ment:2023}. There is also intense community interest in M dwarfs, owing to a shorter-period habitable zone, in which it is easier to find and characterize potentially habitable worlds (e.g., TRAPPIST-1, \citealt{Gillon:2017} and TOI-700, \citealt{Gilbert:2020}). It is therefore crucial to understand both the architectures and compositions of M-dwarf planetary systems, given how they also relate to formation and habitability. 

Most of the existing comparative studies of exoplanet singles and multis have focused on the differences between the orbital and architectural properties --such as eccentricity, orbital period and mutual inclinations. In this paper, we focus on the compositional differences of exoplanets in singles and multis. To our knowledge, such a comparison is yet to be made in the literature, at least for M-dwarf planets. Evidence of a dichotomy in planetary composition between singles and multis could point to different formation or evolution mechanisms. We also consider here how the planet's composition is related to the star's chemical composition (specifically, the [Fe/H] metallicity), and how that might vary between singles and multis. If there are significant differences between the planetary composition of singles versus multiples, then it would be reasonable to expect differences in the properties of the parent stars themselves as well since we assume that they formed from the same protoplanetary disk and thus that planets will inherit some characteristics of their parent stars. On the other hand, similar stellar properties among the hosts of multis and singles would be indicative of a common formation mechanism or origin for these populations. This is something we test.

This paper is organized as follows. In section~\ref{sec:sample}, we describe our sample selection. In sections~\ref{sec:planet_properties} and~\ref{sec:stellar_properties}, we present a comparison between the properties of planets in multis and singles and their host stars. Finally, in section~\ref{sec:discussion}, we discuss the implications of our findings and conclude.

\section{Sample Selection} \label{sec:sample}

\begin{figure}[ht]
\vspace{.1in}
\centering
\includegraphics[width=1\linewidth]{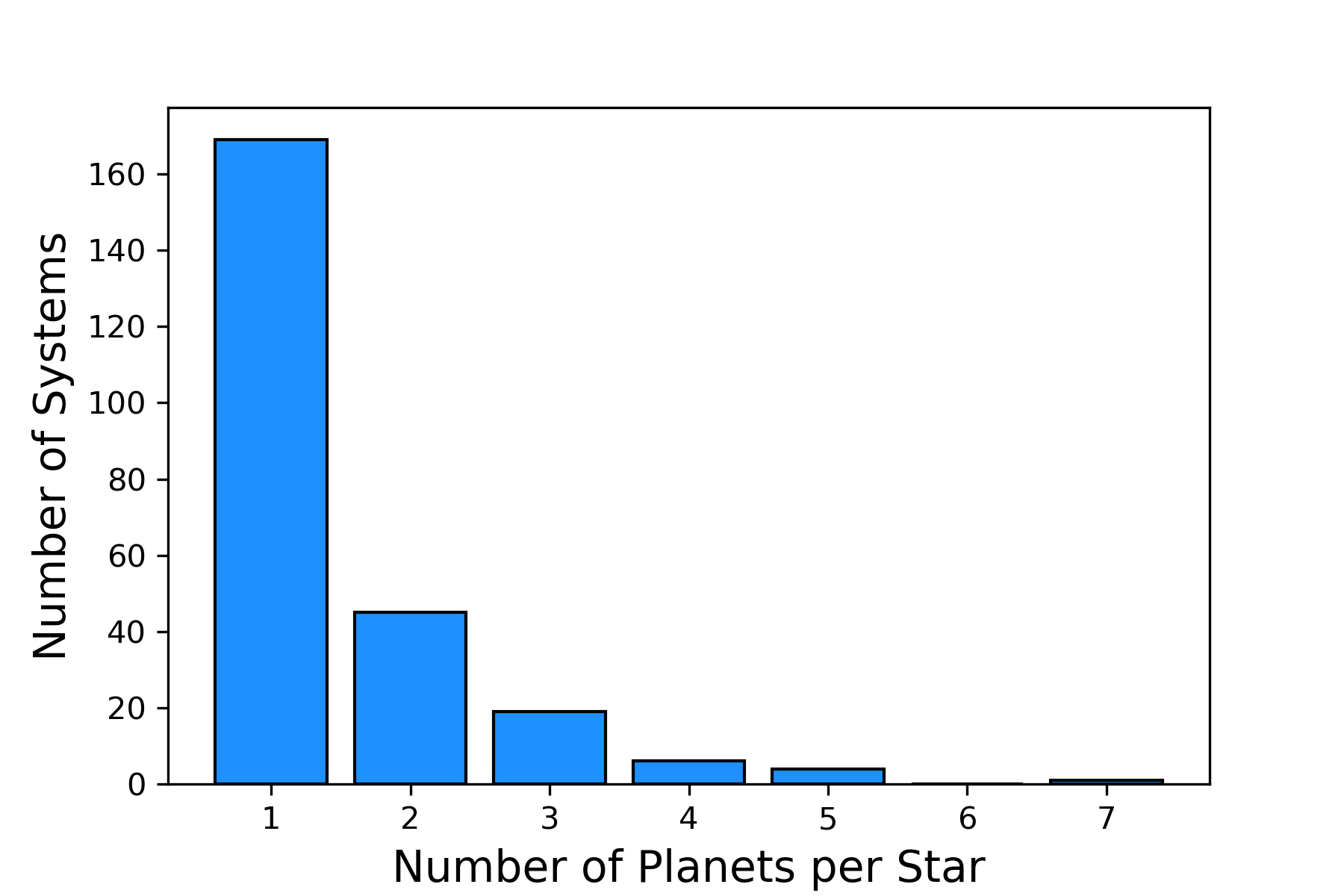}
\caption{Histogram showing the distribution of planet-hosting M dwarfs by the number of planets per star. There are currently 170 systems hosting a single (detected) planet and 75 systems with 2 or more planets. There are currently no known M dwarfs hosting 6 planets.}
\label{fig:total_systems} 
\end{figure}

We selected our targets from the publicly available NASA Exoplanet Archive \citep{Akeson:2013} on July 2022. In this study, we focus on planets orbiting M-dwarf stars, so we began by selecting all the stars with effective temperatures between 2450 K and 4000 K, following the spectral classifications by \citet{Pecaut:2013}. To avoid possible contamination by red giants, we additionally checked that all the stars in our sample had surface gravity values of  log$g_{*}$ $>$ 3.5. This search resulted in 170 M dwarfs hosting a single (detected) planet and 75 multi-planet hosts. Specifically, 45 stars host 2 known planets, 19 host 3, 6 host 4, 4 host 5 planets, and only one hosts 7 (TRAPPIST-1; \citealt{Gillon:2017}). See Figure~\ref{fig:total_systems} for a breakdown of the systems by number of planets.

We applied a series of quality cuts to this sample of M-dwarf planetary systems. First, we selected planets with directly measured masses, and thus we excluded any planets with only a minimum mass ($m\sin i$) or with masses determined with empirical mass-radius relationships. There were 100 exoplanets with direct masses. Of those, we excluded objects with masses above $\sim$13 $M_{\rm Jup}$, which may potentially be brown dwarfs. From the remaining sample of 91 planets, we additionally removed any objects with upper/lower mass limits, and objects with calculated (rather than measured) radii (i.e., non-transiting planets that have radii calculated by the NASA Exoplanet Archive using the mass-radius relationships from \citealt{Chen:2017}). We further discarded planets with unphysical bulk densities, $\rho_{p}$, which we define as anything that falls below the pure iron theoretical mass-radius curve\footnote{These ultra-dense planets could be Super-Mercuries (see, e.g. \citealt{Santerne:2018}, \citealt{Adibekyan:2021}, \citealt{Schulze:2021}, \citealt{RodriguezMartinez:2023}, and \citealt{Barros:2022}), but they have relatively large mass uncertainties so more observations are needed to robustly determine their composition.} (using the planet interior models from \citealt{Zeng:2019}). These planets were: Kepler-54c with $\rho_{p}=$ 58~\dens~ \citep{Steffen:2013}, Kepler-231b with $\rho_{p}=$ 37~\dens~\citep{Rowe:2014}, and Kepler-327c with $\rho_{p}=$ 100~\dens~\citep{Rowe:2014}. This leaves a total of 65 planets. Because of the relatively small sample, we did not impose a cut in planetary mass/radius uncertainty.

\begin{figure}[ht]
\vspace{.1in}
\centering
\includegraphics[width=1\linewidth]{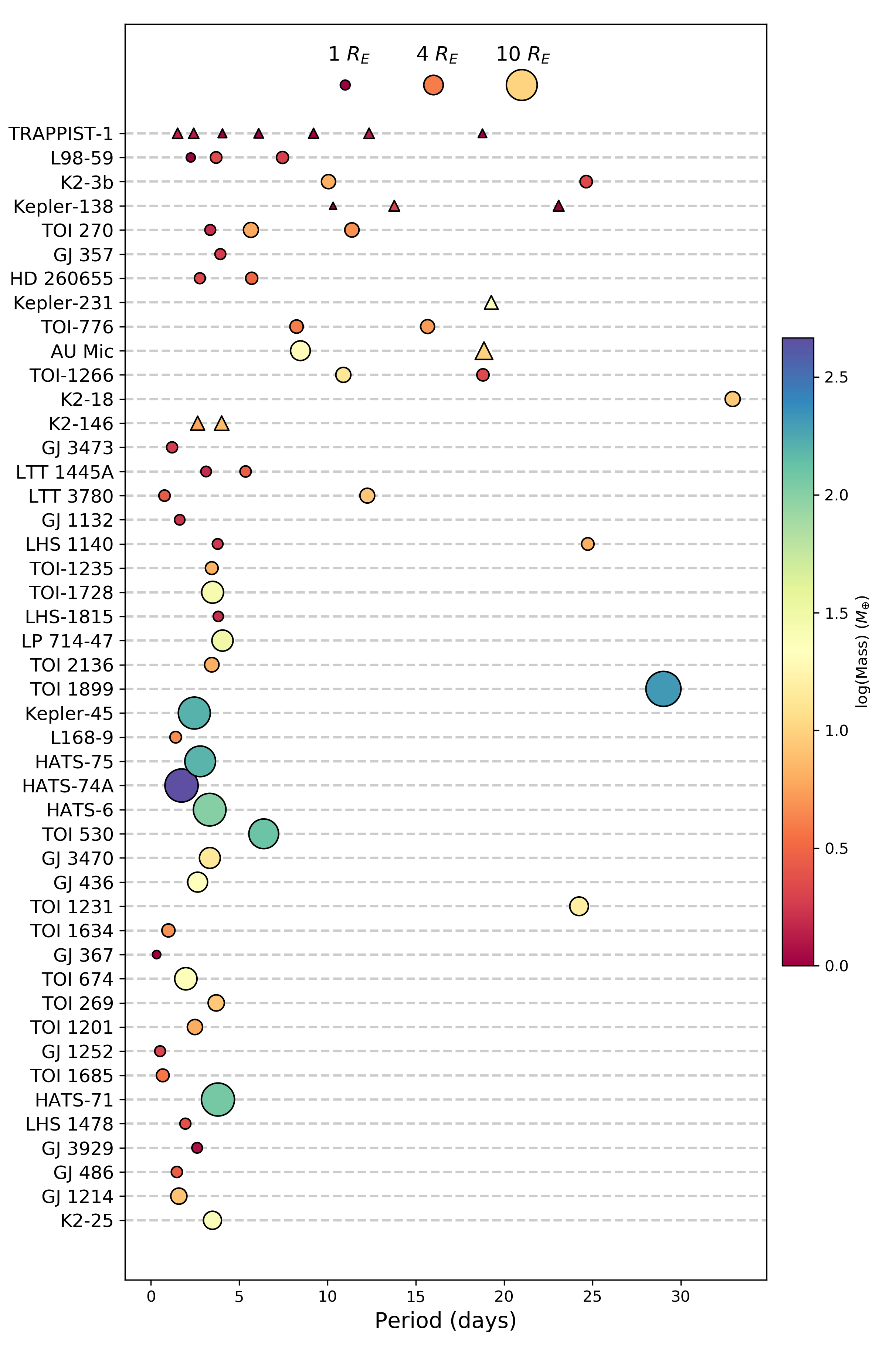}
\caption{Orbital architectures of representative planetary systems in our sample. The x-axis shows the orbital periods. The symbol sizes are proportional to the planets' radii and they are color-coded by planet mass. The circles are planets with masses determined from radial velocity measurements while the triangles represent planets with masses from transit timing variations (TTV). All of them are transiting exoplanets. They are sorted by planet multiplicity from top to bottom, with systems with a higher number of planets at the top.}
\label{fig:abacus} 
\end{figure}

To this sample we added six more exoplanets with revised masses and radii from the sample of \citet{Luque:2022}. This brings the total number of planets in our sample with measured masses and radii to 70. They range in mass from 0.06 to 2002 $M_{\oplus}$ (6.2$M_{Jup}$) and in radius from 0.5 to 14 $R_{\oplus}$. We did not make a cut based on stellar multiplicity. Several planets have outer stellar companions, but we do not have any circumbinary planets in our sample. In addition, K2-18b has another planet candidate in the system (K2-18c) but it has been flagged as ``controversial" by the NASA Exoplanet Archive and therefore, out of caution, we treat K2-18b as a single.\footnote{K2-18c has been recently validated by \citet{Radica:2022}, however, we do not include it in our analysis since it does not have a radius measurement.} We note that for the TOI-1685 system, \citet{Bluhm:2021} find tentative evidence for a second planet, but we also treat this system as a single in our analysis. The same is true for TOI-1201 and GJ 436. Our final sample is shown in Figure~\ref{fig:abacus} and their properties are listed in Table~\ref{table:planet_props}.

Throughout the paper, there are occasions when we make an additional cut of $R_{\rm p}<6R_\oplus$ in the planet sample, such that we can see how the gas giants potentially skew any distributions. As with the planet sample, we sometimes also split the stellar sample up as a function of planet radius, where we separate the $R_{\rm p}<6R_\oplus$ planets.


\section{Comparison of Planetary Properties} \label{sec:planet_properties}

\subsection{Bulk Density} \label{subsec:tables}

\begin{figure*}[!ht]
	\centering\vspace{.0in}
	\includegraphics[width=\columnwidth, clip]{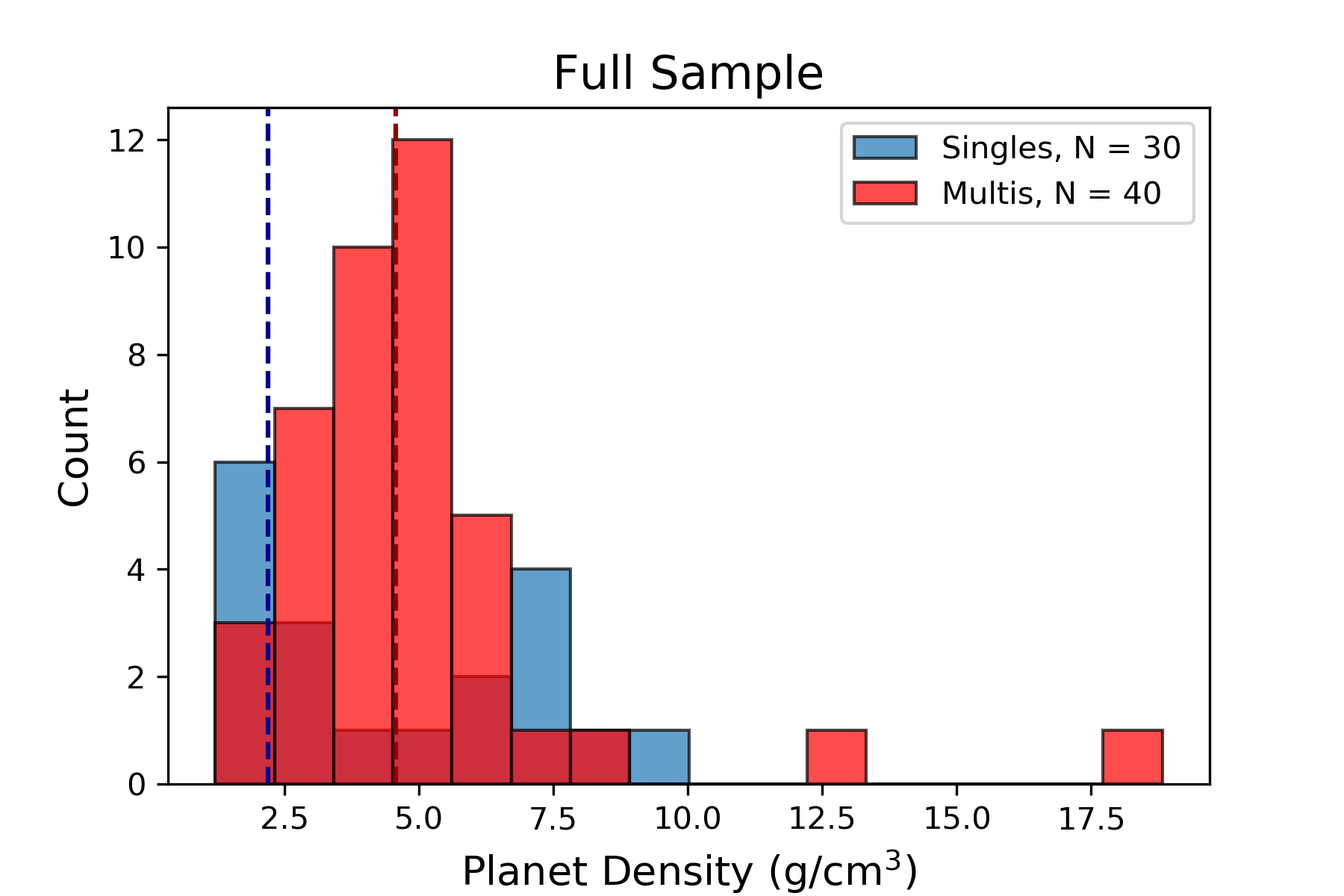}\includegraphics[width=\columnwidth, clip]{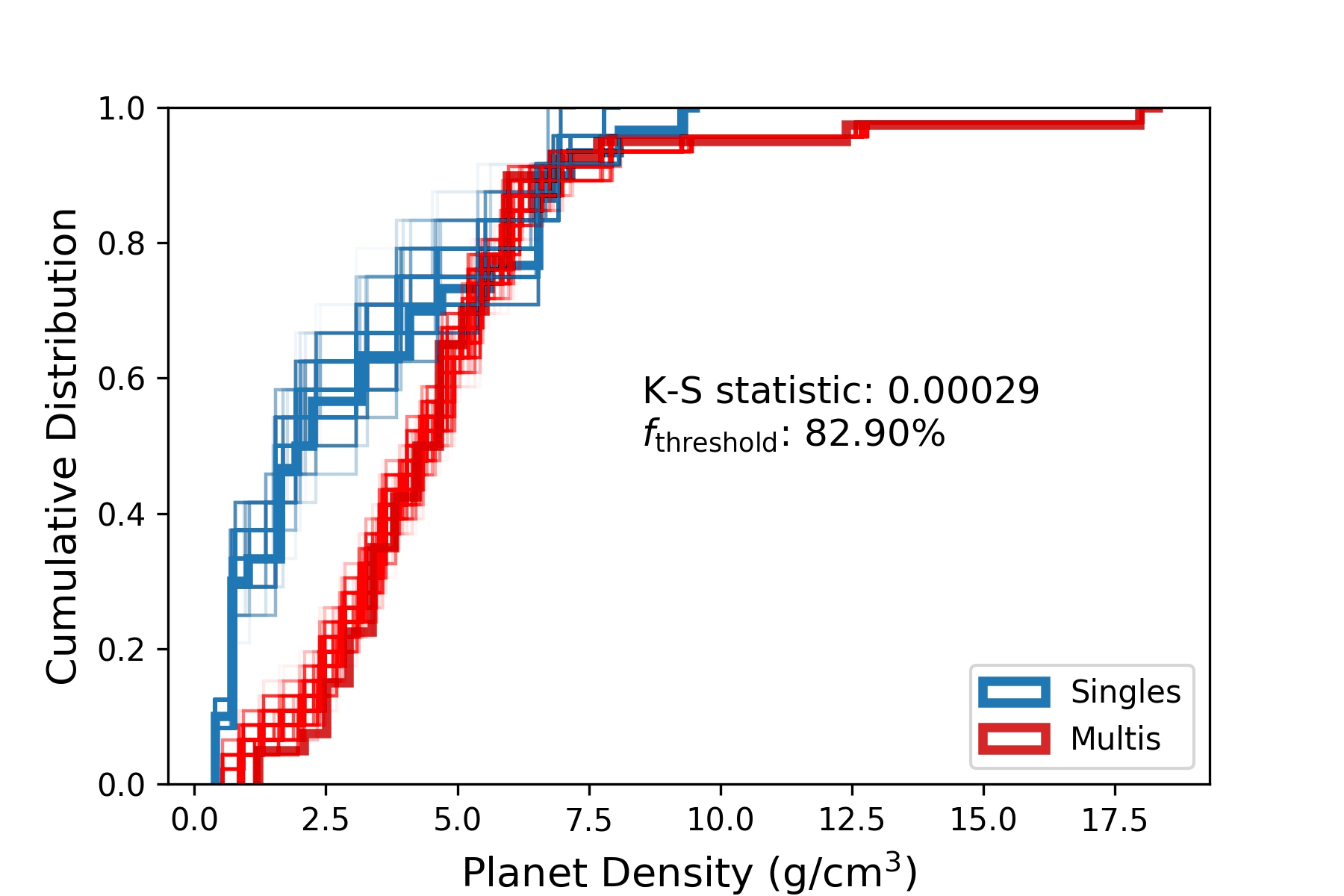} \\
	\includegraphics[width=\columnwidth, clip]{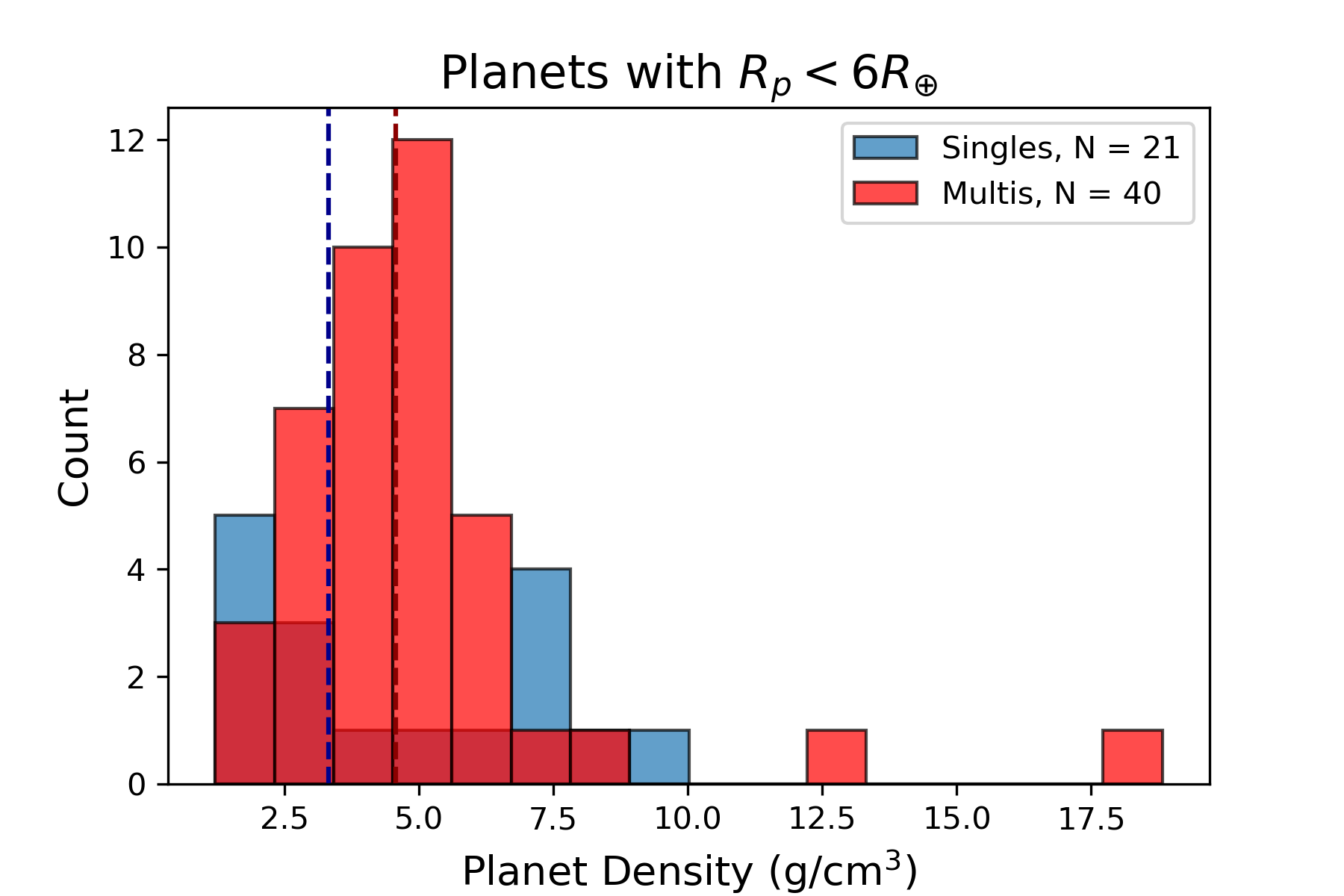}\includegraphics[width=\columnwidth, clip]{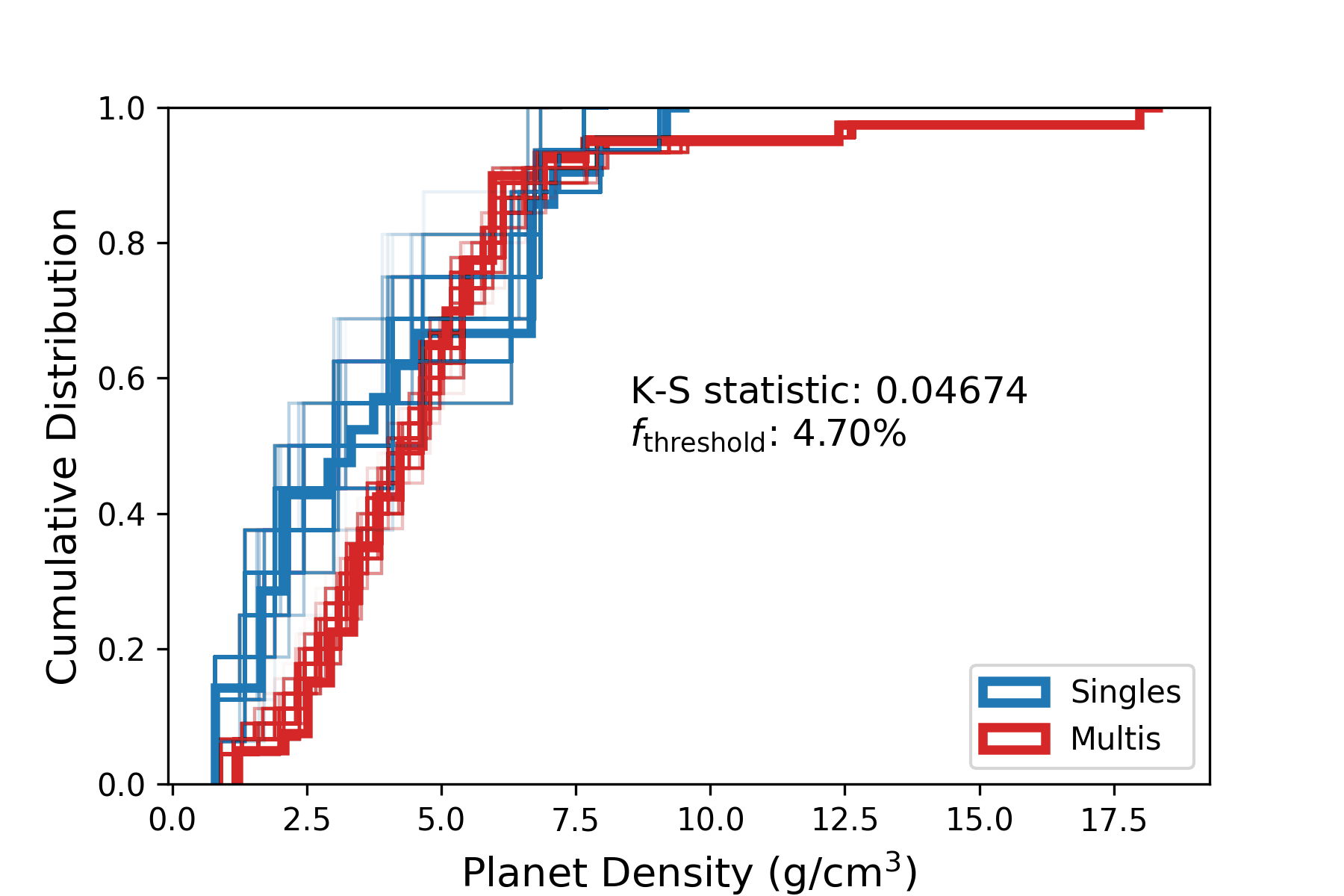}
    \caption{\textbf{Left}: Histograms of the bulk densities of planets in single-planet systems (blue), and multi-planet systems (red) for our entire sample of planets (\textbf{top}) and for planets with $R_{p}<6R_{\oplus}$ (\textbf{bottom}). The red dashed line shows the median density of the multis while the blue dashed line shows the median density of the singles. \textbf{Right}: Bootstrapped empirical cumulative distribution functions (CDF) for our entire sample of planets (\textbf{top}) and for planets with $R_{p}<6R_{\oplus}$ (\textbf{bottom}). The Kolmogorov-Smirnov statistic, or $p$ values, of the original distributions are overplotted. $f_{\rm threshold}$ is explained in Section~\ref{selectionbias}. We note that removal of the dense outliers at $\sim$12~\dens~and $\sim$18~\dens~does not considerably affect the observed differences in the density distributions.}
	\label{fig:densities_cdf}
\end{figure*}

\begin{figure*}[!ht]
	\centering\vspace{.0in}
	\includegraphics[width=\columnwidth, ]{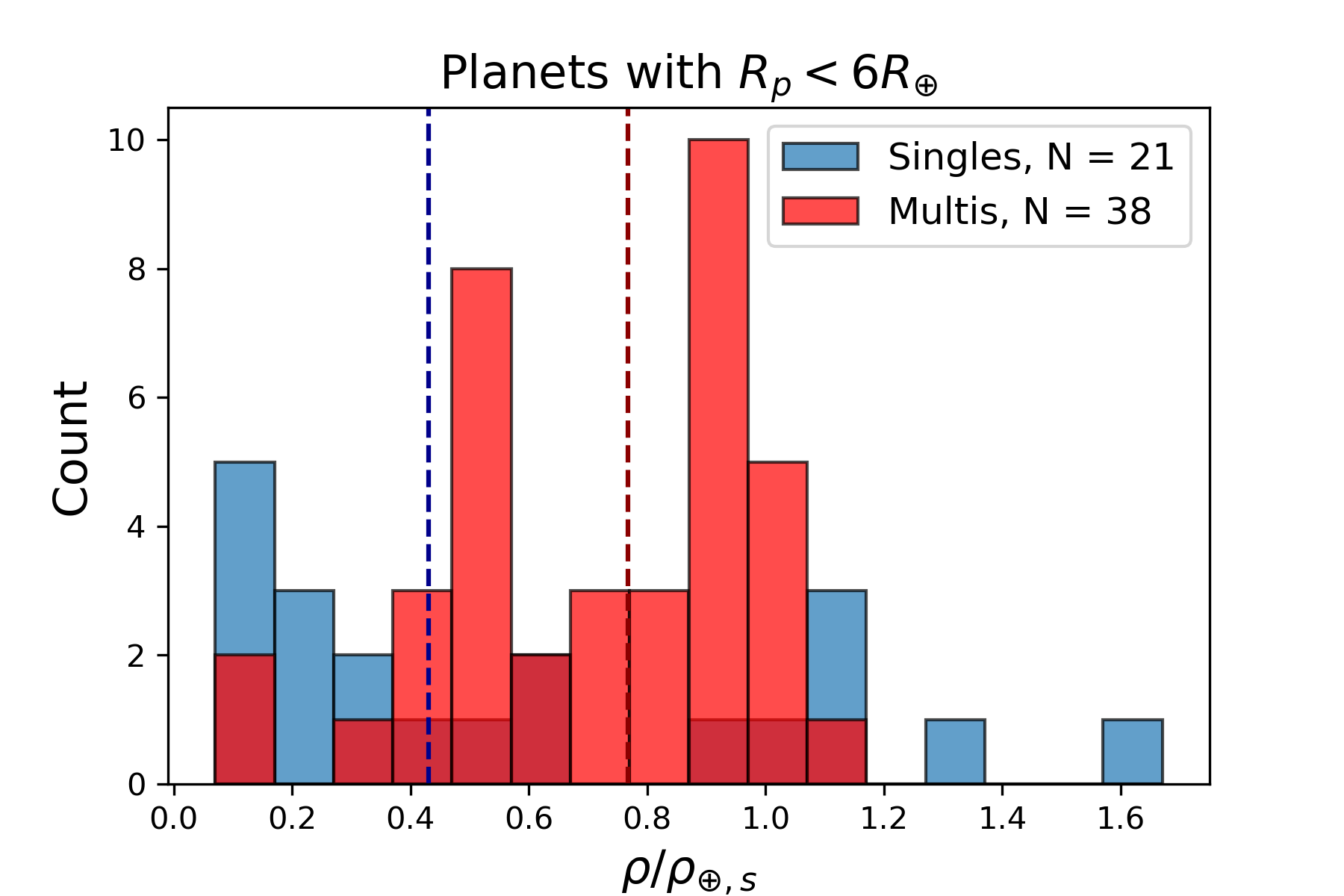}\includegraphics[width=7.3cm, height=5.2cm]{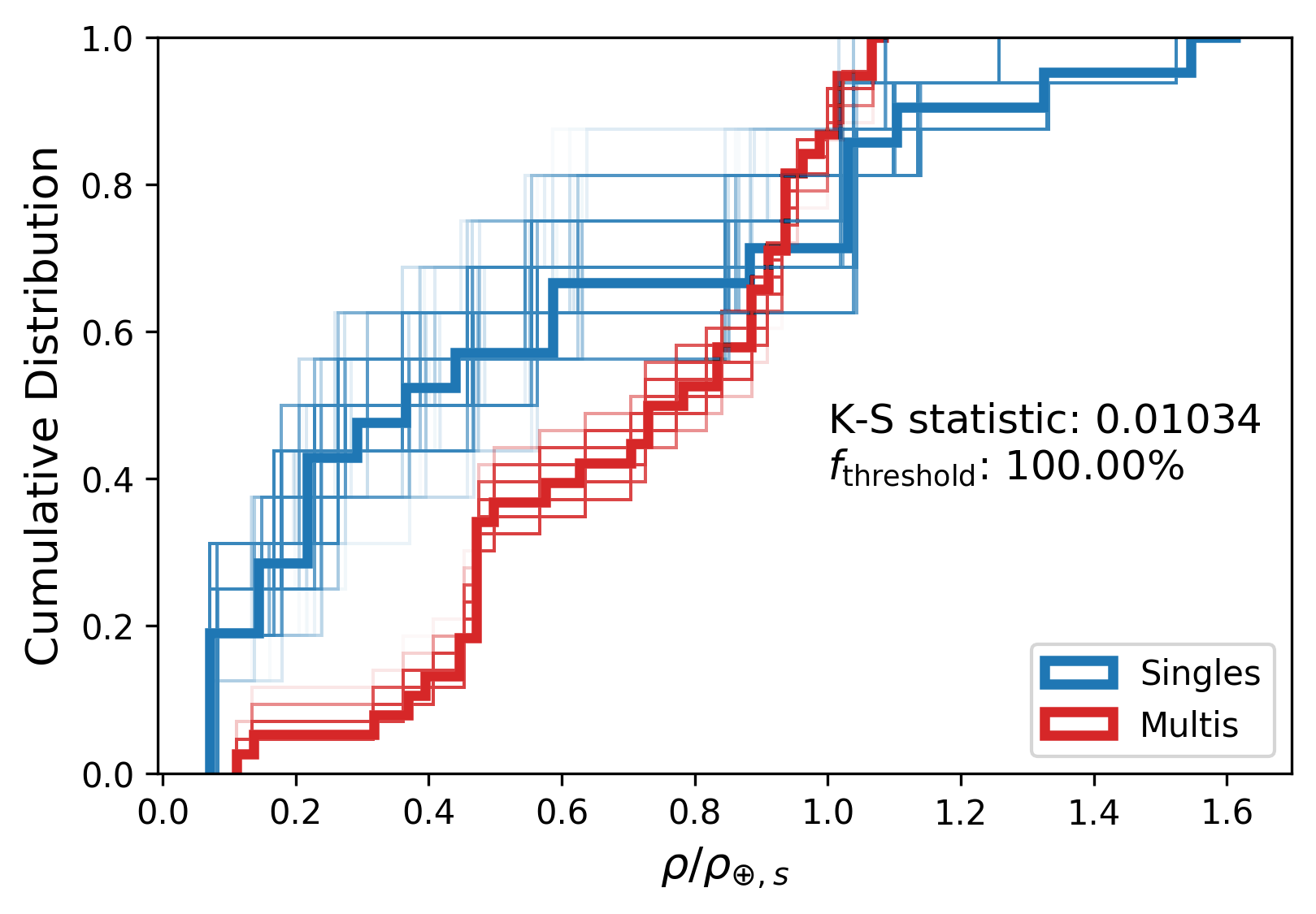} 
    \caption{\textbf{Left}: Histograms of the bulk density normalized by the density that an Earth-like planet (i.e., with $\sim$0.30 core mass fraction) would have at that mass. Single planets are shown in blue and multis in red; all planets have $R_{p}<6R_{\oplus}$. \textbf{Right:} bootstrapped cumulative distributions. $f_{\rm threshold}$ is explained in Section~\ref{selectionbias}.}
	\label{fig:scaled_densities}
\end{figure*}

We analyzed the differences in bulk density --which is a first-order approximation for composition-- of planets in singles and in multi-planet systems. The densities were taken from the NASA Exoplanet Archive and they range from 0.8 to 18~\dens. As shown in the left panel of Figure~\ref{fig:densities_cdf}, there seems to be a clear difference between the distribution of densities of singles and multis: singles have lower densities on average compared to planets in multis. The median density of the singles is $\rho_{p}=$ 2.2~\dens~while the median density of planets in multis is $\rho_{p}=$ 4.57~\dens. For reference, the density of the Earth is $\rho_{\oplus}=$ 5.5~\dens. 

To test whether the difference in the bulk densities of planets in these two architectures is statistically significant, we performed a Kolmogorov-Smirnov (K-S) and an Anderson-Darling (A-D) test for comparison, which test whether two data sets belong to the same underlying normal distributions. With a p-value of $p=0.0002$ from the K-S test and $p=0.001$ from the A-D test, we find that the densities of singles and multis belong to different populations with over 99\% confidence.

We suspect that this big difference between the densities of multis and singles, however, is mostly due to the presence of single giant planets, which can be easily seen in Figure~\ref{fig:abacus}. This may be driving the density distributions because a substantial fraction of single planets are massive and therefore likely to be gaseous and have low densities. We were therefore interested in making the same density comparison but focusing on smaller planets, which can be expected to have similar formation pathways. We restricted the sample to smaller planets with $R_{\rm p} < 6R_{\oplus}$. This radius cut removed 9 planets from our full sample: HATS-71b \citep{Bakos:2020}, COCONUTS-2b \citep{Zhang:2021}, TOI-530b \citep{Gan:2022}, TOI-3714b 
\citep{Canas:2022}, HATS-6b \citep{Hartman:2015}, HATS-74 Ab \citep{Jordan:2022}, HATS-75b 
\citep{Jordan:2022}, Kepler-45b \citep{Johnson:2012}, and TOI-1899b \citep{canas:2020}. For this subset of planets, we find a less strong difference in their bulk density, which is to be expected, since we have now removed larger --and thus generally lower-density-- planets. The planets in multis have a median density of $\rho_{\rm p} = 4.57$ \dens~and the singles have a median of $\rho_{\rm p} = 3.32$ \dens. The A-D and K-S statistics of this restricted sample yield $p$-values of 0.025 and 0.046, respectively. These results are significant at the 2$\sigma$ ($>$95\%) confidence. The right panel of Figure~\ref{fig:densities_cdf} shows their cumulative distributions. We discuss potential selection biases that may affect these differences in density in singles and multis in Section~\ref{selectionbias}. We note that the radius cut at $R_{\rm p} = 6R_{\oplus}$ is rather arbitrary, and we therefore reproduce our analysis with a radius cut at $R_{\rm p} = 4R_{\oplus}$, which is another commonly used size threshold in the literature. Excluding planets with $R_{\rm p} \geq 4R_{\oplus}$ (15 planets), we get a K-S and A-D statistic of 0.12 and 0.22, respectively. In other words, the bulk density distributions of singles and multis smaller than $4R_{\oplus}$ are statistically indistinguishable.

In addition to comparing bulk densities, we also consider another compositional proxy commonly used in the literature. Namely, we calculate the bulk density normalized by an Earth model, denoted $\rho/\rho_{\oplus,s}$. This parameter is defined as the bulk density of the planet divided by the density that an Earth-like planet (i.e., a rocky planet with $\sim$0.32 CMF) would have at the mass of the planet under consideration. We compute these density ratios only for planets in our sample with $R_{p} < 6R_{\oplus}$, since larger planets are unlikely to be terrestrial and therefore a density scaled by an Earth model is not a useful way to characterize them. For this small-planet sample, we obtain scaled density ratios ranging from 0.07 to 1.62. When expressed in these units, the density distributions of multis and singles differ significantly, with K-S and A-D p-values of 0.01 and 0.005, respectively, implying that they belong to different underlying populations. This contrasts with our high p-values when using the original bulk densities. Figure~\ref{fig:scaled_densities} shows the density distribution histograms and the cumulative distributions.

\begin{figure}[ht]
\centering
\includegraphics[width=1\linewidth]{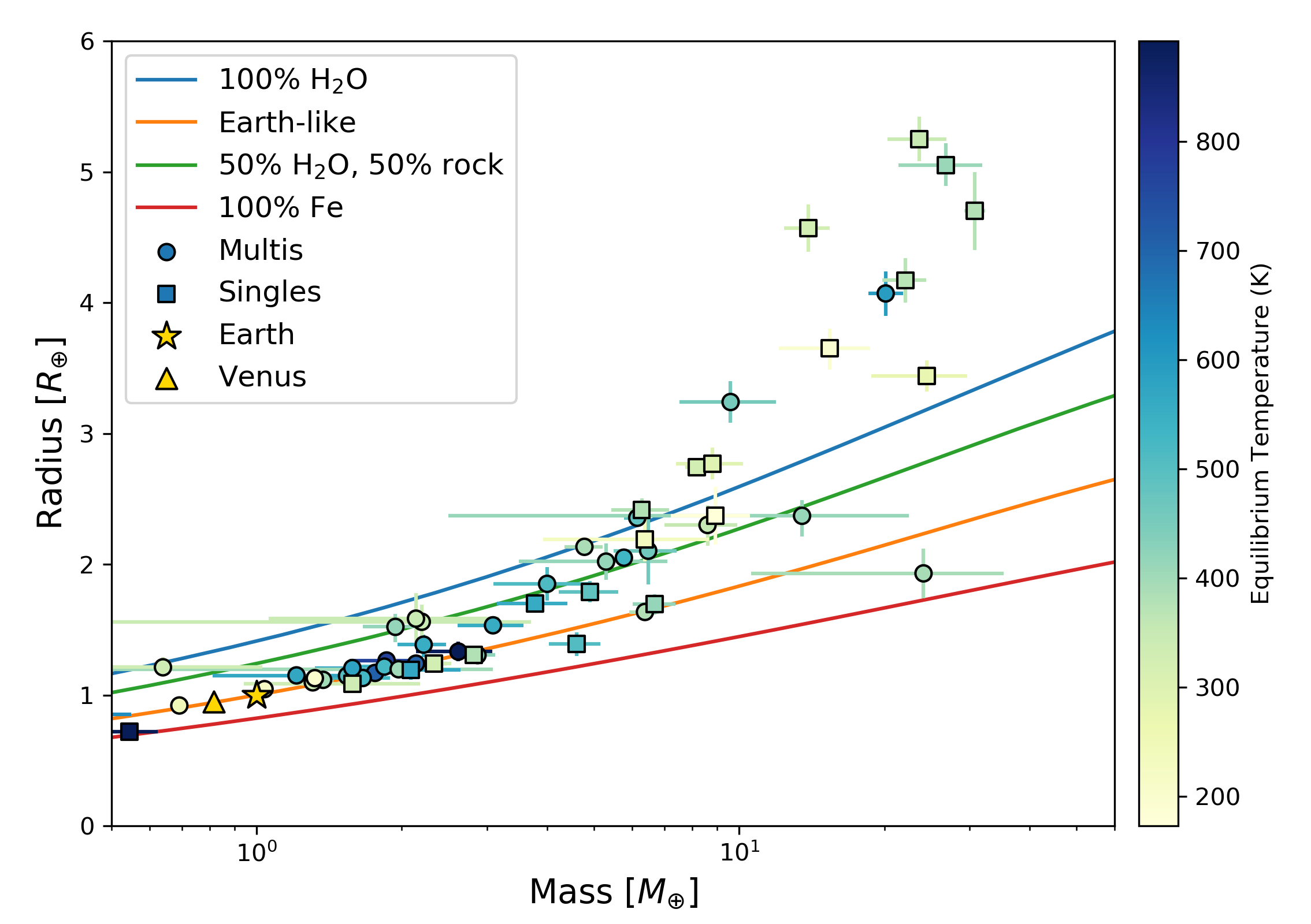}
\caption{Mass-radius diagram of the planets in our sample with $R_{p} < 6R_{\oplus}$. The colored lines represent mass-radius composition curves from \citet{Zeng:2019}. Earth and Venus are also plotted for reference. The circles represent planets in multi-planet systems and the squares are singles. The majority of planets above the 100\% water line are single.}
\label{fig:MRdiagram} 
\end{figure}

Figure~\ref{fig:MRdiagram} shows the mass-radius diagram of this sub-sample of small planets with $R_{p} \lesssim 6 R_{\oplus}$ and $M_{p} \lesssim 100 M_{\oplus}$ and the composition curves from the planet interior models of \citet{Zeng:2019}. The circles represent planets in multis and the squares are apparently single planets. The majority of single planets are clustered near or above the 50\% $\rm H_{2}O$/50\% magnesium silicate (rock) composition lines, likely requiring large fractions of water or substantial H/He envelopes to explain their observed low-density. This is in agreement with our finding that they are generally less dense than multis (see Figure~\ref{fig:densities_cdf}). In contrast, the majority of the planets in multis are either terrestrial or water-worlds (falling neatly on the 50\% $\rm H_{2}O$/50\% or Earth-like lines as was previously shown by \citealt{Luque:2022}). Thus, multis seem more likely to be rocky whereas giant planets or sub-Neptunes tend to be single. This picture is compatible with the phenomenon observed in sunlike stars that hot Jupiters tend to be isolated \citep{Steffen:2012}. 

We note that there were two planets in our sample that have unusually high masses and densities that are not expected for rocky planets from planet formation theories. These are Kepler-54b ($\rho_{\rm p} = 12.8$ \dens, \citealt{Hadden:2014}) and Kepler-231c ($\rho_{\rm p} = 18.4$ \dens, \citealt{Hadden:2014}). These were both discovered and characterized using transit timing variations (TTV) and have relatively large mass uncertainties of $\sim$25\% (Kepler-54b) and $\sim$50\% (Kepler-231c). Thus, it is likely that these planets are actually less massive than observations suggest. Despite the relatively large mass uncertainties, we included these planets in our analysis, since we did not make any cuts based on mass/radius uncertainty given the relatively small sample size. However, we note that removing these two planets from our analysis would not significantly affect our results: without them, the K-S statistic or $p$-value only changes by 0.0001 for the full sample, and 0.01 for the sample of small planets.

\subsection{Planetary Core/Water Mass Fractions}

Previous studies of multi-planet systems have revealed that planets within the same system tend to have similar radii, orbital spacings \citep{Weiss:2018a}, and masses \citep{Millholland:2017, Goyal:2022}, such that they are like ``peas in a pod". Put another way, planets within the same system have similar \textit{bulk densities}. This might seem to imply by extension, that their compositions ought to be similar as well. However, a planet's bulk density is degenerate with respect to composition. This degeneracy is reduced when the planet can be assumed to be rocky (i.e., density greater than pure silicate). For such planets, the primary constraint on bulk density is the planet's mass and core mass fraction \citep{Unterborn:2016}, which is defined as the mass of the core divided by the total mass of the planet as 

\begin{equation}
   \rm CMF = M_{core}/M_{planet}
\end{equation}

The CMF plays an important role in the existence and lifetime of a magnetic field, which is, in turn, important for habitability, as it may shield the planet from high-energy radiation and reduces atmospheric loss \citep{Driscoll:2015,Green:2021}.  Although studies have revealed intrasystem uniformity in terms of mass and radius, whether or not the interior structure of planets within the same system are similar, and how it varies among singles and multis, remains an open question.
\begin{figure}[ht]
\vspace{.1in}
\centering
\includegraphics[width=1\linewidth]{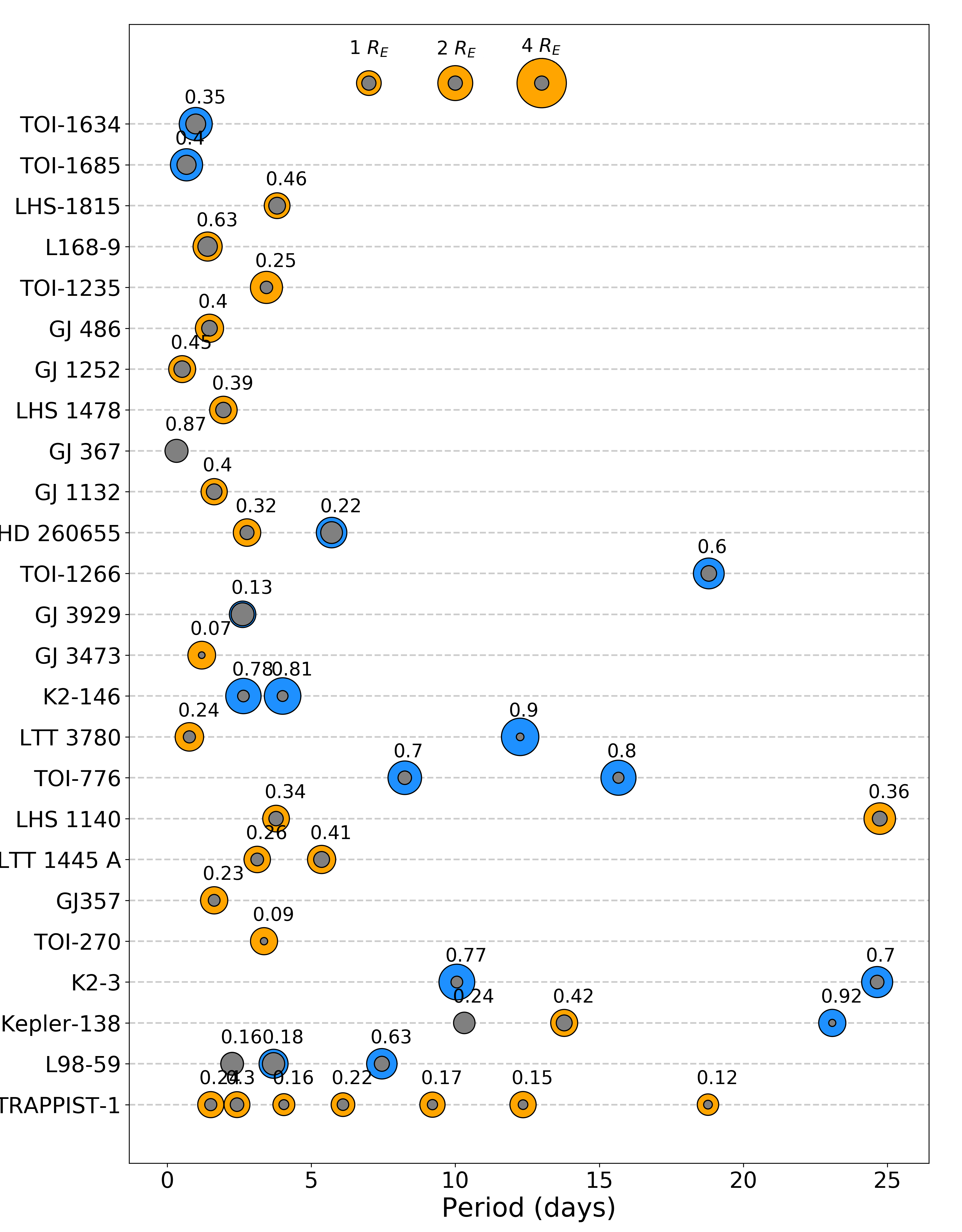}
\caption{Representation of the interior structure and core/water mass fractions for low-mass planets in our sample with $M_{p} < 10M_{\oplus}$. The gray circles are the liquid iron cores and the blue and orange circles represent water and rocky layers, respectively. The relative sizes of the circles are proportional to the planet/core/water masses. The numbers above each planet are the derived water/core mass fractions. The systems are sorted by planet multiplicity from low (top) to high (bottom) multiplicity.}
\label{fig:cmf_abacus} 
\end{figure}

We therefore now turn our attention to this property and calculate the CMFs of all planets in our sample with $M_{\rm p} < 10 M_{\oplus}$ and densities greater than pure silicate at their mass. To do this, we used the open-source ExoPlex package \citep{Unterborn:2018,Unterborn:2019, Unterborn:2023}, which self-consistently solves the equations of planetary structure including the conservation of mass, hydrostatic equilibrium, and the equation of state. We use the ExoPlex parallel mass-radius script, which solves for the Fe/Mg distribution from a planet's observed mass and radius distributions, and calculate the CMF per Eq. 8 of \citet{Unterborn:2023}. We model planets with densities greater than pure silicate with a pure, liquid iron core and an iron-free magnesium silicate ($\rm MgSiO_{3}$) mantle. Planets with densities less than pure silicate are modeled as ``water worlds". This approach is reasonable given the findings of \citet{Luque:2022}, who revised the properties of a large sample of M-dwarf planets and identified a large population of water-rich planets. We solve for the water mass fraction (WMF) of each water world by fixing the rocky interior to have 29\% iron core and 71\% silicate mantle by mass. Given the degeneracies of the three-layer (Fe-core, silicate mantle, water) models, studies generally assume the rocky interiors of terrestrial planets to either be Earth-like or have relative amounts of Fe, Mg, and Si that reflect their host star’s photospheric abundances of these materials (e.g., \citet{Unterborn:2023} and references therein). Most of the stars in our sample do not have abundances beyond [Fe/H] and so we assume their compositions, and therefore the compositions of their planets, reflect the galactic stellar composition, for which we use the values calculated in \citet{Unterborn:2023} using the Hypatia Catalog Database \citep{Hinkel:2014}, which yield an average rocky interior that is 29\% Fe core and 71\% $\rm MgSiO_{3}$ mantle. To constrain WMFs, Exoplex uses the Seafreeze software, which calculates the physical properties of water like the phase and thermoelastic parameters. See section 2.4 of \cite{Unterborn:2023} for a more thorough description. See also \cite{Aguichine:2021} for a more detailed exploration of water-world interior modeling.

We calculate core/water mass fractions for a total of 41 exoplanets in our sample, of which 24 are consistent with being rocky and 17 water-rich. For the likely rocky planets, we find that the core mass fractions range from 0.07 to 0.87. Similarly, for the potential water worlds, we find water mass fractions ranging from 0.13 to 0.92. Our core/water mass fraction values are listed in Table~\ref{table:planet_props}. We note that measuring core mass fractions precisely is difficult, and it is often more reliable for well-measured exoplanets with mass and radius fractional uncertainties of $\Delta m/m \leq 20\%$ and $\Delta r/r \leq 10\%$, as noted in \citet{Schulze:2021}. Since some of the planets in our sample have relatively large mass and radius uncertainties, their CMF estimates have correspondingly significant uncertainties. Figure~\ref{fig:cmf_abacus} shows the orbital architectures and core/water mass fractions of the planets in our sample. The orange and blue circles are rocky and water worlds, respectively. 

We make note of the compositionally diverse system Kepler-138. This multi-planet system consists of three small planets with relatively low densities. We model the innermost planet, Kepler-138b as a rocky planet and obtain a CMF of 0.24. We model planets c and d as water worlds and obtain WMFs of  0.42 and 0.92, respectively. \citet{Jontof-Hutter:2015} determined the masses and radii of these planets and inferred that, given its extremely low density, Kepler-138d must have a deep H/He envelope. Recently, \citet{Piaulet:2023} updated the properties of the system and found that, although Kepler-138d is small (1.5 $R_{\oplus}$), it must have a substantial amount of water, with at least $11^{+3}_{-4}$\% volatiles/water by mass or $\sim$51\% by volume. We find a much higher WMF for Kepler-138d because we adopted the mass and radius of \citet{Jontof-Hutter:2015} instead. Regardless, this is an interesting system that shows that small planets can be significantly volatile-rich and that there can be a wide range of compositions within systems of similarly sized worlds.

We compare the core mass fractions of the singles and multis for the likely terrestrial planets in our sample. Excluding planets with orbital periods longer than 5 days, we find that the error-weighted average of the CMFs of the singles and multis is 0.51 $\pm$ 0.07 and 0.23 $\pm$ 0.02, respectively. This appears to be in slight tension with our finding that the multis are denser than singles on average, given that one would naively expect denser planets to have higher core mass fractions. However, this is because the sample we used to compare bulk densities is a different sample than that of planets with CMF measurements. The former includes planets with $R_p < 6R_{\oplus}$, while the latter includes only low-mass planets with $M_{p}<10M_{\oplus}$, which in our sample corresponds to a radius range of $R_p <1.6 R_{\oplus}$. We do not compare the WMFs of singles and multis, since the vast majority of water worlds reside in multi-planet systems.

\begin{figure*}[!ht]
	\centering\vspace{.0in}
	\includegraphics[width=\columnwidth, clip]{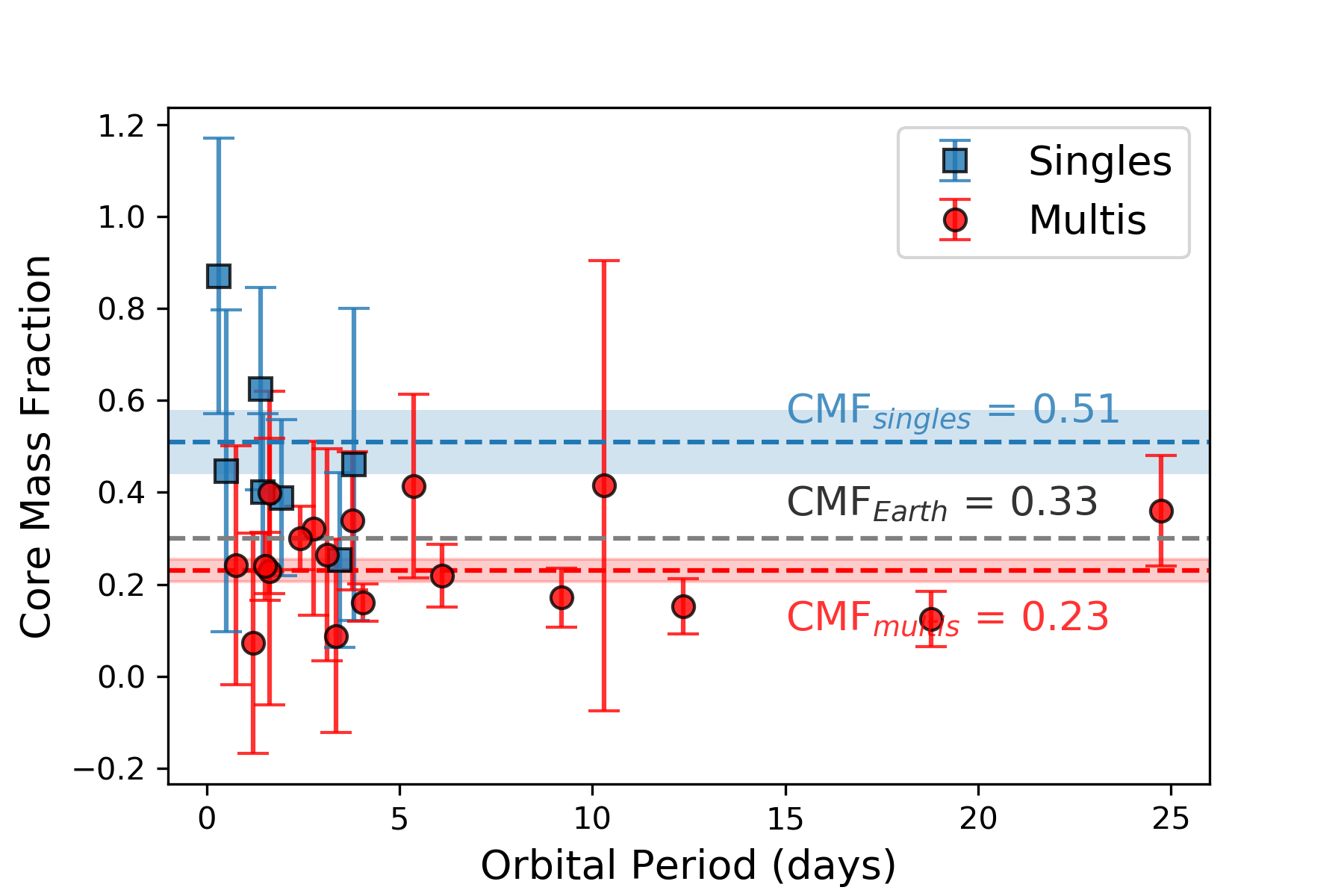}\includegraphics[width=\columnwidth, clip]{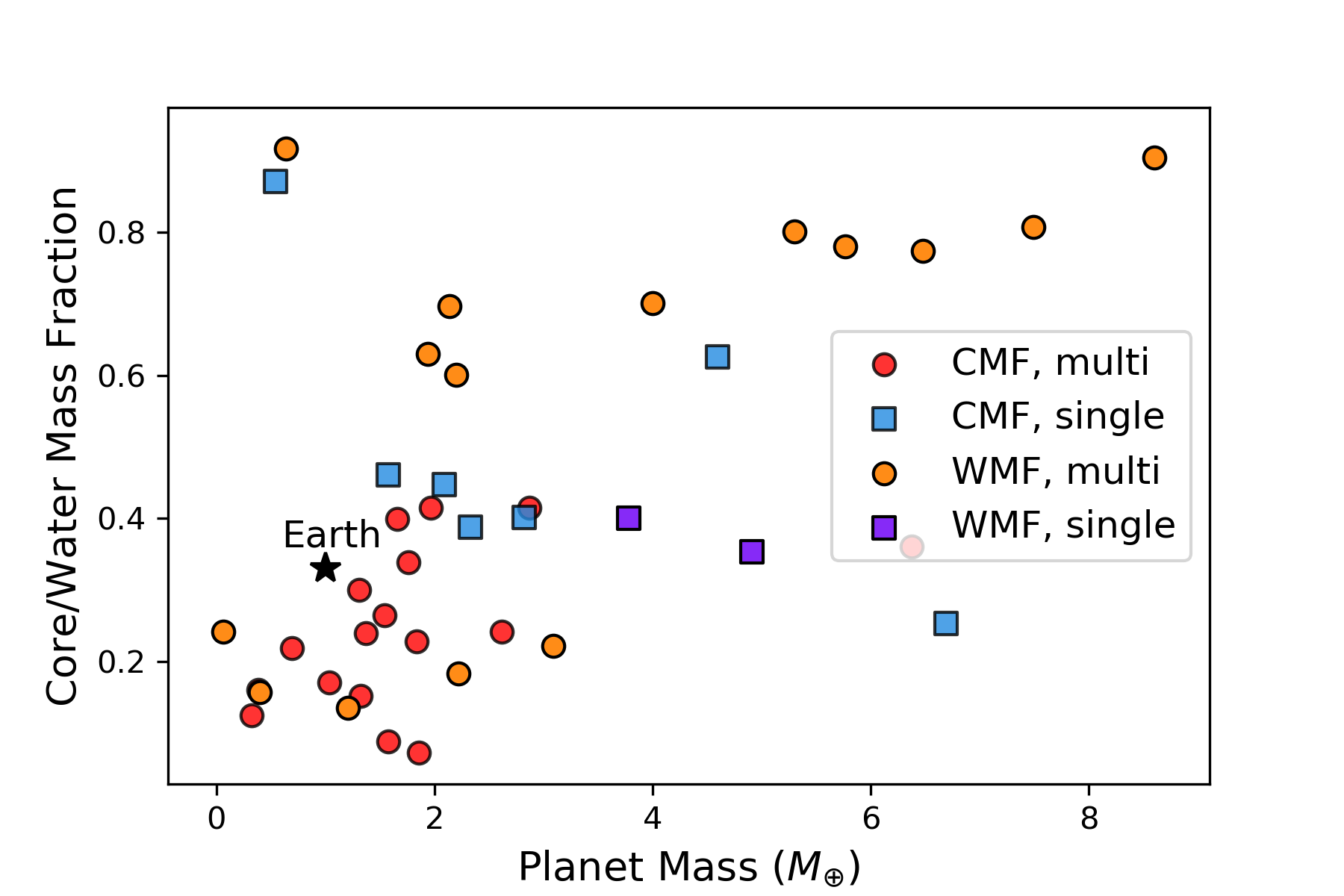}
    \caption{\textbf{Left:} Core mass fraction as a function of orbital period for the 24 likely rocky planets in our sample with $M_{p}<10 M_{\oplus}$. The blue, and red horizontal dashed lines are the weighted average CMFs of the singles and multiples, respectively. We do not find any correlation between CMF and period. The red and blue bands are $1\sigma$ confidence intervals on the mean.  \textbf{Right:} Core/Water mass fractions as a function of planet mass. The circles are multis and squares are singles. The Earth, with a CMF of 0.33, is plotted for reference in both plots.}
	\label{fig:CMF_massperiod}
\end{figure*}

We also explore the correlation between CMF and semi-major axis, or orbital period. In Figure 5 of \citet{RodriguezMartinez:2023}, we show a trend with density and equilibrium temperature of known planets with $M_{\rm p}< 10 M_{\oplus}$: the densest planets appear to have higher equilibrium temperatures (or equivalently, smaller semi-major axes) than volatile-rich planets.  This may be a result of photoevaporation: short-period planets become smaller and denser after their primordial H/He envelopes are evaporated due to the high irradiation from their host stars \citep{Owen:2017}. Alternatively, this may support the theory that the densest planets form in the inner regions of the protoplanetary disk where there may be a greater supply of iron compared to other elements \citep{McDonough:2021}. One potential observational signature of this is that the CMF would be inversely proportional to distance from the host star, since CMF is essentially the amount of iron in a planet. However, as shown in the left panel of Figure~\ref{fig:CMF_massperiod}, we do not see any trends between CMF and orbital period, at least by eye, for either singles or multis. This agrees with the findings of \citet{Plotnykov:2020}, who measured the core mass fractions of a sample of rocky planets and did not see any correlation between CMF and insolation (which is proportional to orbital period), shown in Figure 5 of their paper. One might also expect to see a gradual decrease (or gradient) in iron, or equivalently, in CMF, as a function of distance \textit{within a given planetary system}. In other words, we might expect the innermost planets in multi-planet systems to be more iron-rich than their outer siblings. In fact, \citet{Barros:2022} characterized the HD 23472 system, which contains 5 small planets orbiting a K-dwarf, and they found a clear decrease in the core mass fractions of the planets with increasing period. Similarly, they found that the water and gas mass fraction increases as a function of period such that the innermost planets are dry and dense and the outermost ones have much larger fractions of water and gas.
However, we do not see a trend with CMF and orbital period within individual planetary systems either, as shown in Figure~\ref{fig:cmf_abacus}. Regarding the question of whether there is intra-system uniformity in terms of core mass fractions, (or whether CMFs are similar for exoplanets in the same system), we observe a variety of CMF values within multi-planet systems, which possibly argues against any uniformity (see Figure~\ref{fig:cmf_abacus}). However, we note that there are less than ten muli-planet systems in our sample with two or more CMF or WMF measurements, so it is hard to quantify the uniformity given the small sample size. 

Exploring the connection between CMF and orbital period or insolation may inform theories of planet formation and evolution and therefore it is worth investigating this further, perhaps by extending this analysis to multi-planetary systems orbiting sunlike stars. Such analysis is beyond the scope of this paper, however. Finally, in the right panel of Figure~\ref{fig:CMF_massperiod}, we plot the core/water mass fractions as a function of planet mass. Interestingly, we do not see a strong dependence of CMF/WMF on planet mass for either singles or multis.

\section{Comparison of Stellar Properties} \label{sec:stellar_properties}

\begin{table*}[!ht]
\footnotesize
\setlength{\tabcolsep}{2pt}
\caption{Statistics of Planetary and Stellar Properties \label{table:statistics}} 
\centering
\begin{tabular}{lcccc}
\hline\hline
\textbf{Property} & \textbf{Kolmogorov-Smirnov} & \textbf{Anderson-Darling} & \textbf{Median} & \textbf{$N_{\rm systems}$} \\
&&&(Multi, Single)&\\
Planetary Bulk Density (Full sample) & 0.0002 & 0.001 & (4.57, 2.20 g/$\rm cm^3$) & 70\\ 
Planetary Bulk Density ($R_{\rm p} < 6R_{\oplus}$) & 0.046 & 0.025 & (4.57, 3.32 g/$\rm cm^3$) & 61\\
Stellar Metallicity (All planet hosts) & 0.0009 &  0.001 & ([Fe/H] = $-$0.09, 0.0) & 221 \\ 
Stellar Metallicity (Hosts of $R_{\rm p} < 6R_{\oplus}$) & 0.009 &  0.001 & ([Fe/H] = $-$0.1, $-$0.03)& 192\\ 

Stellar Rotation Period (All planet hosts) & 0.066 & 0.051 & (60, 39 days)&80\\ 
Stellar Rotation Period (Hosts of $R_{\rm p} < 6R_{\oplus}$) & 0.237 & 0.151 & (56, 42 days)& 72\\

\hline
\end{tabular}
\end{table*}

We compared the stellar rotation periods and [Fe/H] metallicities of all M-dwarf hosts of multis and singles with those reported properties in the literature. For each property of interest, we excluded stars for which there were no reported uncertainties in the NASA Exoplanet Archive. The rotation period and metallicity values of our samples have a mean uncertainty of 3.5 days and 0.12 dex, respectively.

\subsection{Stellar Rotation Period}
\label{subsec:rotation}

\begin{figure*}
	\centering\vspace{.0in}
	\includegraphics[width=\columnwidth, clip]{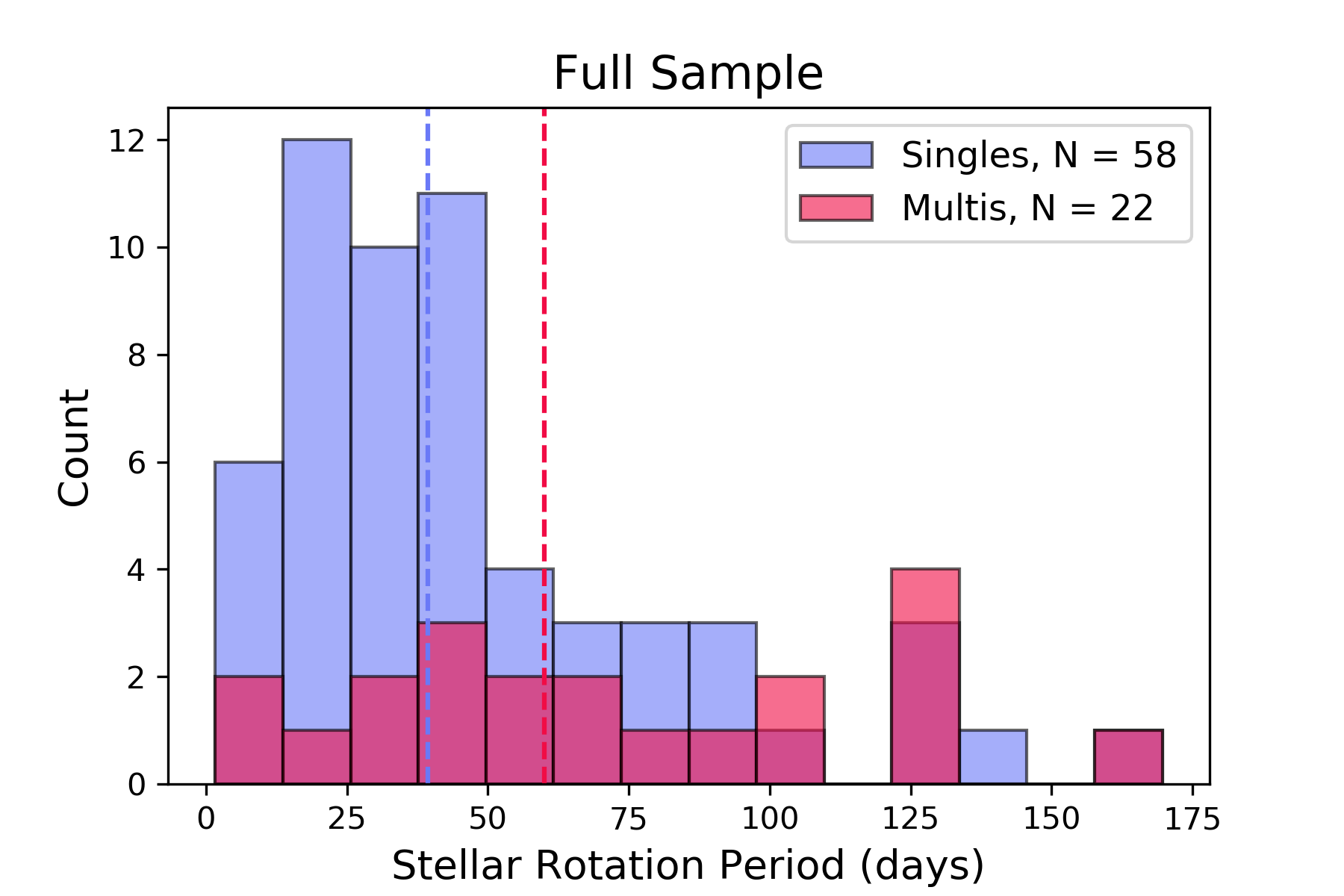}\includegraphics[width=\columnwidth, clip]{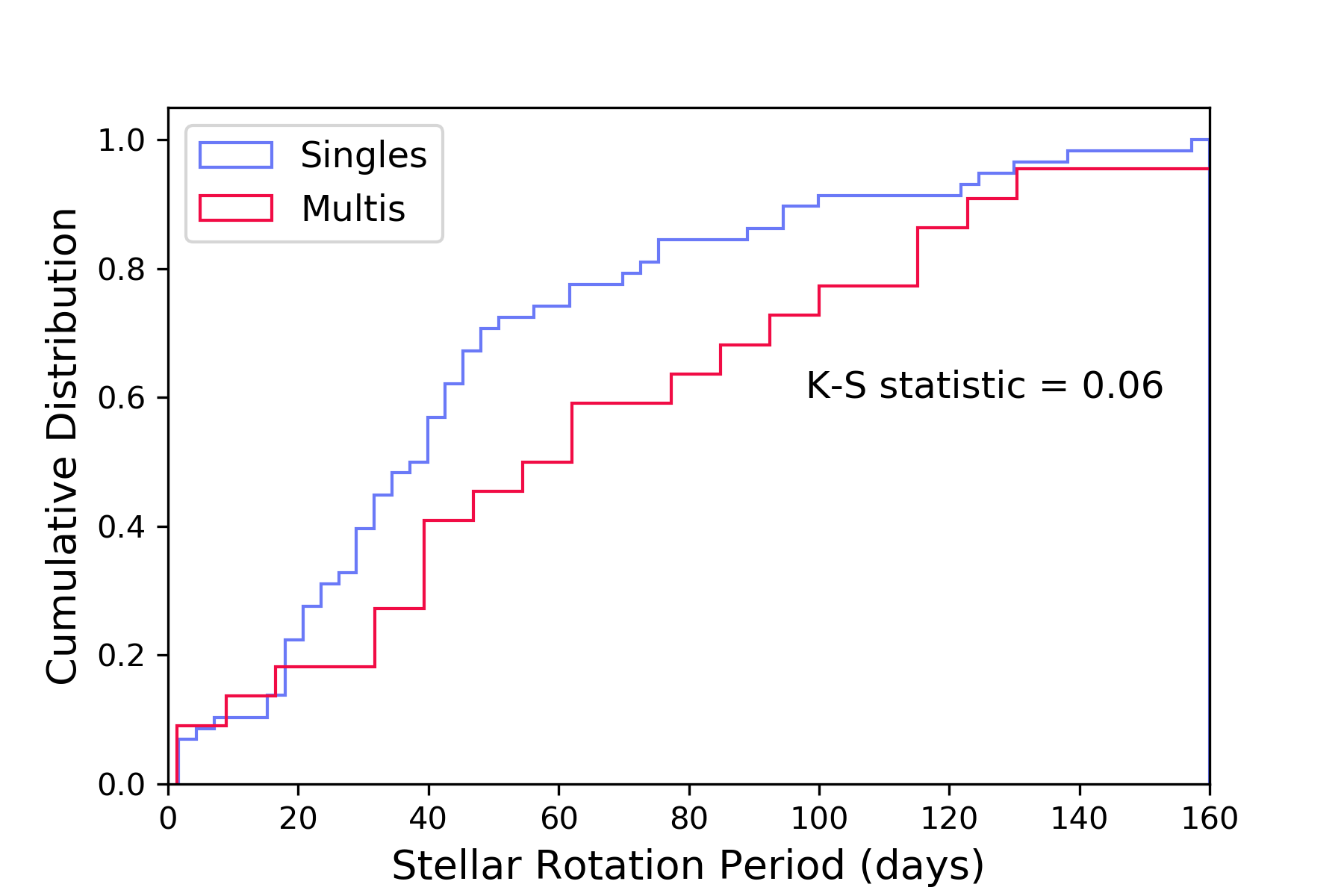}
	\includegraphics[width=\columnwidth, clip]{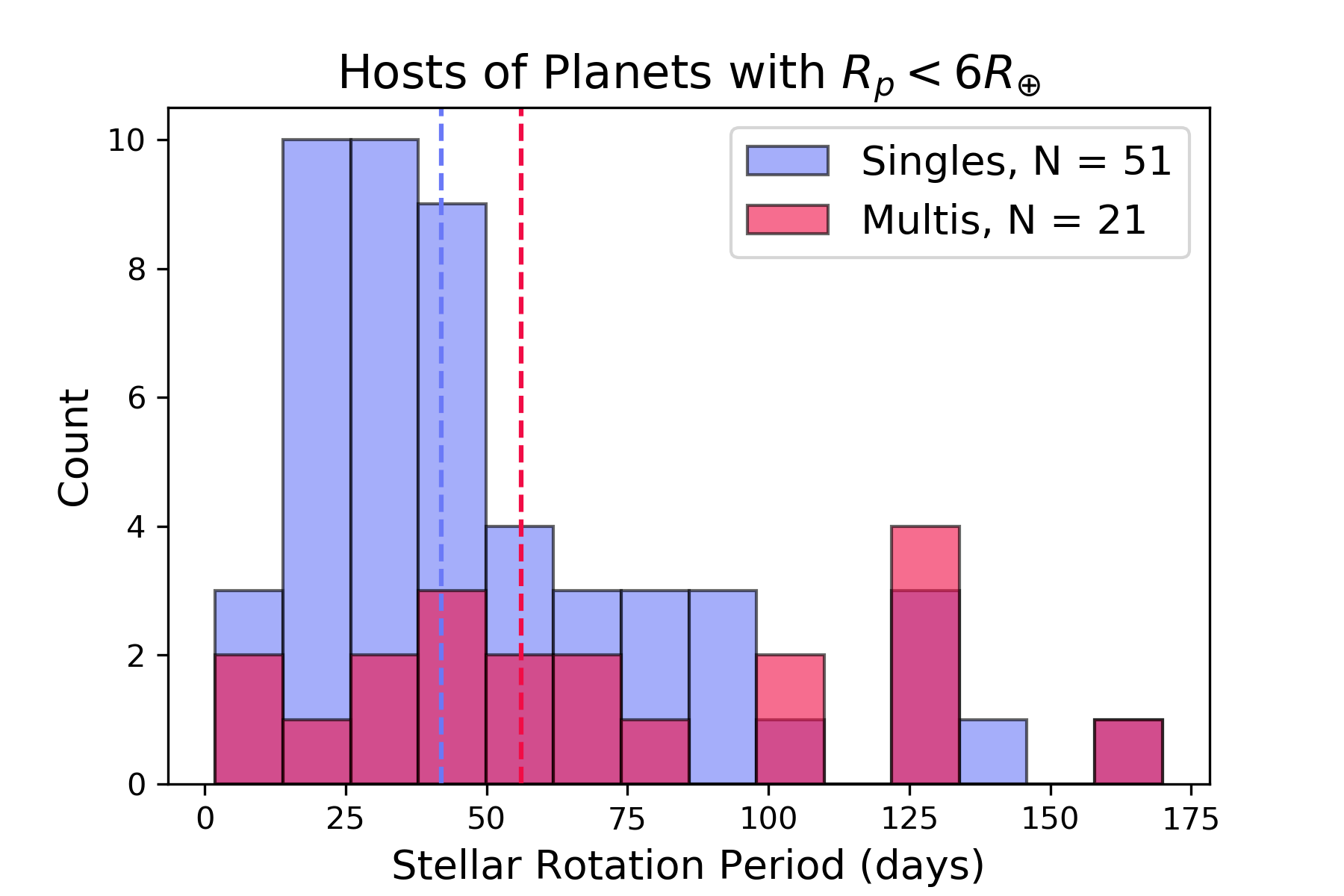}\includegraphics[width=\columnwidth, clip]{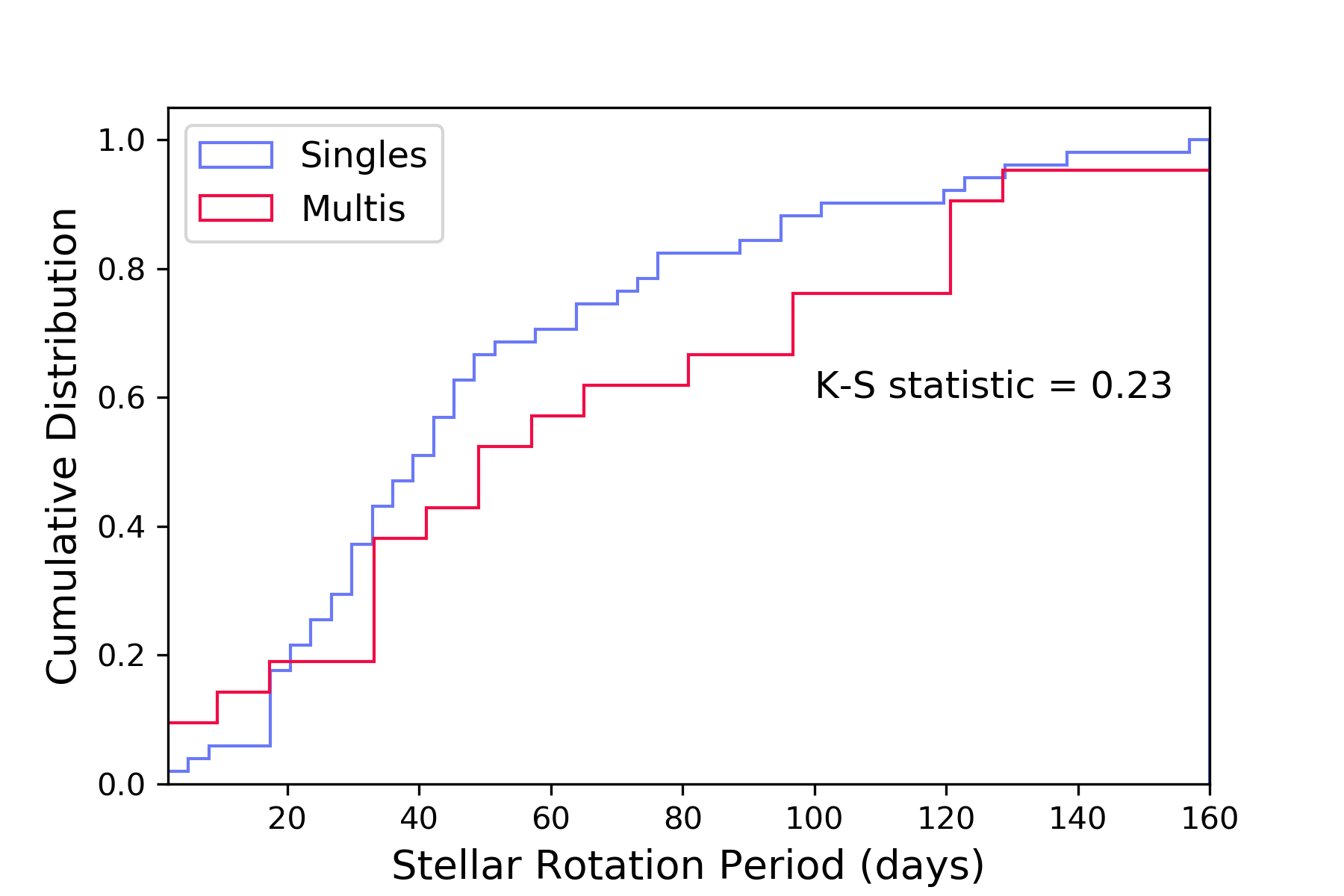}
    \caption{\textbf{Left}: Histograms of the rotation periods of single- (purple) and multi-planet (pink) host stars for the full sample (top) and for the sub-sample excluding $R_{p} \geq 6R_{\oplus}$ planet host stars (bottom). The vertical dashed lines denote the median values of each distribution. \textbf{Right}: Empirical cumulative distribution functions of the samples, with their Kolmogorov-Smirnov statistic values overplotted. $f_{\rm threshold}$ is explained in Section~\ref{selectionbias}.}
	\label{fig:stellar_rotation}
\end{figure*}

We found 80 M dwarfs with rotation period measurements in the literature: 58 hosting single planets and 22 hosting multiple. The rotation periods range from 1.4 to 168.3 days. We find that the multis are slightly slower rotators, with a median rotation period of 60 days, while the singles have a median rotation of 39 days. With a K-S statistic of 0.06, these distributions are only marginally statistically different. Since the stellar rotation period is generally considered a proxy for youth, as stars spin down with age \citep{Skumanich:1972}, multi-planet hosts may be older on average than single-planet hosts. However, there is the caveat that young stars/rapid rotators are also more magnetically active, which in turn can negatively affect the sensitivity to planet detections. Thus, it is possible that the higher prevalence of multis around slow rotators (older stars) is at least partly a consequence of the higher RV sensitivity around those stars compared to faster rotators (younger stars).

If we remove hosts of Neptune-sized and larger planets ($R_{p} \geq 6R_{\oplus}$), we get 72 stars: 51 singles and 21 multis. The median rotation of the singles and multis is 42 and 56 days, respectively. The K-S statistic yields $p= 0.23$, indicating that they are likely drawn from the same distribution and are statistically indistinguishable. The bottom two panels of Figure~\ref{fig:stellar_rotation} show the rotation periods for hosts of singles and multis for the full sample (top) and for the subset of stars hosting planets with $R_{p} < 6R_{\oplus}$ (bottom).



\subsection{Stellar Metallicity}
\label{subsec:metallicity}

\begin{figure*}[!ht]
	\centering\vspace{.0in}
	\includegraphics[width=\columnwidth, clip]{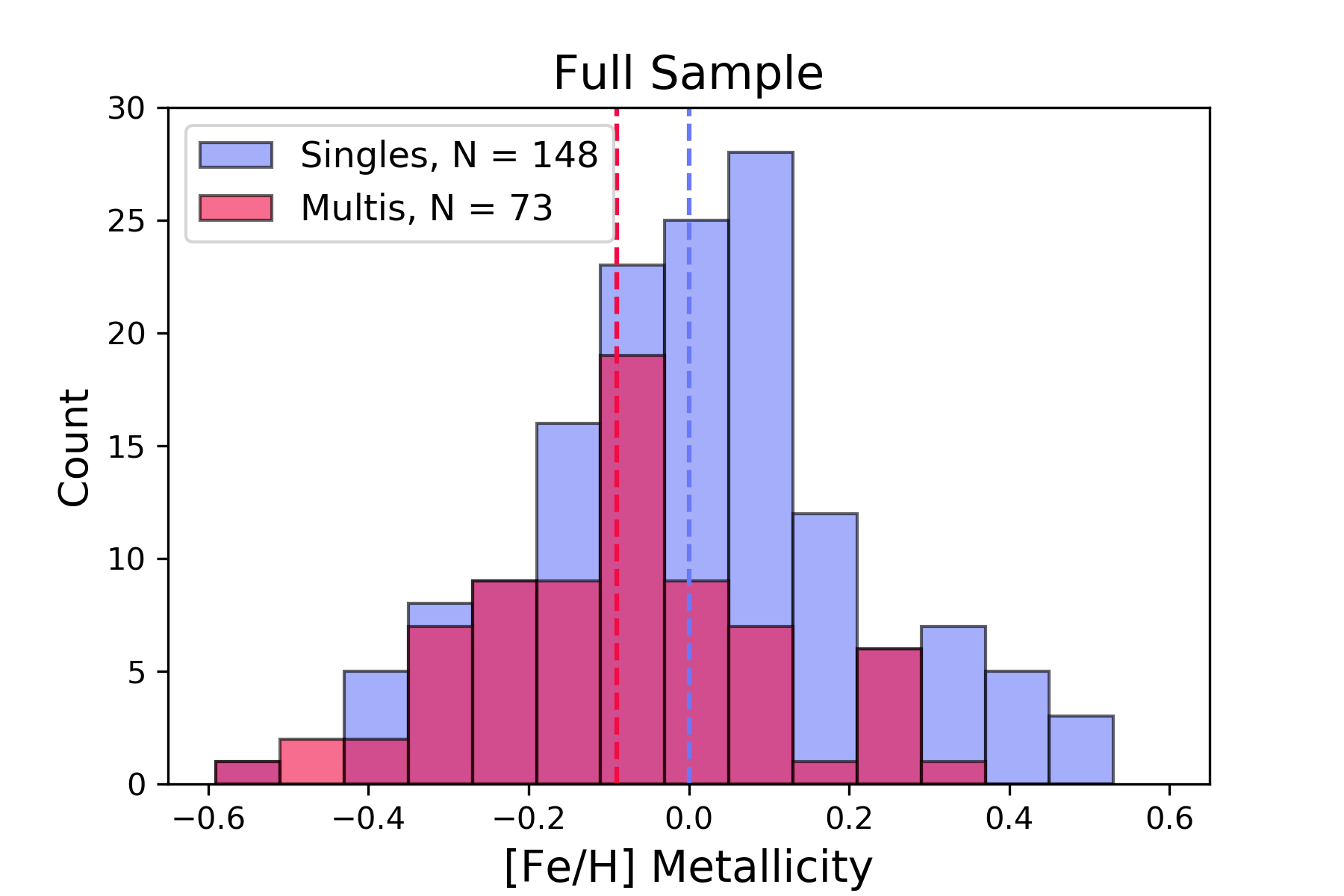}\includegraphics[width=\columnwidth, clip]{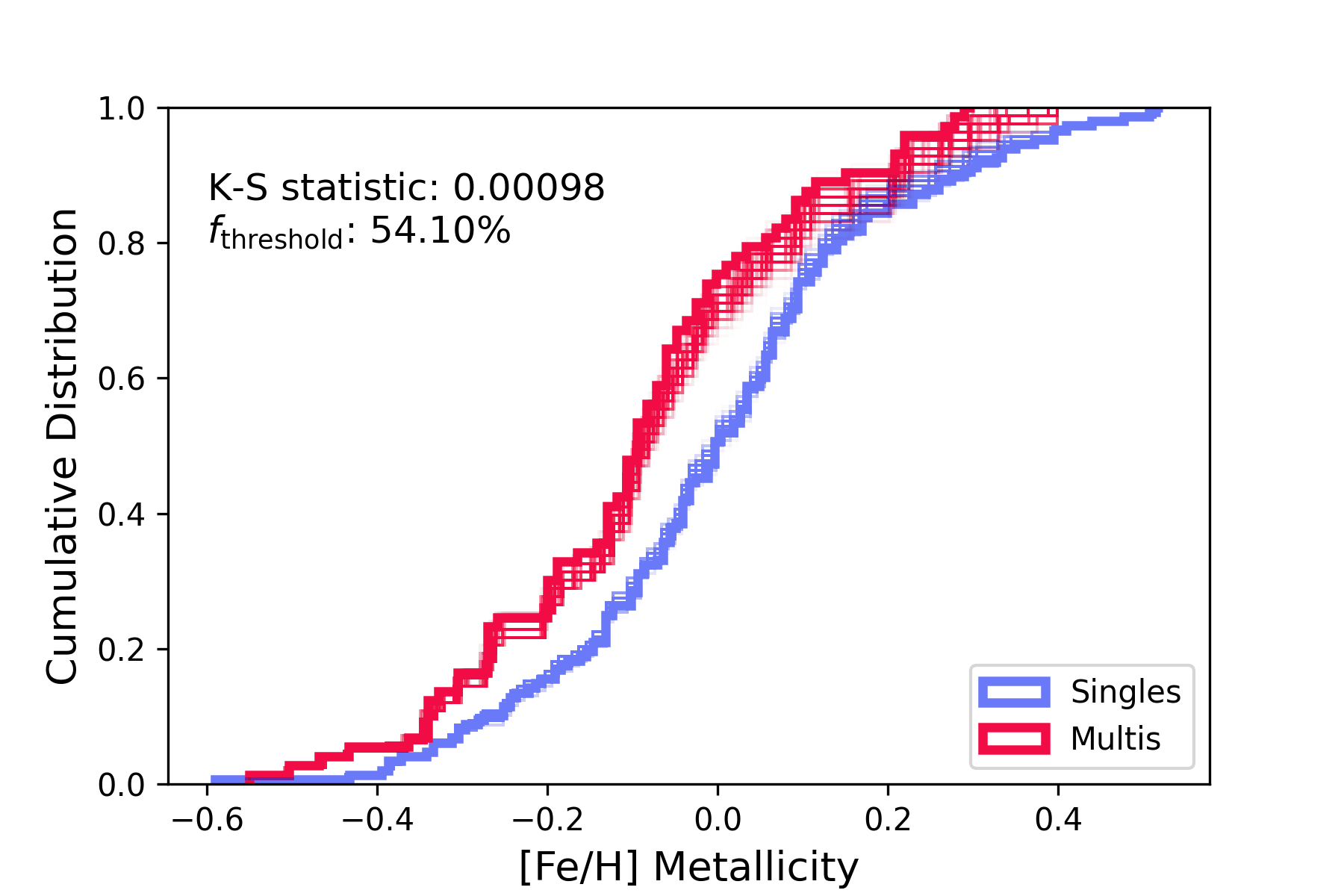}\\
	\includegraphics[width=\columnwidth, clip]{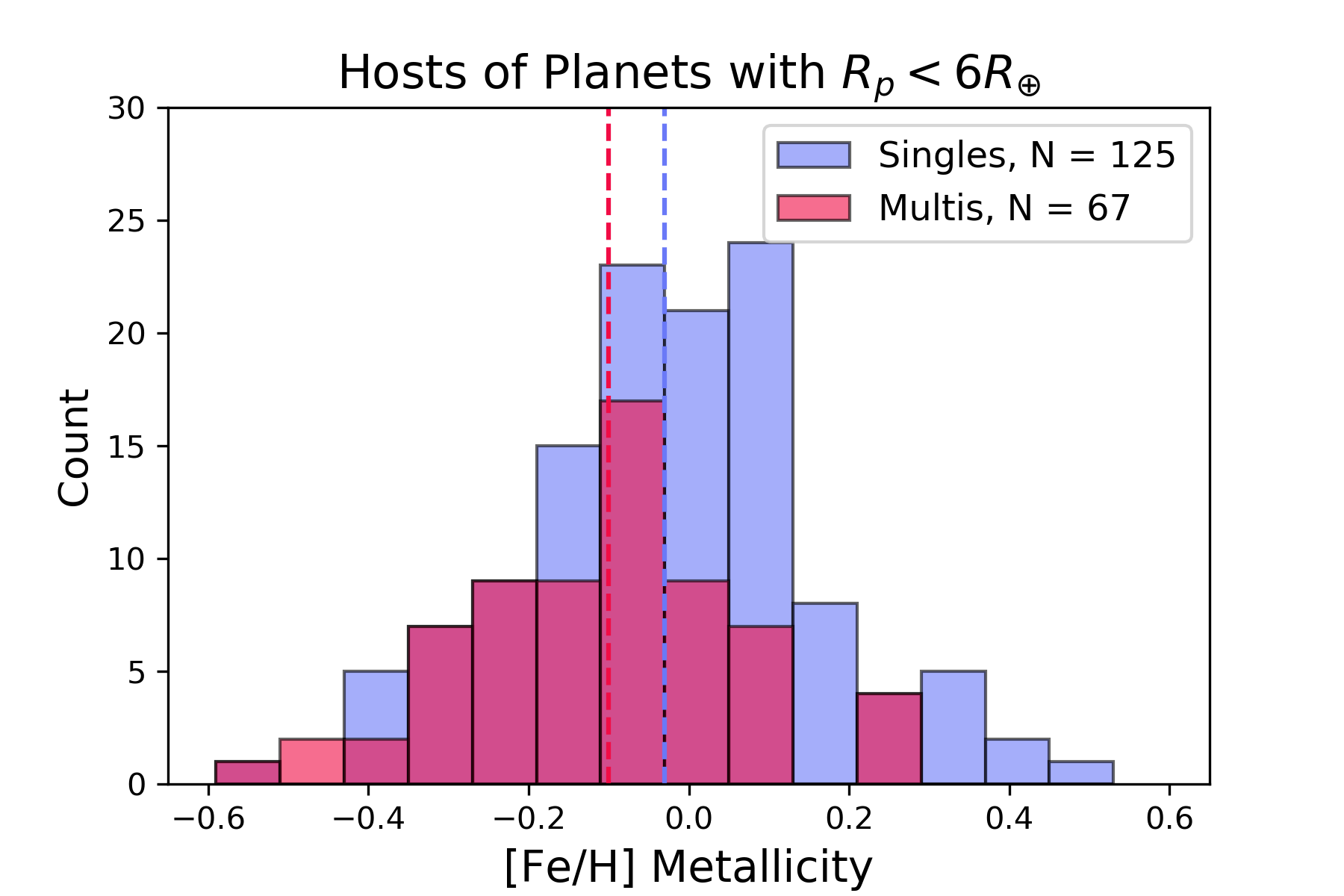}\includegraphics[width=\columnwidth, clip]{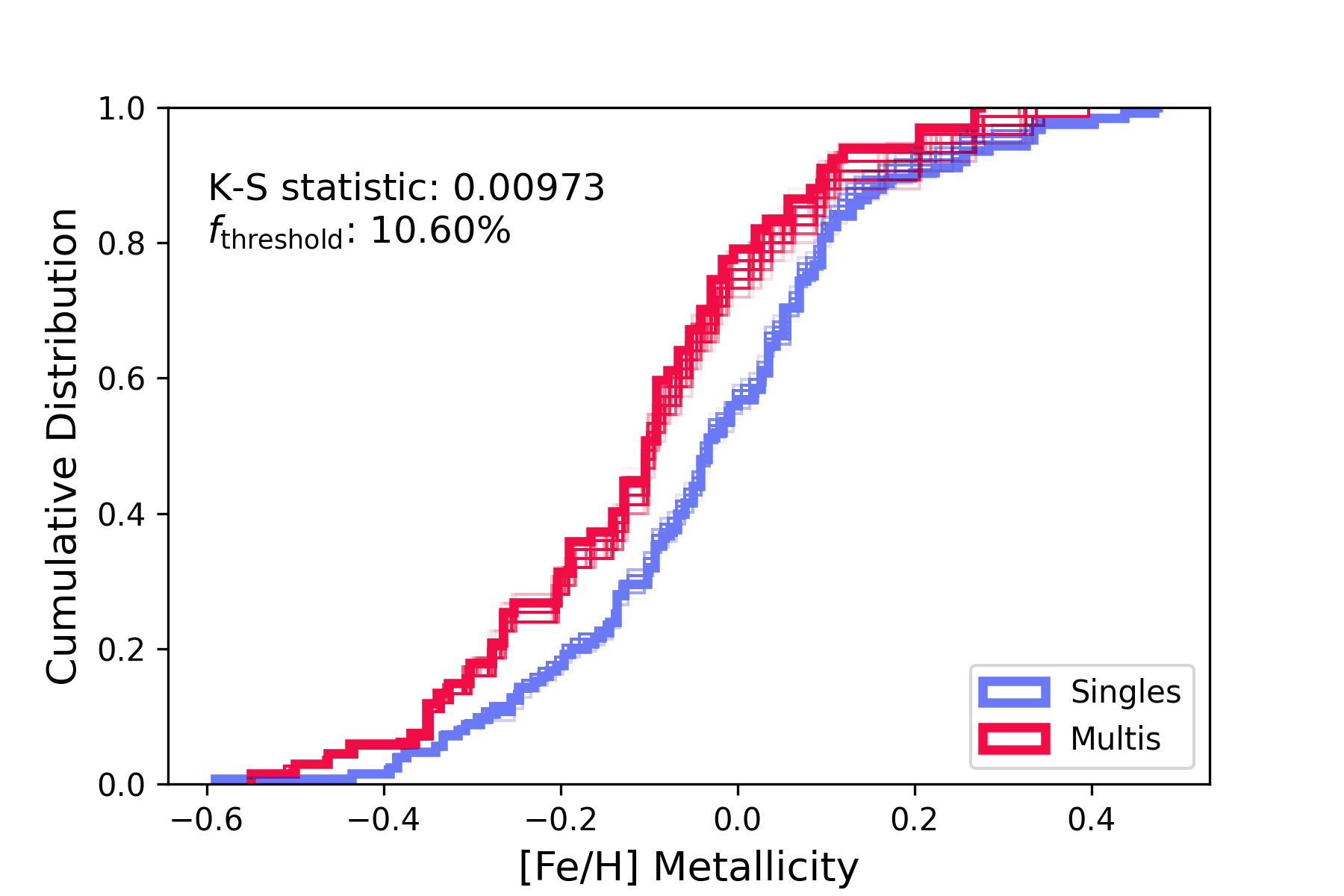}
    \caption{\textbf{Left}: Histograms of the [Fe/H] of the single- (purple) and multi-planet (pink) host stars for the full sample (top) and for the sub-sample excluding $R_{p} \geq 6R_{\oplus}$ planet host stars (bottom). The vertical dashed lines denote the median values of each distribution. \textbf{Right}: Empirical cumulative distribution functions of the samples, with their Kolmogorov-Smirnov statistic values overplotted. $f_{\rm threshold}$ is explained in Section~\ref{selectionbias}.}
	\label{fig:stellar_FeH}
\end{figure*}

\begin{figure*}[!ht]
	\centering\vspace{.0in}
	\includegraphics[width=\columnwidth, clip]{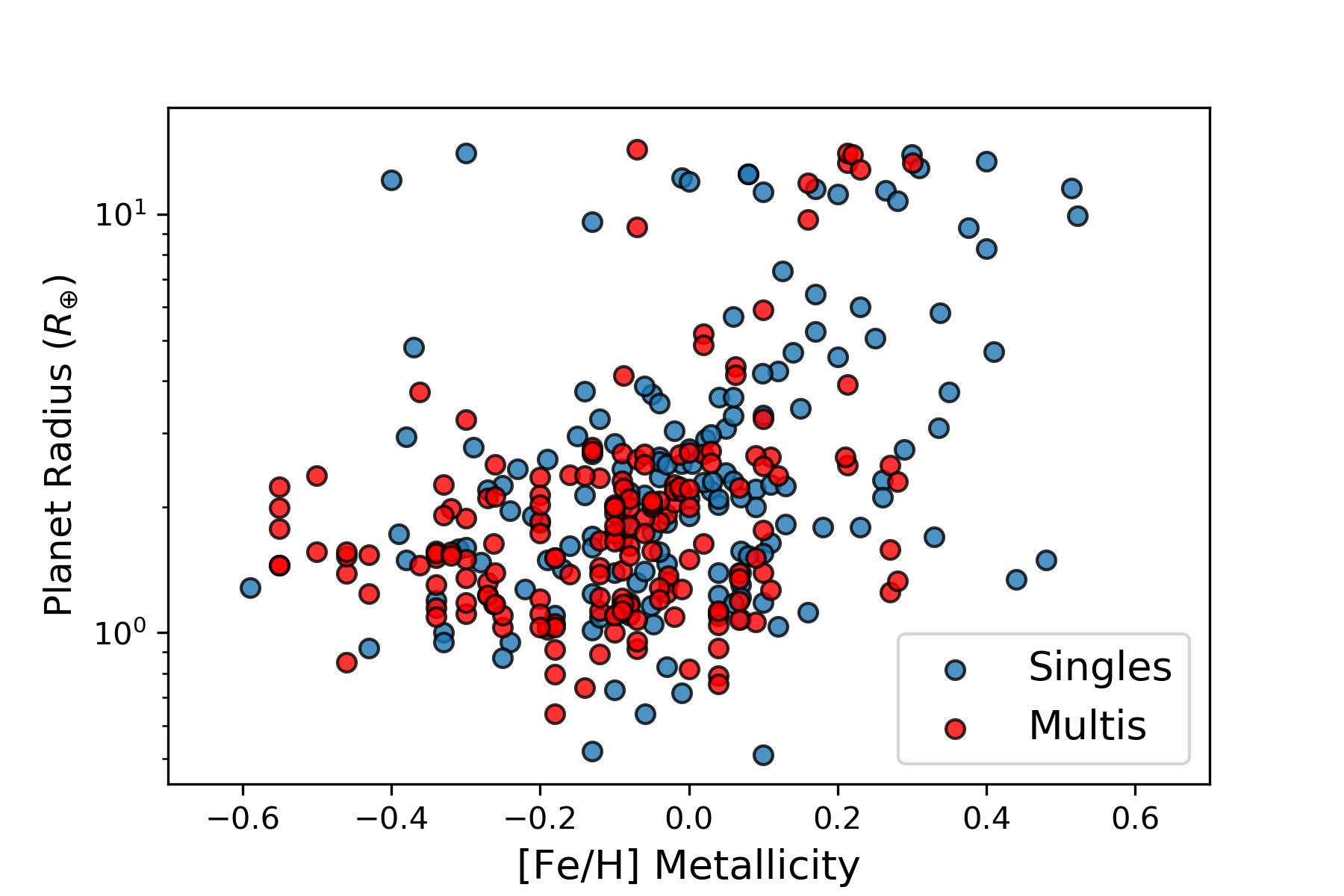}\includegraphics[width=\columnwidth, clip]{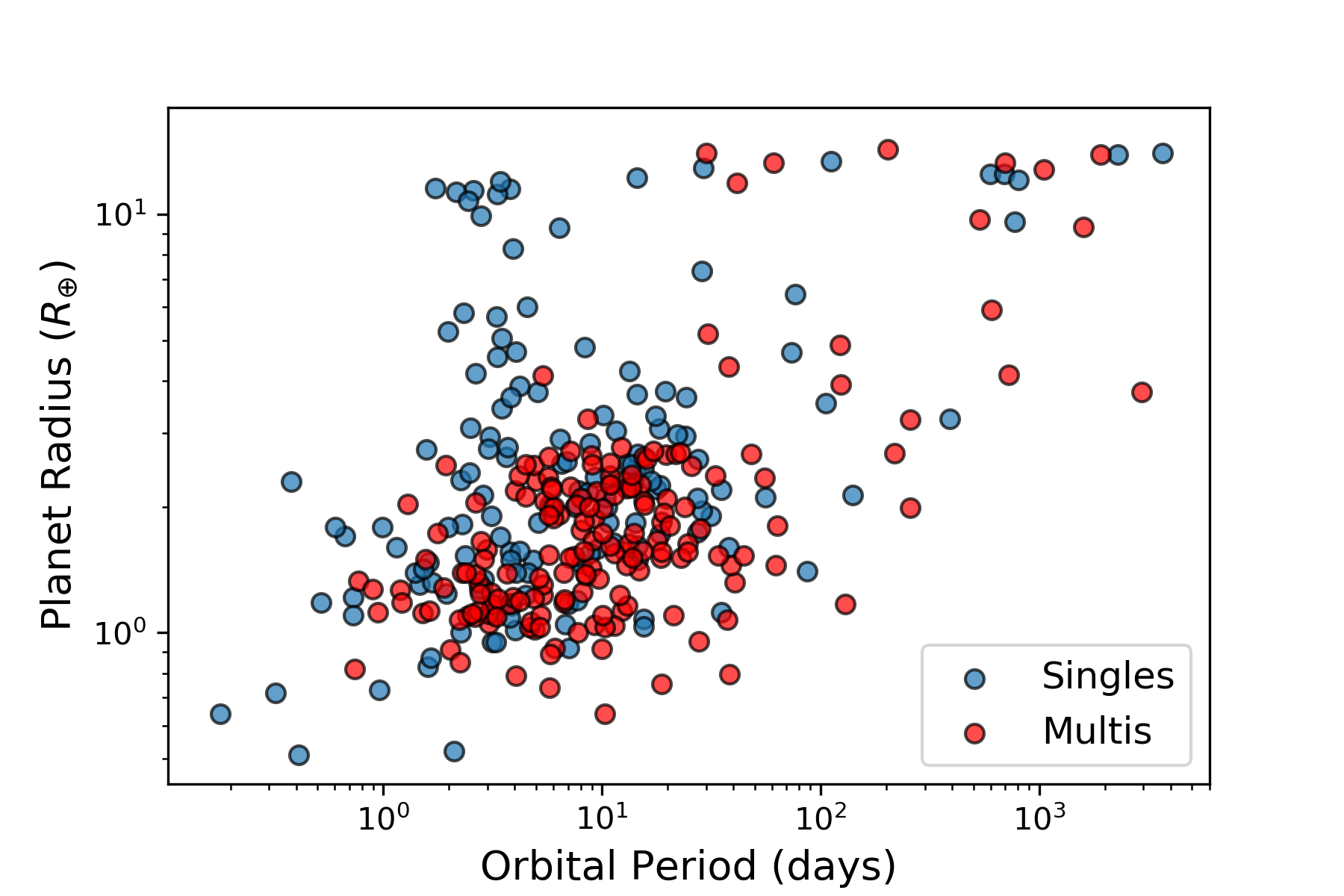}
    \caption{\textbf{Left:} Planet radius as a function of stellar [Fe/H] metallicity for all planet hosts with measured metallicities. The points are color-coded by planet multiplicity.  \textbf{Right:} Planet radius as a function of orbital period for the same sample on the left.}
	\label{fig:radius_metal_multi}
\end{figure*}

We found 221 M dwarfs on the NASA Exoplanet Archive with reported [Fe/H] metallicities ranging from $-$0.59 to +0.52 dex. Of these, 148 host single planets, and 73 host multiples. We find that stars hosting multiple planets are more metal-poor than those hosting only one planet. The multi-planet hosts have a median metallicity of [Fe/H] $=-0.09$ while the single-planet ones have a median of [Fe/H] $=0.0$. This difference is highly statistically significant, with a K-S statistic of 0.0009 and A-D statistic of 0.001. If we exclude hosts of planets with $R_{p} \geq 6R_{\oplus}$, we get 192 stars: 125 hosting single planets and 67 hosting multiples. The multis have a median metallicity of [Fe/H] $=-0.1$ and the singles [Fe/H] $=-0.03$. This difference is less strong than with the full sample, however, with a K-S statistic of 0.009 (see the top two panels of Figure~\ref{fig:stellar_FeH} and Table~\ref{table:statistics} for a summary of both K-S and A-D statistic values). This difference between the full sample and the subset without the giant planets possibly suggests that these larger planets are skewing the observed metallicity distribution in the full sample of planets. This may be hinting at the existence of a scaled-down population of Saturn-mass, short-period giant planets around M dwarfs, analogous to the population of hot Jupiters around sunlike stars, which, as is well known, are isolated and more frequently found around metal-rich stars \citep{Steffen:2012, Fischer:2005}. In fact, we note that all the planets with radii larger than $6R_{\oplus}$ in this sample are seemingly isolated and they have masses and radii between $\sim$ 100-200 $M_{\oplus}$ (except for HATS-74Ab and COCONUTS-2b, which have significantly larger masses), closer to the mass of Saturn (95$M_{\oplus}$). They also orbit predominantly metal-rich (0 $<$ [Fe/H] $<$ 0.5, with a median [Fe/H] of $\sim$0.3), early M dwarfs (M0-M3). This population may also be seen in the left panel of Figure~\ref{fig:radius_metal_multi}, which shows planet radius as a function of metallicity for singles and multis (color-coded in blue and red, respectively).  There are significantly more large, single planets orbiting stars with solar and supersolar metallicities than subsolar metallicity stars. Likewise, there is an apparent lack of multis orbiting metal-rich stars. Beyond [Fe/H] $\sim$ 0.1 dex, the majority of systems are single. The right panel of Figure~\ref{fig:radius_metal_multi} shows the same sample but we plot planet radius as a function of orbital period. It is difficult to draw any robust conclusions yet, however, given the small number of known transiting giants around M dwarfs. The core-accretion theory of planet formation predicts giant planets to be rare around M dwarfs and several studies find low occurrence rates for such planets \citep{Schlecker:2022,Gan:2023,Bryant:2023}. \citet{Gan:2022} explored the correlation between giant planets and stellar metallicity for M dwarfs and compared it to the metallicity correlation for sunlike stars. They found that transiting giants around M dwarfs orbit mostly metal-rich M dwarfs. Overall, including both RV-only and transiting giants, the median metallicity of an M-dwarf hosting a giant is higher than the median metallicity ([Fe/H]$ \sim 0.13$ dex) of a sunlike star hosting a giant (see their Figure 12). They conclude that giants orbit predominantly metal-rich stars, indicating that (transiting) giant planets have a strong host star metallicity dependence, as predicted by core-accretion theory. This is also consistent with the findings of \citet{Maldonado:2020}, who discovered a strong correlation between having a high metallicity and the probability of hosting a giant planet for M dwarfs.

\begin{figure}[!ht]
\centering 
\includegraphics[width=\columnwidth]{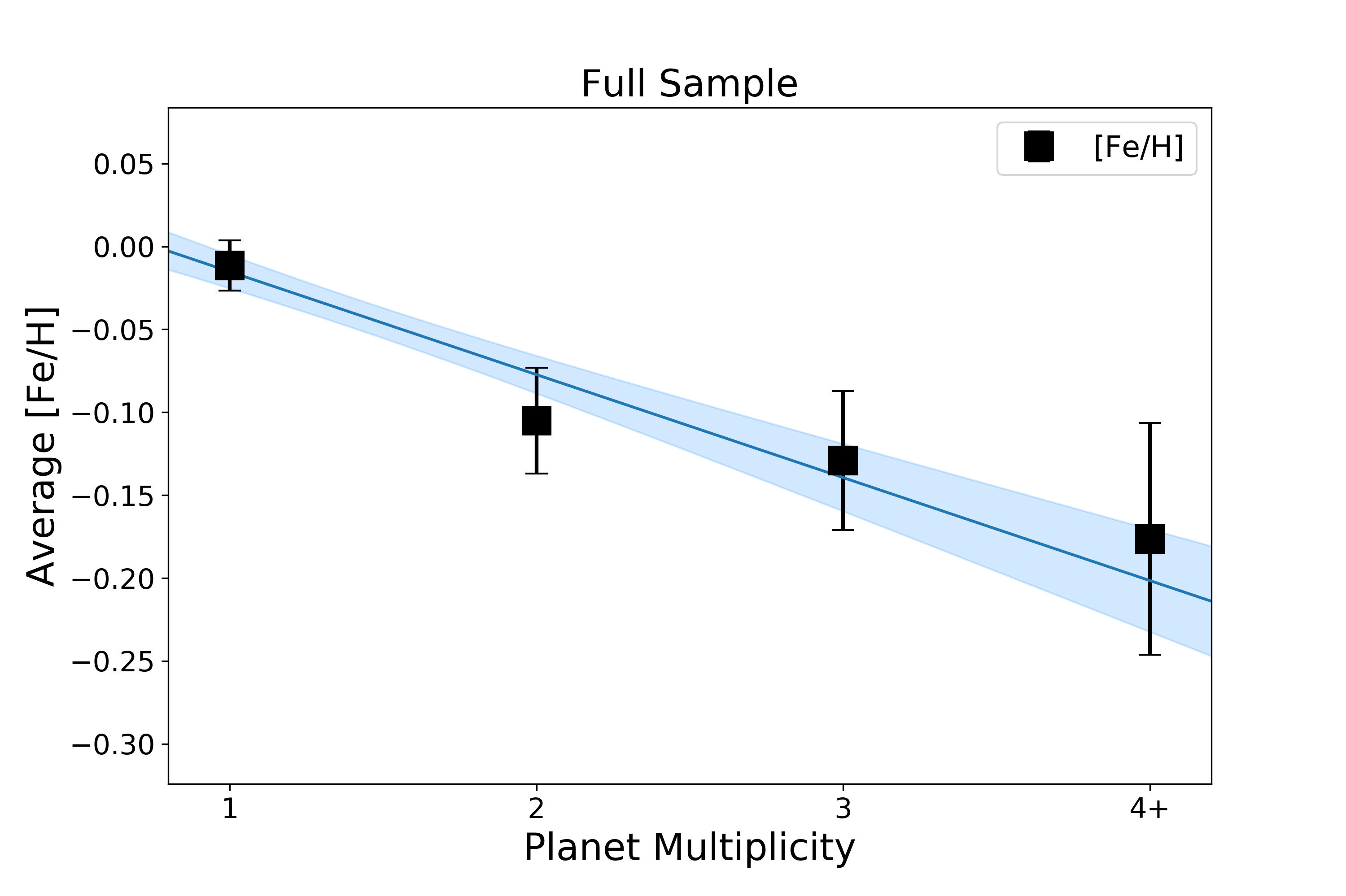}\\
\includegraphics[width=\columnwidth]{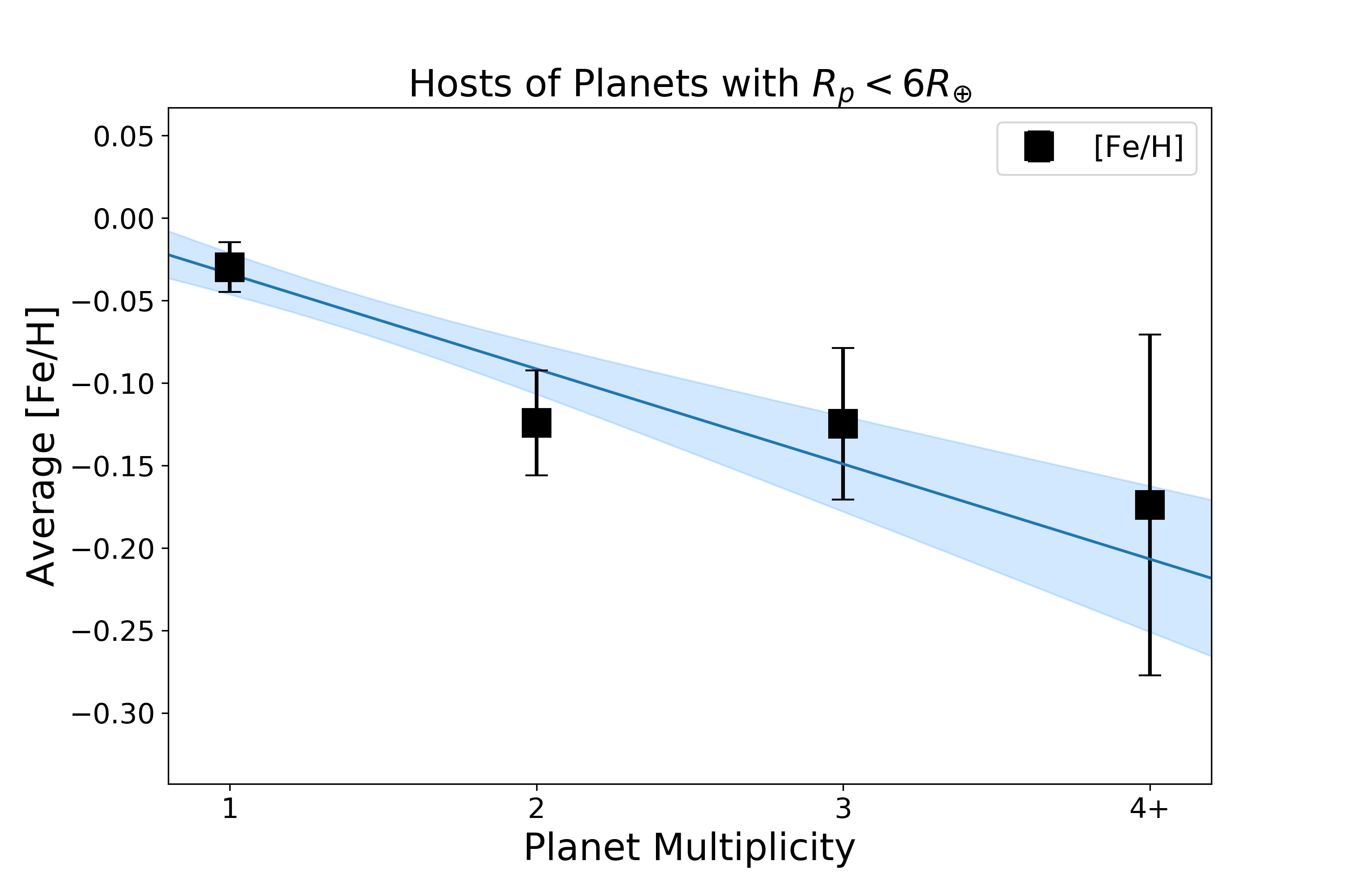}
\caption{Average [Fe/H] metallicity of M dwarfs hosting 1, 2, 3, and 4+ planets for our full sample (\textbf{top}) and for the subset of planets with $R_{p} < 6R_{\oplus}$ (\textbf{bottom}). The error bars show the dispersion in the metallicity for each multiplicity. The shaded blue region shows the confidence interval for the slope.}
\label{fig:metallicity-multiplicity}
\end{figure}

We further investigated stellar [Fe/H] metallicity as a function of planet multiplicity for systems with 1, 2, 3, and 4 or more planets. The error-weighted average [Fe/H] metallicity for 1, 2, 3, and 4+ planet host stars is $-0.01$, $-0.11$, $-0.13$, $-0.18$, dex, respectively. As illustrated in the top panel of Figure~\ref{fig:metallicity-multiplicity}, there is an apparent downward trend of metallicity with planet multiplicity, such that metallicity appears to decrease with increasing number of planets. The error bars show the dispersion in the values of stars with N planets, where N goes from 1 to 4+. We model this relationship with a straight line and find a significant anti-correlation between these two parameters, namely

\begin{equation}
    \rm [Fe/H] = -0.062~(\pm 0.016)\times N_{\rm p} - 0.047~ (\pm 0.026),
\end{equation}
where $N_{\rm p}$ is number of planets. 

We repeated this analysis excluding hosts of planets larger than $6 R_{\oplus}$, and we find a similar negative correlation between metallicity and planet multiplicity (see bottom panel of Figure~\ref{fig:metallicity-multiplicity}). It should be noted that this analysis naturally only includes M dwarfs that have metallicity measurements and therefore these lists are not complete. This fraction of stars with metallicity measurements may be biased in some unknown way. Furthermore, we point out that there is some uncertainty associated with the number of planets itself, as there are several systems in our sample that contain ``controversial" planets in the literature, and whose designation as a candidate or real planet may affect the trends that we see. This uncertainty is very difficult to quantify. 

A comparison between the metallicity of singles and multis has been previously explored by several authors. For example, \citet{Weiss:2018a} compared the metallicities of single- and multiply- hosting FGK stars from the California-Kepler Sample (CKS) and found no significant difference between them (with a $p= 0.29$). Similarly, \citet{Munoz-Romero:2018} compared the metallicities of single and multiple-planet FGK host stars using values from the CKS survey and did not find any significant differences. On the other hand,  \citet{Brewer:2018} found that compact, multi-planet systems occur more frequently around (FGK) stars of lower [Fe/H] metallicities. \citet{Anderson:2021} also found a higher prevalence of compact multis around metal-poor M- and late K-dwarfs. Our results are more in line with \citet{Brewer:2018} and \citet{Anderson:2021}.

\subsection{Consideration of Selection Biases}
\label{selectionbias}

One important caveat of this study is the uncertainty in planet multiplicity; namely, that a non-negligible fraction of the observed singles may be multis in which other hidden planets eluded detection\footnote{This is not a concern with multi-planet systems, however, which we consider to be true multis, following the statistical arguments from \citet{Lissauer:2012,Lissauer:2014,Rowe:2014}.}. \citet{Ballard:2019} found that roughly one third of TESS M-dwarf single-planet systems are expected to have one or more additional planets that are missed by TESS. There are 18 single planets discovered by TESS in our sample (of 30 singles in total). The rest of the singles were discovered by ground-based surveys (9 planets) or by \textit{Kepler}/\textit{K2} (3 planets). Here we analyze the effects that instrument detection completeness may have on our results. Specifically, we investigate what the bulk density and stellar metallicity distributions look like if we assume that a third of the singles are actually multiples. We focused on these properties since those are the ones for which we find a  significant difference between singles and multis. 

We performed a bootstrap analysis in which we randomly changed the multiplicity designation of one third of the TESS singles 1000 times. We then compute the K-S statistic and cumulative distribution for each of the randomly generated samples. The right panels of Figure~\ref{fig:densities_cdf} show all the bootstrapped cumulative distributions for the bulk density of the full sample (top) and the subset of small planets (bottom). In each plot, $f_{\rm threshold}$ is the fraction of bootstrapped values that have a K-S statistic below 0.01 (99\% confidence), which is our conservative threshold for statistical significance. In the case of the bulk density of the full sample, about 83\% of the bootstrapped samples are below 0.01.  This fraction is high enough that we conclude that detection completeness does not significantly affect the observed difference in density between the singles and multis. For the sample of small planets, the K-S statistic of the true, observed distributions is lower (K-S = 0.04), and thus $f_{\rm threshold}$ is lower as well ($f_{\rm threshold} = 4.70\%$) and therefore detection completeness does affect this difference more strongly. 

For the stellar metallicity, we get a lower fraction of values below 0.01 for both the full sample ($f_{\rm threshold} = 54\%$) and for the subset of small-planet hosts ($f_{\rm threshold} = 10.6\%$). See the top two right panels of Figure~\ref{fig:stellar_FeH}. Thus, the difference in metallicity between the singles and multis is less robust than the difference in planet density and should be interpreted more cautiously.

We also consider the extent to which detection and characterization methods bias the observed difference in density in multis and singles. Within our sample of multi-planet systems, roughly half have masses obtained through transit timing variations and half have masses from radial velocities. Within the full sample, 14 have masses from TTVs, and 56 have RV masses. All the single planets are entirely detected through RVs, since TTVs are only measurable in multi-planet systems.

The signal-to-noise ratio of TTV signals has a strong direct dependence on planet radius and orbital period (unlike RV data), and therefore TTVs can measure lower masses than RVs for both larger planets and/or at longer periods. The strong sensitivity to $R_p$ also implies that TTV planets will generally have lower densities than RV planets of the same mass. This is visible in the mass-radius distribution of RV and TTV planets in Figure 1 of \cite{Steffen:2016} (made with data from \citealt{Wolfgang:2015}) and in Figure 5 of \citet{Jontof-Hutter:2015}. However, in this work, we find that planets in multi-planet systems (most of which have TTV masses) are actually {\it denser} than single (RV) planets, which is the opposite of what we would expect if detection biases played a role in the observed density distributions of multis and singles.


On the other hand, although the RV method is more sensitive to more massive planets, we expect lower mass (denser) planets to have been detected if they existed, in systems where only one planet was found via RV (at least within $\sim$30 days, which is the longest period for the planets in our sample). We have already tested the former assumption with the bootstrap analysis discussed above, and we find that even if we assumed that there were additional planets in one third of the TESS singles (all of which have RV masses and which comprise the majority of the singles in our sample), it would not significantly affect the observed difference in bulk density between the singles and multis. In summary, we conclude that the observed differences in the planetary bulk density and stellar metallicity we find here (at least for the full samples of planets and stars) cannot be attributed to detection completeness or to detection/characterization methods.

Finally, we recognize that Figure~\ref{fig:abacus} shows a different orbital period distribution of singles and multis, which will likely be a combination of the true distributions in nature and a reflection of our observational biases. Most of the singles have a period between 1 and 5 days. Only two singles have periods longer than 10 days. Most of the multis have an innermost planet between 1 and 5 days, but the majority of multis include at least one planet beyond 10 days. 

A possible ``natural'' explanation is that the M-dwarf singles contain more gas giants, which seem to be lonely, as is the case around FGK stars. If their formation mechanism includes high eccentricity migration, either from planet scattering or Kozai-Lidov cycles, then this may destabilize any nearby planets at longer periods. Another natural explanation is that multis are believed to migrate in resonant chains, which are typically then disrupted but sometimes preserved (as is the case for TRAPPIST-1). This resonant chain migration would lead to a short-period inner planet, like the ones seen between 1 and 5 days, and a string of planets out beyond 10 days. With respect to observing biases, as previously mentioned, TTVs are generally more sensitive than RVs to planets at a given mass and period. This would bias the singles to shorter periods.

If there is a bias in the orbital period distribution between the singles and multis, then it is natural to consider what effect it would have on our results. According to the `peas in a pod'' paradigm for multi-planet systems, there is intra-system radius uniformity. This means that comparing singles, with periods $\sim$1-5 days, to multis with periods $\sim$1-25 days, may be reasonable because the long-period planets in multis are similar to the short-period planets. With the recent advent of spectrographs with redder wavelength sensitivity, such as KPF \citep{Gibson:2016}, HPF \citep{Mahadevan:2012}, NIRPS \citep{Bouchy:2022}, MAROON-X \citep{Seifahrt:2018} and SPIRou \citep{Artigau:2014}, that are more suitable for M dwarfs, it is hoped that the population of longer period single planets around M dwarfs will grow.

\section{Discussion and Conclusions} \label{sec:discussion}

In this paper, we have calculated and compared the properties and interior composition of planets in single- versus multiple-planet systems around M dwarfs. In addition, we have compared the fundamental properties of M dwarfs hosting a single versus multiple planets. Our main conclusions are as follows: 

1. The bulk density of planets in multi-planet systems is significantly higher than that of singly-transiting planets (with a $p$-value of 0.0002). If we remove larger planets with  $R_{\rm p} \geq 6R_{\oplus}$, this difference remains but is less significant ($p = 0.046$). We cannot attribute these differences in composition to selection biases.

2. We measured the core mass fractions of the likely rocky, low-mass planets in our sample ($M_{\rm p} < 10M_{\oplus}$, which corresponds to $R_p \lessapprox 1.6 R_{\oplus}$ in our sample) and we find that on average, single planets have considerably higher core mass fractions than the multis (0.51 $\pm$ 0.07 versus 0.23 $\pm$ 0.02). We do not see correlations between core/water mass fraction (CMF/WMF) and orbital period nor between CMF/WMF and distance from the host star within individual planetary systems. 

3. We compare the [Fe/H] metallicity and rotation period of all single- and multi-planet M-dwarf hosts with such measurements in the literature. We find a statistically significant difference between the metallicity distributions of singles and multis; namely, that the hosts of singles are significantly more metal-rich than those hosting multiple planets ($p = 0.0009$). If we remove larger planets ($R_{\rm p} \geq 6R_{\oplus}$), this difference is less strong ($p = 0.009$) and less robust to selection effects and detection completeness. On the other hand, we find only a moderate difference between the rotation period distributions, with multis being slightly slower rotators on average ($p = 0.06$). 


4. All the transiting planets in our sample larger than $6R_{\oplus}$ are single and they orbit early type (M0-M3), metal-rich (median of [Fe/H] = 0.3 dex) M dwarfs. This metallicity is higher than the average metallicity of a sunlike star ([Fe/H] $\sim$ 0.13 dex) hosting a giant, per \citet{Gan:2022}. This possibly indicates that the strong metallicity dependence observed for sunlike stars \citep{Fischer:2005,Santos:2004,Wang:2015} also holds for M-dwarf giant planets. This agrees with results from previous authors \citep{Maldonado:2020,Gan:2022}.

Thus, our work reveals two notable features of the M-dwarf planetary population. First, there appears to be a clear difference in the bulk density of M-dwarf planets such that single planets are significantly less dense than those in multi-planet systems. Second, the metallicity of multi-planet host stars is lower on average than those of singles. Moreover, the number of observed planets increases with decreasing host star metallicity. These two observations combined are suggestive of a divergence in the formation mechanisms of multis and singles. 

One possibility we consider is that metal-rich hosts give rise more often to the formation of giant planets, whereas metal-poor M dwarfs could predominantly favor the formation of low-mass, rocky planets in flat (coplanar), compact, and dynamically “quiet” configurations that are stable over long timescales. This could be because higher metallicity protoplanetary disks are more massive, and thus there is more available material to form more massive, giant planets. The idea that multis and singles come from different underlying populations (at least for sunlike stars) is also supported by previous authors who have found significant differences in their obliquities \citep{Morton:2014} and eccentricities \citep{Limbach:2015, Xie:2016,VanEylen:2019,Sagear:2023}. In our own sample, single planets have higher eccentricities on average than those in multiples ($e_{\rm singles} = 0.14$ and $e_{\rm multis} = 0.08$), in agreement with the literature. Alternatively, it is possible that multis and singles may have a common physical origin, or come from the same underlying population, but evolve in vastly different ways, given their significantly different planetary architectures and compositions. However, this possibility seems less likely given that singles and multis have markedly different host star metallicity distributions, as shown here. The trends and conclusions in this paper could be better tested with a larger sample of well-characterized M-dwarf planets. There are currently $\sim$500 TESS planet candidates orbiting M dwarfs, in agreement with predictions \citep{Barclay:2018,Ballard:2019}. The validation and characterization of these candidates will enable much stronger conclusions about the differences between singles and multis. In addition, it remains to be seen whether the observed trends in M-dwarf planets also hold for sunlike stars, and the role that stellar metallicity plays in such trends. It would also be interesting to see whether these correlations in [Fe/H] metallicity are also seen with other rock-forming, refractory elements, such as with Si and Mg. We could not make these comparisons because most of the stars in our sample do not have reported Si and Mg abundances in the literature. Nevertheless, the trends shown here begin to offer clues about the formation and evolution of single- and multi-planet systems and motivate further theoretical work into this area. 


\begin{acknowledgments}

RRM and BSG were supported by the Thomas Jefferson Chair for Space Exploration endowment from The Ohio State University. RRM acknowledges support from the Presidential Fellowship granted by The Ohio State University. RRM acknowledges Jack Neustadt for insightful conversations that improved this manuscript. Support for DVM was provided by NASA through the NASA Hubble Fellowship grant HF2-51464 awarded by the Space Telescope Science Institute, which is operated by the Association of Universities for Research in Astronomy, Inc., for NASA, under contract NAS5-26555. APA acknowledges the support of the NASA NSTGRO fellowship under contract number 80NSSC22K1197. This research has made use of the NASA Exoplanet Archive, which is operated by the California Institute of Technology, under contract with the National Aeronautics and Space Administration under the Exoplanet Exploration Program.

\end{acknowledgments}



\bibliography{Mdwarfs}{}

\begin{thebibliography}{}
\expandafter\ifx\csname natexlab\endcsname\relax\def\natexlab#1{#1}\fi
\providecommand{\url}[1]{\href{#1}{#1}}
\providecommand{\dodoi}[1]{doi:~\href{http://doi.org/#1}{\nolinkurl{#1}}}
\providecommand{\doeprint}[1]{\href{http://ascl.net/#1}{\nolinkurl{http://ascl.net/#1}}}
\providecommand{\doarXiv}[1]{\href{https://arxiv.org/abs/#1}{\nolinkurl{https://arxiv.org/abs/#1}}}

\bibitem[{{Adibekyan} {et~al.}(2021){Adibekyan}, {Dorn}, {Sousa}, {Santos},
  {Bitsch}, {Israelian}, {Mordasini}, {Barros}, {Delgado Mena}, {Demangeon},
  {Faria}, {Figueira}, {Hakobyan}, {Oshagh}, {Soares}, {Kunitomo}, {Takeda},
  {Jofr{\'e}}, {Petrucci}, \& {Martioli}}]{Adibekyan:2021}
{Adibekyan}, V., {Dorn}, C., {Sousa}, S.~G., {et~al.} 2021, Science, 374, 330,
  \dodoi{10.1126/science.abg8794}

\bibitem[{{Agol} {et~al.}(2021){Agol}, {Dorn}, {Grimm}, {Turbet}, {Ducrot},
  {Delrez}, {Gillon}, {Demory}, {Burdanov}, {Barkaoui}, {Benkhaldoun},
  {Bolmont}, {Burgasser}, {Carey}, {de Wit}, {Fabrycky}, {Foreman-Mackey},
  {Haldemann}, {Hernandez}, {Ingalls}, {Jehin}, {Langford}, {Leconte},
  {Lederer}, {Luger}, {Malhotra}, {Meadows}, {Morris}, {Pozuelos}, {Queloz},
  {Raymond}, {Selsis}, {Sestovic}, {Triaud}, \& {Van Grootel}}]{Agol:2021}
{Agol}, E., {Dorn}, C., {Grimm}, S.~L., {et~al.} 2021, \psj, 2, 1,
  \dodoi{10.3847/PSJ/abd022}

\bibitem[{{Aguichine} {et~al.}(2021){Aguichine}, {Mousis}, {Deleuil}, \&
  {Marcq}}]{Aguichine:2021}
{Aguichine}, A., {Mousis}, O., {Deleuil}, M., \& {Marcq}, E. 2021, \apj, 914,
  84, \dodoi{10.3847/1538-4357/abfa99}

\bibitem[{{Akeson} {et~al.}(2013){Akeson}, {Chen}, {Ciardi}, {Crane}, {Good},
  {Harbut}, {Jackson}, {Kane}, {Laity}, {Leifer}, {Lynn}, {McElroy}, {Papin},
  {Plavchan}, {Ram{\'\i}rez}, {Rey}, {von Braun}, {Wittman}, {Abajian}, {Ali},
  {Beichman}, {Beekley}, {Berriman}, {Berukoff}, {Bryden}, {Chan}, {Groom},
  {Lau}, {Payne}, {Regelson}, {Saucedo}, {Schmitz}, {Stauffer}, {Wyatt}, \&
  {Zhang}}]{Akeson:2013}
{Akeson}, R.~L., {Chen}, X., {Ciardi}, D., {et~al.} 2013, \pasp, 125, 989,
  \dodoi{10.1086/672273}

\bibitem[{{Anderson} {et~al.}(2021){Anderson}, {Dittmann}, {Ballard}, \&
  {Bedell}}]{Anderson:2021}
{Anderson}, S.~G., {Dittmann}, J.~A., {Ballard}, S., \& {Bedell}, M. 2021, \aj,
  161, 203, \dodoi{10.3847/1538-3881/abe70b}

\bibitem[{{Artigau} {et~al.}(2014){Artigau}, {Kouach}, {Donati}, {Doyon},
  {Delfosse}, {Baratchart}, {Lacombe}, {Moutou}, {Rabou}, {Par{\`e}s},
  {Micheau}, {Thibault}, {Reshetov}, {Dubois}, {Hernandez}, {Vall{\'e}e},
  {Wang}, {Dolon}, {Pepe}, {Bouchy}, {Striebig}, {H{\'e}nault}, {Loop},
  {Saddlemyer}, {Barrick}, {Vermeulen}, {Dupieux}, {H{\'e}brard}, {Boisse},
  {Martioli}, {Alencar}, {do Nascimento}, \& {Figueira}}]{Artigau:2014}
{Artigau}, {\'E}., {Kouach}, D., {Donati}, J.-F., {et~al.} 2014, in Society of
  Photo-Optical Instrumentation Engineers (SPIE) Conference Series, Vol. 9147,
  Ground-based and Airborne Instrumentation for Astronomy V, ed. S.~K.
  {Ramsay}, I.~S. {McLean}, \& H.~{Takami}, 914715, \dodoi{10.1117/12.2055663}

\bibitem[{{Astudillo-Defru} {et~al.}(2020){Astudillo-Defru}, {Cloutier},
  {Wang}, {Teske}, {Brahm}, {Hellier}, {Ricker}, {Vanderspek}, {Latham},
  {Seager}, {Winn}, {Jenkins}, {Collins}, {Stassun}, {Ziegler}, {Almenara},
  {Anderson}, {Artigau}, {Bonfils}, {Bouchy}, {Brice{\~n}o}, {Butler},
  {Charbonneau}, {Conti}, {Crane}, {Crossfield}, {Davies}, {Delfosse},
  {D{\'\i}az}, {Doyon}, {Dragomir}, {Eastman}, {Espinoza}, {Essack}, {Feng},
  {Figueira}, {Forveille}, {Gan}, {Glidden}, {Guerrero}, {Hart}, {Henning},
  {Horch}, {Isopi}, {Jenkins}, {Jord{\'a}n}, {Kielkopf}, {Law}, {Lovis},
  {Mallia}, {Mann}, {de Medeiros}, {Melo}, {Mennickent}, {Mignon}, {Murgas},
  {Nusdeo}, {Pepe}, {Relles}, {Rose}, {Santos}, {S{\'e}gransan}, {Shectman},
  {Shporer}, {Smith}, {Torres}, {Udry}, {Villasenor}, {Winters}, \&
  {Zhou}}]{Astudillo-Defru:2020}
{Astudillo-Defru}, N., {Cloutier}, R., {Wang}, S.~X., {et~al.} 2020, \aap, 636,
  A58, \dodoi{10.1051/0004-6361/201937179}

\bibitem[{{Bakos} {et~al.}(2020){Bakos}, {Bayliss}, {Bento}, {Bhatti}, {Brahm},
  {Csubry}, {Espinoza}, {Hartman}, {Henning}, {Jord{\'a}n}, {Mancini}, {Penev},
  {Rabus}, {Sarkis}, {Suc}, {de Val-Borro}, {Zhou}, {Butler}, {Crane},
  {Durkan}, {Shectman}, {Kim}, {L{\'a}z{\'a}r}, {Papp}, {S{\'a}ri}, {Ricker},
  {Vanderspek}, {Latham}, {Seager}, {Winn}, {Jenkins}, {Chacon},
  {F{\H{u}}r{\'e}sz}, {Goeke}, {Li}, {Quinn}, {Quintana}, {Tenenbaum}, {Teske},
  {Vezie}, {Yu}, {Stockdale}, {Evans}, \& {Relles}}]{Bakos:2020}
{Bakos}, G.~{\'A}., {Bayliss}, D., {Bento}, J., {et~al.} 2020, \aj, 159, 267,
  \dodoi{10.3847/1538-3881/ab8ad1}

\bibitem[{{Ballard}(2019)}]{Ballard:2019}
{Ballard}, S. 2019, \aj, 157, 113, \dodoi{10.3847/1538-3881/aaf477}

\bibitem[{{Ballard} \& {Johnson}(2016)}]{Ballard:2016}
{Ballard}, S., \& {Johnson}, J.~A. 2016, \apj, 816, 66,
  \dodoi{10.3847/0004-637X/816/2/66}

\bibitem[{{Barclay} {et~al.}(2018){Barclay}, {Pepper}, \&
  {Quintana}}]{Barclay:2018}
{Barclay}, T., {Pepper}, J., \& {Quintana}, E.~V. 2018, \apjs, 239, 2,
  \dodoi{10.3847/1538-4365/aae3e9}

\bibitem[{{Barros} {et~al.}(2022){Barros}, {Demangeon}, {Alibert}, {Leleu},
  {Adibekyan}, {Lovis}, {Bossini}, {Sousa}, {Hara}, {Bouchy}, {Lavie},
  {Rodrigues}, {Gomes da Silva}, {Lillo-Box}, {Pepe}, {Tabernero}, {Zapatero
  Osorio}, {Sozzetti}, {Su{\'a}rez Mascare{\~n}o}, {Micela}, {Allende Prieto},
  {Cristiani}, {Damasso}, {Di Marcantonio}, {Ehrenreich}, {Faria}, {Figueira},
  {Gonz{\'a}lez Hern{\'a}ndez}, {Jenkins}, {Lo Curto}, {Martins}, {Micela},
  {Nunes}, {Pall{\'e}}, {Santos}, {Rebolo}, {Seager}, {Twicken}, {Udry},
  {Vanderspek}, \& {Winn}}]{Barros:2022}
{Barros}, S.~C.~C., {Demangeon}, O.~D.~S., {Alibert}, Y., {et~al.} 2022, \aap,
  665, A154, \dodoi{10.1051/0004-6361/202244293}

\bibitem[{{Bluhm} {et~al.}(2020){Bluhm}, {Luque}, {Espinoza}, {Pall{\'e}},
  {Caballero}, {Dreizler}, {Livingston}, {Mathur}, {Quirrenbach}, {Stock}, {Van
  Eylen}, {Nowak}, {L{\'o}pez}, {Csizmadia}, {Zapatero Osorio}, {Sch{\"o}fer},
  {Lillo-Box}, {Oshagh}, {Gonz{\'a}lez-{\'A}lvarez}, {Amado}, {Barrado},
  {B{\'e}jar}, {Cale}, {Chaturvedi}, {Cifuentes}, {Cochran}, {Collins},
  {Collins}, {Cort{\'e}s-Contreras}, {D{\'\i}ez Alonso}, {El Mufti},
  {Ercolino}, {Fridlund}, {Gaidos}, {Garc{\'\i}a}, {Georgieva},
  {Gonz{\'a}lez-Cuesta}, {Guerra}, {Hatzes}, {Henning}, {Herrero}, {Hidalgo},
  {Isopi}, {Jeffers}, {Jenkins}, {Jensen}, {K{\'a}bath}, {Kaminski}, {Kemmer},
  {Korth}, {Kossakowski}, {K{\"u}rster}, {Lafarga}, {Mallia}, {Montes},
  {Morales}, {Morales-Calder{\'o}n}, {Murgas}, {Narita}, {Passegger}, {Pedraz},
  {Persson}, {Plavchan}, {Rauer}, {Redfield}, {Reffert}, {Reiners}, {Ribas},
  {Ricker}, {Rodr{\'\i}guez-L{\'o}pez}, {Santos}, {Seager}, {Schlecker},
  {Schweitzer}, {Shan}, {Soto}, {Subjak}, {Tal-Or}, {Trifonov}, {Vanaverbeke},
  {Vanderspek}, {Wittrock}, {Zechmeister}, \& {Zohrabi}}]{Bluhm:2020}
{Bluhm}, P., {Luque}, R., {Espinoza}, N., {et~al.} 2020, \aap, 639, A132,
  \dodoi{10.1051/0004-6361/202038160}

\bibitem[{{Bluhm} {et~al.}(2021){Bluhm}, {Pall{\'e}}, {Molaverdikhani},
  {Kemmer}, {Hatzes}, {Kossakowski}, {Stock}, {Caballero}, {Lillo-Box},
  {B{\'e}jar}, {Soto}, {Amado}, {Brown}, {Cadieux}, {Cloutier}, {Collins},
  {Collins}, {Cort{\'e}s-Contreras}, {Doyon}, {Dreizler}, {Espinoza}, {Fukui},
  {Gonz{\'a}lez-{\'A}lvarez}, {Henning}, {Horne}, {Jeffers}, {Jenkins},
  {Jensen}, {Kaminski}, {Kielkopf}, {Kusakabe}, {K{\"u}rster},
  {Lafreni{\`e}re}, {Luque}, {Murgas}, {Montes}, {Morales}, {Narita},
  {Passegger}, {Quirrenbach}, {Sch{\"o}fer}, {Reffert}, {Reiners}, {Ribas},
  {Ricker}, {Seager}, {Schweitzer}, {Schwarz}, {Tamura}, {Trifonov},
  {Vanderspek}, {Winn}, {Zechmeister}, \& {Zapatero Osorio}}]{Bluhm:2021}
{Bluhm}, P., {Pall{\'e}}, E., {Molaverdikhani}, K., {et~al.} 2021, \aap, 650,
  A78, \dodoi{10.1051/0004-6361/202140688}

\bibitem[{{Bonfils} {et~al.}(2018){Bonfils}, {Almenara}, {Cloutier},
  {W{\"u}nsche}, {Astudillo-Defru}, {Berta-Thompson}, {Bouchy}, {Charbonneau},
  {Delfosse}, {D{\'\i}az}, {Dittmann}, {Doyon}, {Forveille}, {Irwin}, {Lovis},
  {Mayor}, {Menou}, {Murgas}, {Newton}, {Pepe}, {Santos}, \&
  {Udry}}]{Bonfils:2018}
{Bonfils}, X., {Almenara}, J.~M., {Cloutier}, R., {et~al.} 2018, \aap, 618,
  A142, \dodoi{10.1051/0004-6361/201731884}

\bibitem[{{Borucki} {et~al.}(2010){Borucki}, {Koch}, {Basri}, {Batalha},
  {Brown}, {Caldwell}, {Caldwell}, {Christensen-Dalsgaard}, {Cochran},
  {DeVore}, {Dunham}, {Dupree}, {Gautier}, {Geary}, {Gilliland}, {Gould},
  {Howell}, {Jenkins}, {Kondo}, {Latham}, {Marcy}, {Meibom}, {Kjeldsen},
  {Lissauer}, {Monet}, {Morrison}, {Sasselov}, {Tarter}, {Boss}, {Brownlee},
  {Owen}, {Buzasi}, {Charbonneau}, {Doyle}, {Fortney}, {Ford}, {Holman},
  {Seager}, {Steffen}, {Welsh}, {Rowe}, {Anderson}, {Buchhave}, {Ciardi},
  {Walkowicz}, {Sherry}, {Horch}, {Isaacson}, {Everett}, {Fischer}, {Torres},
  {Johnson}, {Endl}, {MacQueen}, {Bryson}, {Dotson}, {Haas}, {Kolodziejczak},
  {Van Cleve}, {Chandrasekaran}, {Twicken}, {Quintana}, {Clarke}, {Allen},
  {Li}, {Wu}, {Tenenbaum}, {Verner}, {Bruhweiler}, {Barnes}, \&
  {Prsa}}]{Borucki:2010}
{Borucki}, W.~J., {Koch}, D., {Basri}, G., {et~al.} 2010, Science, 327, 977,
  \dodoi{10.1126/science.1185402}

\bibitem[{{Bouchy} {et~al.}(2022){Bouchy}, {Wildi}, \& {Gonz{\'a}lez
  Hern{\'a}ndez}}]{Bouchy:2022}
{Bouchy}, F., {Wildi}, F., \& {Gonz{\'a}lez Hern{\'a}ndez}, J.~I. 2022, in
  European Planetary Science Congress, EPSC2022--937,
  \dodoi{10.5194/epsc2022-937}

\bibitem[{{Brewer} {et~al.}(2018){Brewer}, {Wang}, {Fischer}, \&
  {Foreman-Mackey}}]{Brewer:2018}
{Brewer}, J.~M., {Wang}, S., {Fischer}, D.~A., \& {Foreman-Mackey}, D. 2018,
  \apjl, 867, L3, \dodoi{10.3847/2041-8213/aae710}

\bibitem[{{Bryant} {et~al.}(2023){Bryant}, {Bayliss}, \& {Van
  Eylen}}]{Bryant:2023}
{Bryant}, E.~M., {Bayliss}, D., \& {Van Eylen}, V. 2023, \mnras,
  \dodoi{10.1093/mnras/stad626}

\bibitem[{{Burt} {et~al.}(2021){Burt}, {Dragomir}, {Molli{\`e}re},
  {Youngblood}, {Garc{\'\i}a Mu{\~n}oz}, {McCann}, {Kreidberg}, {Huang},
  {Collins}, {Eastman}, {Abe}, {Almenara}, {Crossfield}, {Ziegler},
  {Rodriguez}, {Mamajek}, {Stassun}, {Halverson}, {Villanueva}, {Butler},
  {Wang}, {Schwarz}, {Ricker}, {Vanderspek}, {Latham}, {Seager}, {Winn},
  {Jenkins}, {Agabi}, {Bonfils}, {Ciardi}, {Cointepas}, {Crane}, {Crouzet},
  {Dransfield}, {Feng}, {Furlan}, {Guillot}, {Gupta}, {Howell}, {Jensen},
  {Law}, {Mann}, {Marie-Sainte}, {Matson}, {Matthews}, {M{\'e}karnia},
  {Pepper}, {Scott}, {Shectman}, {Schlieder}, {Schmider}, {Stevens}, {Teske},
  {Triaud}, {Charbonneau}, {Berta-Thompson}, {Burke}, {Daylan}, {Barclay},
  {Wohler}, \& {Brasseur}}]{Burt:2021}
{Burt}, J.~A., {Dragomir}, D., {Molli{\`e}re}, P., {et~al.} 2021, \aj, 162, 87,
  \dodoi{10.3847/1538-3881/ac0432}

\bibitem[{{Ca{\~n}as} {et~al.}(2020){Ca{\~n}as}, {Stefansson}, {Kanodia},
  {Mahadevan}, {Cochran}, {Endl}, {Robertson}, {Bender}, {Ninan}, {Beard},
  {Lubin}, {Gupta}, {Everett}, {Monson}, {Wilson}, {Lewis}, {Brewer},
  {Majewski}, {Hebb}, {Dawson}, {Diddams}, {Ford}, {Fredrick}, {Halverson},
  {Hearty}, {Lin}, {Metcalf}, {Rajagopal}, {Ramsey}, {Roy}, {Schwab},
  {Terrien}, \& {Wright}}]{canas:2020}
{Ca{\~n}as}, C.~I., {Stefansson}, G., {Kanodia}, S., {et~al.} 2020, \aj, 160,
  147, \dodoi{10.3847/1538-3881/abac67}

\bibitem[{{Ca{\~n}as} {et~al.}(2022){Ca{\~n}as}, {Kanodia}, {Bender},
  {Mahadevan}, {Stef{\'a}nsson}, {Cochran}, {Lin}, {Hwang}, {Powers}, {Monson},
  {Green}, {Parker}, {Swaby}, {Kobulnicky}, {Wisniewski}, {Gupta}, {Everett},
  {Jones}, {Anjakos}, {Beard}, {Blake}, {Diddams}, {Dong}, {Fredrick},
  {Hakemiamjad}, {Hebb}, {Libby-Roberts}, {Logsdon}, {McElwain}, {Metcalf},
  {Ninan}, {Rajagopal}, {Ramsey}, {Robertson}, {Roy}, {Ruhle}, {Schwab},
  {Terrien}, \& {Wright}}]{Canas:2022}
{Ca{\~n}as}, C.~I., {Kanodia}, S., {Bender}, C.~F., {et~al.} 2022, \aj, 164,
  50, \dodoi{10.3847/1538-3881/ac7804}

\bibitem[{{Cale} {et~al.}(2021){Cale}, {Reefe}, {Plavchan}, {Tanner}, {Gaidos},
  {Gagn{\'e}}, {Gao}, {Kane}, {B{\'e}jar}, {Lodieu}, {Anglada-Escud{\'e}},
  {Ribas}, {Pall{\'e}}, {Quirrenbach}, {Amado}, {Reiners}, {Caballero}, {Rosa
  Zapatero Osorio}, {Dreizler}, {Howard}, {Fulton}, {Xuesong Wang}, {Collins},
  {El Mufti}, {Wittrock}, {Gilbert}, {Barclay}, {Klein}, {Martioli},
  {Wittenmyer}, {Wright}, {Addison}, {Hirano}, {Tamura}, {Kotani}, {Narita},
  {Vermilion}, {Lee}, {Geneser}, {Teske}, {Quinn}, {Latham}, {Esquerdo},
  {Calkins}, {Berlind}, {Zohrabi}, {Stibbards}, {Kotnana}, {Jenkins},
  {Twicken}, {Henze}, {Kidwell}, {Burke}, {Villase{\~n}or}, \&
  {Boyd}}]{Cale:2021}
{Cale}, B.~L., {Reefe}, M., {Plavchan}, P., {et~al.} 2021, \aj, 162, 295,
  \dodoi{10.3847/1538-3881/ac2c80}

\bibitem[{{Chen} \& {Kipping}(2017)}]{Chen:2017}
{Chen}, J., \& {Kipping}, D. 2017, \apj, 834, 17,
  \dodoi{10.3847/1538-4357/834/1/17}

\bibitem[{{Cloutier} {et~al.}(2021{\natexlab{a}}){Cloutier}, {Charbonneau},
  {Deming}, {Bonfils}, \& {Astudillo-Defru}}]{Cloutier:2021}
{Cloutier}, R., {Charbonneau}, D., {Deming}, D., {Bonfils}, X., \&
  {Astudillo-Defru}, N. 2021{\natexlab{a}}, \aj, 162, 174,
  \dodoi{10.3847/1538-3881/ac1584}

\bibitem[{{Cloutier} {et~al.}(2021{\natexlab{b}}){Cloutier}, {Charbonneau},
  {Stassun}, {Murgas}, {Mortier}, {Massey}, {Lissauer}, {Latham}, {Irwin},
  {Haywood}, {Guerra}, {Girardin}, {Giacalone}, {Bosch-Cabot}, {Bieryla},
  {Winn}, {Watson}, {Vanderspek}, {Udry}, {Tamura}, {Sozzetti}, {Shporer},
  {S{\'e}gransan}, {Seager}, {Savel}, {Sasselov}, {Rose}, {Ricker}, {Rice},
  {Quintana}, {Quinn}, {Piotto}, {Phillips}, {Pepe}, {Pedani}, {Parviainen},
  {Palle}, {Narita}, {Molinari}, {Micela}, {McDermott}, {Mayor}, {Matson},
  {Martinez Fiorenzano}, {Lovis}, {L{\'o}pez-Morales}, {Kusakabe}, {Jensen},
  {Jenkins}, {Huang}, {Howell}, {Harutyunyan}, {F{\H{u}}r{\'e}sz}, {Fukui},
  {Esquerdo}, {Esparza-Borges}, {Dumusque}, {Dressing}, {Fabrizio}, {Collins},
  {Cameron}, {Christiansen}, {Cecconi}, {Buchhave}, {Boschin}, \&
  {Andreuzzi}}]{Cloutier1634:2021}
{Cloutier}, R., {Charbonneau}, D., {Stassun}, K.~G., {et~al.}
  2021{\natexlab{b}}, \aj, 162, 79, \dodoi{10.3847/1538-3881/ac0157}

\bibitem[{{Cointepas} {et~al.}(2021){Cointepas}, {Almenara}, {Bonfils},
  {Bouchy}, {Astudillo-Defru}, {Murgas}, {Otegi}, {Wyttenbach}, {Anderson},
  {Artigau}, {Canto Martins}, {Charbonneau}, {Collins}, {Collins}, {Correia},
  {Curaba}, {Delboulb{\'e}}, {Delfosse}, {D{\'\i}az}, {Dorn}, {Doyon},
  {Feautrier}, {Figueira}, {Forveille}, {Gaisne}, {Gan}, {Gluck}, {Helled},
  {Hellier}, {Jocou}, {Kern}, {Lafrasse}, {Law}, {Le{\~a}o}, {Lovis},
  {Magnard}, {Mann}, {Maurel}, {de Medeiros}, {Melo}, {Moulin}, {Pepe},
  {Rabou}, {Rochat}, {Rodriguez}, {Roux}, {Santos}, {S{\'e}gransan}, {Stadler},
  {Ting}, {Twicken}, {Udry}, {Waalkes}, {West}, {W{\"u}nsche}, {Ziegler},
  {Ricker}, {Vanderspek}, {Latham}, {Seager}, {Winn}, \&
  {Jenkins}}]{Cointepas:2021}
{Cointepas}, M., {Almenara}, J.~M., {Bonfils}, X., {et~al.} 2021, \aap, 650,
  A145, \dodoi{10.1051/0004-6361/202140328}

\bibitem[{{Crossfield} {et~al.}(2015){Crossfield}, {Petigura}, {Schlieder},
  {Howard}, {Fulton}, {Aller}, {Ciardi}, {L{\'e}pine}, {Barclay}, {de Pater},
  {de Kleer}, {Quintana}, {Christiansen}, {Schlafly}, {Kaltenegger}, {Crepp},
  {Henning}, {Obermeier}, {Deacon}, {Weiss}, {Isaacson}, {Hansen}, {Liu},
  {Greene}, {Howell}, {Barman}, \& {Mordasini}}]{Crossfield:2015}
{Crossfield}, I. J.~M., {Petigura}, E., {Schlieder}, J.~E., {et~al.} 2015,
  \apj, 804, 10, \dodoi{10.1088/0004-637X/804/1/10}

\bibitem[{{Demangeon} {et~al.}(2021){Demangeon}, {Zapatero Osorio}, {Alibert},
  {Barros}, {Adibekyan}, {Tabernero}, {Antoniadis-Karnavas}, {Camacho},
  {Su{\'a}rez Mascare{\~n}o}, {Oshagh}, {Micela}, {Sousa}, {Lovis}, {Pepe},
  {Rebolo}, {Cristiani}, {Santos}, {Allart}, {Allende Prieto}, {Bossini},
  {Bouchy}, {Cabral}, {Damasso}, {Di Marcantonio}, {D'Odorico}, {Ehrenreich},
  {Faria}, {Figueira}, {G{\'e}nova Santos}, {Haldemann}, {Hara}, {Gonz{\'a}lez
  Hern{\'a}ndez}, {Lavie}, {Lillo-Box}, {Lo Curto}, {Martins}, {M{\'e}gevand},
  {Mehner}, {Molaro}, {Nunes}, {Pall{\'e}}, {Pasquini}, {Poretti}, {Sozzetti},
  \& {Udry}}]{Demangeon:2021}
{Demangeon}, O.~D.~S., {Zapatero Osorio}, M.~R., {Alibert}, Y., {et~al.} 2021,
  \aap, 653, A41, \dodoi{10.1051/0004-6361/202140728}

\bibitem[{{Demory} {et~al.}(2013){Demory}, {Torres}, {Neves}, {Rogers},
  {Gillon}, {Horch}, {Sullivan}, {Bonfils}, {Delfosse}, {Forveille}, {Lovis},
  {Mayor}, {Santos}, {Seager}, {Smalley}, \& {Udry}}]{Demory:2013}
{Demory}, B.-O., {Torres}, G., {Neves}, V., {et~al.} 2013, \apj, 768, 154,
  \dodoi{10.1088/0004-637X/768/2/154}

\bibitem[{{Demory} {et~al.}(2020){Demory}, {Pozuelos}, {G{\'o}mez Maqueo Chew},
  {Sabin}, {Petrucci}, {Schroffenegger}, {Grimm}, {Sestovic}, {Gillon},
  {McCormac}, {Barkaoui}, {Benz}, {Bieryla}, {Bouchy}, {Burdanov}, {Collins},
  {de Wit}, {Dressing}, {Garcia}, {Giacalone}, {Guerra}, {Haldemann}, {Heng},
  {Jehin}, {Jofr{\'e}}, {Kane}, {Lillo-Box}, {Maign{\'e}}, {Mordasini},
  {Morris}, {Niraula}, {Queloz}, {Rackham}, {Savel}, {Soubkiou}, {Srdoc},
  {Stassun}, {Triaud}, {Zambelli}, {Ricker}, {Latham}, {Seager}, {Winn},
  {Jenkins}, {Calvario-Vel{\'a}squez}, {Franco Herrera}, {Colorado}, {Cadena
  Zepeda}, {Figueroa}, {Watson}, {Lugo-Ibarra}, {Carigi}, {Guisa}, {Herrera},
  {Sierra D{\'\i}az}, {Su{\'a}rez}, {Barrado}, {Batalha}, {Benkhaldoun},
  {Chontos}, {Dai}, {Essack}, {Ghachoui}, {Huang}, {Huber}, {Isaacson},
  {Lissauer}, {Morales-Calder{\'o}n}, {Robertson}, {Roy}, {Twicken},
  {Vanderburg}, \& {Weiss}}]{Demory:2020}
{Demory}, B.~O., {Pozuelos}, F.~J., {G{\'o}mez Maqueo Chew}, Y., {et~al.} 2020,
  \aap, 642, A49, \dodoi{10.1051/0004-6361/202038616}

\bibitem[{{Dreizler} {et~al.}(2020){Dreizler}, {Crossfield}, {Kossakowski},
  {Plavchan}, {Jeffers}, {Kemmer}, {Luque}, {Espinoza}, {Pall{\'e}}, {Stassun},
  {Matthews}, {Cale}, {Caballero}, {Schlecker}, {Lillo-Box}, {Zechmeister},
  {Lalitha}, {Reiners}, {Soubkiou}, {Bitsch}, {Zapatero Osorio}, {Chaturvedi},
  {Hatzes}, {Ricker}, {Vanderspek}, {Latham}, {Seager}, {Winn}, {Jenkins},
  {Aceituno}, {Amado}, {Barkaoui}, {Barbieri}, {Batalha}, {Bauer}, {Benneke},
  {Benkhaldoun}, {Beichman}, {Berberian}, {Burt}, {Butler}, {Caldwell},
  {Chintada}, {Chontos}, {Christiansen}, {Ciardi}, {Cifuentes}, {Collins},
  {Collins}, {Combs}, {Cort{\'e}s-Contreras}, {Crane}, {Daylan}, {Dragomir},
  {Esparza-Borges}, {Evans}, {Feng}, {Flowers}, {Fukui}, {Fulton}, {Furlan},
  {Gaidos}, {Geneser}, {Giacalone}, {Gillon}, {Gonzales}, {Gorjian}, {Hellier},
  {Hidalgo}, {Howard}, {Howell}, {Huber}, {Isaacson}, {Jehin}, {Jensen},
  {Kaminski}, {Kane}, {Kawauchi}, {Kielkopf}, {Klahr}, {Kosiarek}, {Kreidberg},
  {K{\"u}rster}, {Lafarga}, {Livingston}, {Louie}, {Mann}, {Madrigal-Aguado},
  {Matson}, {Mocnik}, {Morales}, {Muirhead}, {Murgas}, {Nandakumar}, {Narita},
  {Nowak}, {Oshagh}, {Parviainen}, {Passegger}, {Pollacco}, {Pozuelos},
  {Quirrenbach}, {Reefe}, {Ribas}, {Robertson}, {Rodr{\'\i}guez-L{\'o}pez},
  {Rose}, {Roy}, {Schweitzer}, {Schlieder}, {Shectman}, {Tanner},
  {{\c{S}}enavc{\i}}, {Teske}, {Twicken}, {Villasenor}, {Wang}, {Weiss},
  {Wittrock}, {Y{\i}lmaz}, \& {Zohrabi}}]{Dreizler:2020}
{Dreizler}, S., {Crossfield}, I.~J.~M., {Kossakowski}, D., {et~al.} 2020, \aap,
  644, A127, \dodoi{10.1051/0004-6361/202038016}

\bibitem[{{Dressing} \& {Charbonneau}(2015)}]{Dressing:2015}
{Dressing}, C.~D., \& {Charbonneau}, D. 2015, \apj, 807, 45,
  \dodoi{10.1088/0004-637X/807/1/45}

\bibitem[{{Driscoll} \& {Barnes}(2015)}]{Driscoll:2015}
{Driscoll}, P.~E., \& {Barnes}, R. 2015, Astrobiology, 15, 739,
  \dodoi{10.1089/ast.2015.1325}

\bibitem[{{Fischer} \& {Valenti}(2005)}]{Fischer:2005}
{Fischer}, D.~A., \& {Valenti}, J. 2005, \apj, 622, 1102,
  \dodoi{10.1086/428383}

\bibitem[{{Gan} {et~al.}(2020){Gan}, {Shporer}, {Livingston}, {Collins}, {Mao},
  {Trani}, {Gandolfi}, {Hirano}, {Luque}, {Stassun}, {Ziegler}, {Howell},
  {Hellier}, {Irwin}, {Winters}, {Anderson}, {Brice{\~n}o}, {Law}, {Mann},
  {Bonfils}, {Astudillo-Defru}, {Jensen}, {Anglada-Escud{\'e}}, {Ricker},
  {Vanderspek}, {Latham}, {Seager}, {Winn}, {Jenkins}, {Furesz}, {Guerrero},
  {Quintana}, {Twicken}, {Caldwell}, {Tenenbaum}, {Huang}, {Rowden}, \&
  {Rojas-Ayala}}]{Gan:2020}
{Gan}, T., {Shporer}, A., {Livingston}, J.~H., {et~al.} 2020, \aj, 159, 160,
  \dodoi{10.3847/1538-3881/ab775a}

\bibitem[{{Gan} {et~al.}(2022{\natexlab{a}}){Gan}, {Lin}, {Wang}, {Mao},
  {Fouqu{\'e}}, {Fan}, {Bedell}, {Stassun}, {Giacalone}, {Fukui}, {Murgas},
  {Ciardi}, {Howell}, {Collins}, {Shporer}, {Arnold}, {Barclay}, {Charbonneau},
  {Christiansen}, {Crossfield}, {Dressing}, {Elliott}, {Esparza-Borges},
  {Evans}, {Gnilka}, {Gonzales}, {Howard}, {Isogai}, {Kawauchi}, {Kurita},
  {Liu}, {Livingston}, {Matson}, {Narita}, {Palle}, {Parviainen}, {Rackham},
  {Rodriguez}, {Rose}, {Rudat}, {Schlieder}, {Scott}, {Vezie}, {Ricker},
  {Vanderspek}, {Latham}, {Seager}, {Winn}, \& {Jenkins}}]{Gan:2022}
{Gan}, T., {Lin}, Z., {Wang}, S.~X., {et~al.} 2022{\natexlab{a}}, \mnras, 511,
  83, \dodoi{10.1093/mnras/stab3708}

\bibitem[{{Gan} {et~al.}(2022{\natexlab{b}}){Gan}, {Soubkiou}, {Wang},
  {Benkhaldoun}, {Mao}, {Artigau}, {Fouqu{\'e}}, {Arnold}, {Giacalone},
  {Theissen}, {Aganze}, {Burgasser}, {Collins}, {Shporer}, {Barkaoui},
  {Ghachoui}, {Howell}, {Lamman}, {Demangeon}, {Burdanov}, {Cadieux},
  {Chouqar}, {Collins}, {Cook}, {Delrez}, {Demory}, {Doyon}, {Dransfield},
  {Dressing}, {Ducrot}, {Fan}, {Garcia}, {Gill}, {Gillon}, {Gnilka}, {G{\'o}mez
  Maqueo Chew}, {G{\"u}nther}, {Henze}, {Huang}, {Jehin}, {Jensen}, {Lin},
  {Manset}, {McCormac}, {Murray}, {Niraula}, {Pedersen}, {Pozuelos}, {Queloz},
  {Rackham}, {Savel}, {Schanche}, {Schwarz}, {Sebastian}, {Thompson},
  {Timmermans}, {Triaud}, {Vezie}, {Wells}, {de Wit}, {Ricker}, {Vanderspek},
  {Latham}, {Seager}, {Winn}, \& {Jenkins}}]{Gan2136:2022}
{Gan}, T., {Soubkiou}, A., {Wang}, S.~X., {et~al.} 2022{\natexlab{b}}, \mnras,
  514, 4120, \dodoi{10.1093/mnras/stac1448}

\bibitem[{{Gan} {et~al.}(2022{\natexlab{c}}){Gan}, {Lin}, {Wang}, {Mao},
  {Fouqu{\'e}}, {Fan}, {Bedell}, {Stassun}, {Giacalone}, {Fukui}, {Murgas},
  {Ciardi}, {Howell}, {Collins}, {Shporer}, {Arnold}, {Barclay}, {Charbonneau},
  {Christiansen}, {Crossfield}, {Dressing}, {Elliott}, {Esparza-Borges},
  {Evans}, {Gnilka}, {Gonzales}, {Howard}, {Isogai}, {Kawauchi}, {Kurita},
  {Liu}, {Livingston}, {Matson}, {Narita}, {Palle}, {Parviainen}, {Rackham},
  {Rodriguez}, {Rose}, {Rudat}, {Schlieder}, {Scott}, {Vezie}, {Ricker},
  {Vanderspek}, {Latham}, {Seager}, {Winn}, \& {Jenkins}}]{Gan530:2022}
{Gan}, T., {Lin}, Z., {Wang}, S.~X., {et~al.} 2022{\natexlab{c}}, \mnras, 511,
  83, \dodoi{10.1093/mnras/stab3708}

\bibitem[{{Gan} {et~al.}(2023){Gan}, {Wang}, {Wang}, {Mao}, {Huang}, {Collins},
  {Stassun}, {Shporer}, {Zhu}, {Ricker}, {Vanderspek}, {Latham}, {Seager},
  {Winn}, {Jenkins}, {Barkaoui}, {Belinski}, {Ciardi}, {Evans}, {Girardin},
  {Maslennikova}, {Mazeh}, {Panahi}, {Pozuelos}, {Radford}, {Schwarz},
  {Twicken}, {W{\"u}nsche}, \& {Zucker}}]{Gan:2023}
{Gan}, T., {Wang}, S.~X., {Wang}, S., {et~al.} 2023, \aj, 165, 17,
  \dodoi{10.3847/1538-3881/ac9b12}

\bibitem[{{Gibson} {et~al.}(2016){Gibson}, {Howard}, {Marcy}, {Edelstein},
  {Wishnow}, \& {Poppett}}]{Gibson:2016}
{Gibson}, S.~R., {Howard}, A.~W., {Marcy}, G.~W., {et~al.} 2016, in Society of
  Photo-Optical Instrumentation Engineers (SPIE) Conference Series, Vol. 9908,
  Ground-based and Airborne Instrumentation for Astronomy VI, ed. C.~J.
  {Evans}, L.~{Simard}, \& H.~{Takami}, 990870, \dodoi{10.1117/12.2233334}

\bibitem[{{Gilbert} {et~al.}(2020){Gilbert}, {Barclay}, {Schlieder},
  {Quintana}, {Hord}, {Kostov}, {Lopez}, {Rowe}, {Hoffman}, {Walkowicz},
  {Silverstein}, {Rodriguez}, {Vanderburg}, {Suissa}, {Airapetian}, {Clement},
  {Raymond}, {Mann}, {Kruse}, {Lissauer}, {Col{\'o}n}, {Kopparapu},
  {Kreidberg}, {Zieba}, {Collins}, {Quinn}, {Howell}, {Ziegler}, {Vrijmoet},
  {Adams}, {Arney}, {Boyd}, {Brande}, {Burke}, {Cacciapuoti}, {Chance},
  {Christiansen}, {Covone}, {Daylan}, {Dineen}, {Dressing}, {Essack},
  {Fauchez}, {Galgano}, {Howe}, {Kaltenegger}, {Kane}, {Lam}, {Lee}, {Lewis},
  {Logsdon}, {Mandell}, {Monsue}, {Mullally}, {Mullally}, {Paudel},
  {Pidhorodetska}, {Plavchan}, {Reyes}, {Rinehart}, {Rojas-Ayala}, {Smith},
  {Stassun}, {Tenenbaum}, {Vega}, {Villanueva}, {Wolf}, {Youngblood}, {Ricker},
  {Vanderspek}, {Latham}, {Seager}, {Winn}, {Jenkins}, {Bakos}, {Brice{\~n}o},
  {Ciardi}, {Cloutier}, {Conti}, {Couperus}, {Di Sora}, {Eisner}, {Everett},
  {Gan}, {Hartman}, {Henry}, {Isopi}, {Jao}, {Jensen}, {Law}, {Mallia},
  {Matson}, {Shappee}, {Le Wood}, \& {Winters}}]{Gilbert:2020}
{Gilbert}, E.~A., {Barclay}, T., {Schlieder}, J.~E., {et~al.} 2020, \aj, 160,
  116, \dodoi{10.3847/1538-3881/aba4b2}

\bibitem[{{Gilbert} {et~al.}(2022){Gilbert}, {Barclay}, {Quintana},
  {Walkowicz}, {Vega}, {Schlieder}, {Monsue}, {Cale}, {Collins}, {Gaidos}, {El
  Mufti}, {Reefe}, {Plavchan}, {Tanner}, {Wittenmyer}, {Wittrock}, {Jenkins},
  {Latham}, {Ricker}, {Rose}, {Seager}, {Vanderspek}, \& {Winn}}]{Gilbert:2022}
{Gilbert}, E.~A., {Barclay}, T., {Quintana}, E.~V., {et~al.} 2022, \aj, 163,
  147, \dodoi{10.3847/1538-3881/ac23ca}

\bibitem[{{Gillon} {et~al.}(2017){Gillon}, {Triaud}, {Demory}, {Jehin}, {Agol},
  {Deck}, {Lederer}, {de Wit}, {Burdanov}, {Ingalls}, {Bolmont}, {Leconte},
  {Raymond}, {Selsis}, {Turbet}, {Barkaoui}, {Burgasser}, {Burleigh}, {Carey},
  {Chaushev}, {Copperwheat}, {Delrez}, {Fernandes}, {Holdsworth}, {Kotze}, {Van
  Grootel}, {Almleaky}, {Benkhaldoun}, {Magain}, \& {Queloz}}]{Gillon:2017}
{Gillon}, M., {Triaud}, A. H.~M.~J., {Demory}, B.-O., {et~al.} 2017, \nat, 542,
  456, \dodoi{10.1038/nature21360}

\bibitem[{{Goyal} \& {Wang}(2022)}]{Goyal:2022}
{Goyal}, A.~V., \& {Wang}, S. 2022, \apj, 933, 162,
  \dodoi{10.3847/1538-4357/ac7562}

\bibitem[{{Green} {et~al.}(2021){Green}, {Boardsen}, \& {Dong}}]{Green:2021}
{Green}, J., {Boardsen}, S., \& {Dong}, C. 2021, \apjl, 907, L45,
  \dodoi{10.3847/2041-8213/abd93a}

\bibitem[{{Hadden} \& {Lithwick}(2014)}]{Hadden:2014}
{Hadden}, S., \& {Lithwick}, Y. 2014, \apj, 787, 80,
  \dodoi{10.1088/0004-637X/787/1/80}

\bibitem[{{Hamann} {et~al.}(2019){Hamann}, {Montet}, {Fabrycky}, {Agol}, \&
  {Kruse}}]{Hamann:2019}
{Hamann}, A., {Montet}, B.~T., {Fabrycky}, D.~C., {Agol}, E., \& {Kruse}, E.
  2019, \aj, 158, 133, \dodoi{10.3847/1538-3881/ab32e3}

\bibitem[{{Hartman} {et~al.}(2015){Hartman}, {Bayliss}, {Brahm}, {Bakos},
  {Mancini}, {Jord{\'a}n}, {Penev}, {Rabus}, {Zhou}, {Butler}, {Espinoza}, {de
  Val-Borro}, {Bhatti}, {Csubry}, {Ciceri}, {Henning}, {Schmidt}, {Arriagada},
  {Shectman}, {Crane}, {Thompson}, {Suc}, {Cs{\'a}k}, {Tan}, {Noyes},
  {L{\'a}z{\'a}r}, {Papp}, \& {S{\'a}ri}}]{Hartman:2015}
{Hartman}, J.~D., {Bayliss}, D., {Brahm}, R., {et~al.} 2015, \aj, 149, 166,
  \dodoi{10.1088/0004-6256/149/5/166}

\bibitem[{{Hinkel} {et~al.}(2014){Hinkel}, {Timmes}, {Young}, {Pagano}, \&
  {Turnbull}}]{Hinkel:2014}
{Hinkel}, N.~R., {Timmes}, F.~X., {Young}, P.~A., {Pagano}, M.~D., \&
  {Turnbull}, M.~C. 2014, \aj, 148, 54, \dodoi{10.1088/0004-6256/148/3/54}

\bibitem[{{Howard} {et~al.}(2012){Howard}, {Marcy}, {Bryson}, {Jenkins},
  {Rowe}, {Batalha}, {Borucki}, {Koch}, {Dunham}, {Gautier}, {Van Cleve},
  {Cochran}, {Latham}, {Lissauer}, {Torres}, {Brown}, {Gilliland}, {Buchhave},
  {Caldwell}, {Christensen-Dalsgaard}, {Ciardi}, {Fressin}, {Haas}, {Howell},
  {Kjeldsen}, {Seager}, {Rogers}, {Sasselov}, {Steffen}, {Basri},
  {Charbonneau}, {Christiansen}, {Clarke}, {Dupree}, {Fabrycky}, {Fischer},
  {Ford}, {Fortney}, {Tarter}, {Girouard}, {Holman}, {Johnson}, {Klaus},
  {Machalek}, {Moorhead}, {Morehead}, {Ragozzine}, {Tenenbaum}, {Twicken},
  {Quinn}, {Isaacson}, {Shporer}, {Lucas}, {Walkowicz}, {Welsh}, {Boss},
  {Devore}, {Gould}, {Smith}, {Morris}, {Prsa}, {Morton}, {Still}, {Thompson},
  {Mullally}, {Endl}, \& {MacQueen}}]{Howard:2012}
{Howard}, A.~W., {Marcy}, G.~W., {Bryson}, S.~T., {et~al.} 2012, \apjs, 201,
  15, \dodoi{10.1088/0067-0049/201/2/15}

\bibitem[{{Johnson} {et~al.}(2012){Johnson}, {Gazak}, {Apps}, {Muirhead},
  {Crepp}, {Crossfield}, {Boyajian}, {von Braun}, {Rojas-Ayala}, {Howard},
  {Covey}, {Schlawin}, {Hamren}, {Morton}, {Marcy}, \& {Lloyd}}]{Johnson:2012}
{Johnson}, J.~A., {Gazak}, J.~Z., {Apps}, K., {et~al.} 2012, \aj, 143, 111,
  \dodoi{10.1088/0004-6256/143/5/111}

\bibitem[{{Jontof-Hutter} {et~al.}(2015){Jontof-Hutter}, {Rowe}, {Lissauer},
  {Fabrycky}, \& {Ford}}]{Jontof-Hutter:2015}
{Jontof-Hutter}, D., {Rowe}, J.~F., {Lissauer}, J.~J., {Fabrycky}, D.~C., \&
  {Ford}, E.~B. 2015, \nat, 522, 321, \dodoi{10.1038/nature14494}

\bibitem[{{Jord{\'a}n} {et~al.}(2022){Jord{\'a}n}, {Hartman}, {Bayliss},
  {Bakos}, {Brahm}, {Bryant}, {Csubry}, {Henning}, {Hobson}, {Mancini},
  {Penev}, {Rabus}, {Suc}, {de Val-Borro}, {Wallace}, {Barkaoui}, {Ciardi},
  {Collins}, {Esparza-Borges}, {Furlan}, {Gan}, {Benkhaldoun}, {Ghachoui},
  {Gillon}, {Howell}, {Jehin}, {Fukui}, {Kawauchi}, {Livingston}, {Luque},
  {Matson}, {Matthews}, {Osborn}, {Murgas}, {Narita}, {Palle}, {Parvianen}, \&
  {Waalkes}}]{Jordan:2022}
{Jord{\'a}n}, A., {Hartman}, J.~D., {Bayliss}, D., {et~al.} 2022, \aj, 163,
  125, \dodoi{10.3847/1538-3881/ac4a77}

\bibitem[{{Kanodia} {et~al.}(2020){Kanodia}, {Ca{\~n}as}, {Stefansson},
  {Ninan}, {Hebb}, {Lin}, {Baran}, {Maney}, {Terrien}, {Mahadevan}, {Cochran},
  {Endl}, {Dong}, {Bender}, {Diddams}, {Ford}, {Fredrick}, {Halverson},
  {Hearty}, {Metcalf}, {Monson}, {Ramsey}, {Robertson}, {Roy}, {Schwab}, \&
  {Wright}}]{Kanodia:2020}
{Kanodia}, S., {Ca{\~n}as}, C.~I., {Stefansson}, G., {et~al.} 2020, \apj, 899,
  29, \dodoi{10.3847/1538-4357/aba0a2}

\bibitem[{{Kemmer} {et~al.}(2020){Kemmer}, {Stock}, {Kossakowski}, {Kaminski},
  {Molaverdikhani}, {Schlecker}, {Caballero}, {Amado}, {Astudillo-Defru},
  {Bonfils}, {Ciardi}, {Collins}, {Espinoza}, {Fukui}, {Hirano}, {Jenkins},
  {Latham}, {Matthews}, {Narita}, {Pall{\'e}}, {Parviainen}, {Quirrenbach},
  {Reiners}, {Ribas}, {Ricker}, {Schlieder}, {Seager}, {Vanderspek}, {Winn},
  {Almenara}, {B{\'e}jar}, {Bluhm}, {Bouchy}, {Boyd}, {Christiansen},
  {Cifuentes}, {Cloutier}, {Collins}, {Cort{\'e}s-Contreras}, {Crossfield},
  {Crouzet}, {de Leon}, {Della-Rose}, {Delfosse}, {Dreizler}, {Esparza-Borges},
  {Essack}, {Forveille}, {Figueira}, {Galad{\'\i}-Enr{\'\i}quez}, {Gan},
  {Glidden}, {Gonzales}, {Guerra}, {Harakawa}, {Hatzes}, {Henning}, {Herrero},
  {Hodapp}, {Hori}, {Howell}, {Ikoma}, {Isogai}, {Jeffers}, {K{\"u}rster},
  {Kawauchi}, {Kimura}, {Klagyivik}, {Kotani}, {Kurokawa}, {Kusakabe},
  {Kuzuhara}, {Lafarga}, {Livingston}, {Luque}, {Matson}, {Morales}, {Mori},
  {Muirhead}, {Murgas}, {Nishikawa}, {Nishiumi}, {Omiya}, {Reffert},
  {Rodr{\'\i}guez L{\'o}pez}, {Santos}, {Sch{\"o}fer}, {Schwarz}, {Shiao},
  {Tamura}, {Terada}, {Twicken}, {Ueda}, {Vievard}, {Watanabe}, \&
  {Zechmeister}}]{Kemmer:2020}
{Kemmer}, J., {Stock}, S., {Kossakowski}, D., {et~al.} 2020, \aap, 642, A236,
  \dodoi{10.1051/0004-6361/202038967}

\bibitem[{{Kemmer} {et~al.}(2022){Kemmer}, {Dreizler}, {Kossakowski}, {Stock},
  {Quirrenbach}, {Caballero}, {Amado}, {Collins}, {Espinoza}, {Herrero},
  {Jenkins}, {Latham}, {Lillo-Box}, {Narita}, {Pall{\'e}}, {Reiners}, {Ribas},
  {Ricker}, {Rodr{\'\i}guez}, {Seager}, {Vanderspek}, {Wells}, {Winn},
  {Aceituno}, {B{\'e}jar}, {Barclay}, {Bluhm}, {Chaturvedi}, {Cifuentes},
  {Collins}, {Cort{\'e}s-Contreras}, {Demory}, {Fausnaugh}, {Fukui}, {G{\'o}mez
  Maqueo Chew}, {Galad{\'\i}-Enr{\'\i}quez}, {Gan}, {Gillon}, {Golovin},
  {Hatzes}, {Henning}, {Huang}, {Jeffers}, {Kaminski}, {Kunimoto},
  {K{\"u}rster}, {L{\'o}pez-Gonz{\'a}lez}, {Lafarga}, {Luque}, {McCormac},
  {Molaverdikhani}, {Montes}, {Morales}, {Passegger}, {Reffert}, {Sabin},
  {Sch{\"o}fer}, {Schanche}, {Schlecker}, {Schroffenegger}, {Schwarz},
  {Schweitzer}, {Sota}, {Tenenbaum}, {Trifonov}, {Vanaverbeke}, \&
  {Zechmeister}}]{Kemmer:2022}
{Kemmer}, J., {Dreizler}, S., {Kossakowski}, D., {et~al.} 2022, \aap, 659, A17,
  \dodoi{10.1051/0004-6361/202142653}

\bibitem[{{Kossakowski} {et~al.}(2021){Kossakowski}, {Kemmer}, {Bluhm},
  {Stock}, {Caballero}, {B{\'e}jar}, {Guill{\'e}n}, {Lodieu}, {Collins},
  {Oshagh}, {Schlecker}, {Espinoza}, {Pall{\'e}}, {Henning}, {Kreidberg},
  {K{\"u}rster}, {Amado}, {Anderson}, {Morales}, {Cartwright}, {Charbonneau},
  {Chaturvedi}, {Cifuentes}, {Conti}, {Cort{\'e}s-Contreras}, {Dreizler},
  {Galad{\'\i}-Enr{\'\i}quez}, {Guerra}, {Hart}, {Hellier}, {Henze}, {Herrero},
  {Jeffers}, {Jenkins}, {Jensen}, {Kaminski}, {Kielkopf}, {Kunimoto},
  {Lafarga}, {Latham}, {Lillo-Box}, {Luque}, {Molaverdikhani}, {Montes},
  {Morello}, {Morgan}, {Nowak}, {Pavlov}, {Perger}, {Quintana}, {Quirrenbach},
  {Reffert}, {Reiners}, {Ricker}, {Ribas}, {L{\'o}pez}, {Osorio}, {Seager},
  {Sch{\"o}fer}, {Schweitzer}, {Trifonov}, {Vanaverbeke}, {Vanderspek}, {West},
  {Winn}, \& {Zechmeister}}]{Kossakowski:2021}
{Kossakowski}, D., {Kemmer}, J., {Bluhm}, P., {et~al.} 2021, \aap, 656, A124,
  \dodoi{10.1051/0004-6361/202141587}

\bibitem[{{Lam} {et~al.}(2021){Lam}, {Csizmadia}, {Astudillo-Defru}, {Bonfils},
  {Gandolfi}, {Padovan}, {Esposito}, {Hellier}, {Hirano}, {Livingston},
  {Murgas}, {Smith}, {Collins}, {Mathur}, {Garcia}, {Howell}, {Santos}, {Dai},
  {Ricker}, {Vanderspek}, {Latham}, {Seager}, {Winn}, {Jenkins}, {Albrecht},
  {Almenara}, {Artigau}, {Barrag{\'a}n}, {Bouchy}, {Cabrera}, {Charbonneau},
  {Chaturvedi}, {Chaushev}, {Christiansen}, {Cochran}, {De Meideiros},
  {Delfosse}, {D{\'\i}az}, {Doyon}, {Eigm{\"u}ller}, {Figueira}, {Forveille},
  {Fridlund}, {Gaisn{\'e}}, {Goffo}, {Georgieva}, {Grziwa}, {Guenther},
  {Hatzes}, {Johnson}, {Kab{\'a}th}, {Knudstrup}, {Korth}, {Lewin}, {Lissauer},
  {Lovis}, {Luque}, {Melo}, {Morgan}, {Morris}, {Mayor}, {Narita}, {Osborne},
  {Palle}, {Pepe}, {Persson}, {Quinn}, {Rauer}, {Redfield}, {Schlieder},
  {S{\'e}gransan}, {Serrano}, {Smith}, {{\v{S}}ubjak}, {Twicken}, {Udry}, {Van
  Eylen}, \& {Vezie}}]{Lam:2021}
{Lam}, K. W.~F., {Csizmadia}, S., {Astudillo-Defru}, N., {et~al.} 2021,
  Science, 374, 1271, \dodoi{10.1126/science.aay3253}

\bibitem[{{Latham} {et~al.}(2011){Latham}, {Rowe}, {Quinn}, {Batalha},
  {Borucki}, {Brown}, {Bryson}, {Buchhave}, {Caldwell}, {Carter},
  {Christiansen}, {Ciardi}, {Cochran}, {Dunham}, {Fabrycky}, {Ford}, {Gautier},
  {Gilliland}, {Holman}, {Howell}, {Ibrahim}, {Isaacson}, {Jenkins}, {Koch},
  {Lissauer}, {Marcy}, {Quintana}, {Ragozzine}, {Sasselov}, {Shporer},
  {Steffen}, {Welsh}, \& {Wohler}}]{Latham:2011}
{Latham}, D.~W., {Rowe}, J.~F., {Quinn}, S.~N., {et~al.} 2011, \apjl, 732, L24,
  \dodoi{10.1088/2041-8205/732/2/L24}

\bibitem[{{Lillo-Box} {et~al.}(2020){Lillo-Box}, {Figueira}, {Leleu},
  {Acu{\~n}a}, {Faria}, {Hara}, {Santos}, {Correia}, {Robutel}, {Deleuil},
  {Barrado}, {Sousa}, {Bonfils}, {Mousis}, {Almenara}, {Astudillo-Defru},
  {Marcq}, {Udry}, {Lovis}, \& {Pepe}}]{LilloBox}
{Lillo-Box}, J., {Figueira}, P., {Leleu}, A., {et~al.} 2020, \aap, 642, A121,
  \dodoi{10.1051/0004-6361/202038922}

\bibitem[{{Limbach} \& {Turner}(2015)}]{Limbach:2015}
{Limbach}, M.~A., \& {Turner}, E.~L. 2015, Proceedings of the National Academy
  of Science, 112, 20, \dodoi{10.1073/pnas.1406545111}

\bibitem[{{Lissauer} {et~al.}(2012){Lissauer}, {Marcy}, {Rowe}, {Bryson},
  {Adams}, {Buchhave}, {Ciardi}, {Cochran}, {Fabrycky}, {Ford}, {Fressin},
  {Geary}, {Gilliland}, {Holman}, {Howell}, {Jenkins}, {Kinemuchi}, {Koch},
  {Morehead}, {Ragozzine}, {Seader}, {Tanenbaum}, {Torres}, \&
  {Twicken}}]{Lissauer:2012}
{Lissauer}, J.~J., {Marcy}, G.~W., {Rowe}, J.~F., {et~al.} 2012, \apj, 750,
  112, \dodoi{10.1088/0004-637X/750/2/112}

\bibitem[{{Lissauer} {et~al.}(2014){Lissauer}, {Marcy}, {Bryson}, {Rowe},
  {Jontof-Hutter}, {Agol}, {Borucki}, {Carter}, {Ford}, {Gilliland}, {Kolbl},
  {Star}, {Steffen}, \& {Torres}}]{Lissauer:2014}
{Lissauer}, J.~J., {Marcy}, G.~W., {Bryson}, S.~T., {et~al.} 2014, \apj, 784,
  44, \dodoi{10.1088/0004-637X/784/1/44}

\bibitem[{{Luque} \& {Pall{\'e}}(2022)}]{Luque:2022}
{Luque}, R., \& {Pall{\'e}}, E. 2022, arXiv e-prints, arXiv:2209.03871.
\newblock \doarXiv{2209.03871}

\bibitem[{{Luque} {et~al.}(2019){Luque}, {Pall{\'e}}, {Kossakowski},
  {Dreizler}, {Kemmer}, {Espinoza}, {Burt}, {Anglada-Escud{\'e}}, {B{\'e}jar},
  {Caballero}, {Collins}, {Collins}, {Cort{\'e}s-Contreras},
  {D{\'\i}ez-Alonso}, {Feng}, {Hatzes}, {Hellier}, {Henning}, {Jeffers},
  {Kaltenegger}, {K{\"u}rster}, {Madden}, {Molaverdikhani}, {Montes}, {Narita},
  {Nowak}, {Ofir}, {Oshagh}, {Parviainen}, {Quirrenbach}, {Reffert}, {Reiners},
  {Rodr{\'\i}guez-L{\'o}pez}, {Schlecker}, {Stock}, {Trifonov}, {Winn},
  {Zapatero Osorio}, {Zechmeister}, {Amado}, {Anderson}, {Batalha}, {Bauer},
  {Bluhm}, {Burke}, {Butler}, {Caldwell}, {Chen}, {Crane}, {Dragomir},
  {Dressing}, {Dynes}, {Jenkins}, {Kaminski}, {Klahr}, {Kotani}, {Lafarga},
  {Latham}, {Lewin}, {McDermott}, {Monta{\~n}{\'e}s-Rodr{\'\i}guez}, {Morales},
  {Murgas}, {Nagel}, {Pedraz}, {Ribas}, {Ricker}, {Rowden}, {Seager},
  {Shectman}, {Tamura}, {Teske}, {Twicken}, {Vanderspeck}, {Wang}, \&
  {Wohler}}]{Luque:2019}
{Luque}, R., {Pall{\'e}}, E., {Kossakowski}, D., {et~al.} 2019, \aap, 628, A39,
  \dodoi{10.1051/0004-6361/201935801}

\bibitem[{{Luque} {et~al.}(2021){Luque}, {Serrano}, {Molaverdikhani}, {Nixon},
  {Livingston}, {Guenther}, {Pall{\'e}}, {Madhusudhan}, {Nowak}, {Korth},
  {Cochran}, {Hirano}, {Chaturvedi}, {Goffo}, {Albrecht}, {Barrag{\'a}n},
  {Brice{\~n}o}, {Cabrera}, {Charbonneau}, {Cloutier}, {Collins}, {Collins},
  {Col{\'o}n}, {Crossfield}, {Csizmadia}, {Dai}, {Deeg}, {Esposito},
  {Fridlund}, {Gandolfi}, {Georgieva}, {Glidden}, {Goeke}, {Grziwa}, {Hatzes},
  {Henze}, {Howell}, {Irwin}, {Jenkins}, {Jensen}, {K{\'a}bath}, {Kidwell},
  {Kielkopf}, {Knudstrup}, {Lam}, {Latham}, {Lissauer}, {Mann}, {Matthews},
  {Mireles}, {Narita}, {Paegert}, {Persson}, {Redfield}, {Ricker}, {Rodler},
  {Schlieder}, {Scott}, {Seager}, {{\v{S}}ubjak}, {Tan}, {Ting}, {Vanderspek},
  {Van Eylen}, {Winn}, \& {Ziegler}}]{Luque:2021}
{Luque}, R., {Serrano}, L.~M., {Molaverdikhani}, K., {et~al.} 2021, \aap, 645,
  A41, \dodoi{10.1051/0004-6361/202039455}

\bibitem[{{Maciejewski} {et~al.}(2014){Maciejewski}, {Niedzielski}, {Nowak},
  {Pall{\'e}}, {Tingley}, {Errmann}, \& {Neuh{\"a}user}}]{Maciejewski:2014}
{Maciejewski}, G., {Niedzielski}, A., {Nowak}, G., {et~al.} 2014, \actaa, 64,
  323.
\newblock \doarXiv{1501.02711}

\bibitem[{{Mahadevan} {et~al.}(2012){Mahadevan}, {Ramsey}, {Bender}, {Terrien},
  {Wright}, {Halverson}, {Hearty}, {Nelson}, {Burton}, {Redman}, {Osterman},
  {Diddams}, {Kasting}, {Endl}, \& {Deshpande}}]{Mahadevan:2012}
{Mahadevan}, S., {Ramsey}, L., {Bender}, C., {et~al.} 2012, in Society of
  Photo-Optical Instrumentation Engineers (SPIE) Conference Series, Vol. 8446,
  Ground-based and Airborne Instrumentation for Astronomy IV, ed. I.~S.
  {McLean}, S.~K. {Ramsay}, \& H.~{Takami}, 84461S, \dodoi{10.1117/12.926102}

\bibitem[{{Maldonado} {et~al.}(2020){Maldonado}, {Micela}, {Baratella},
  {D'Orazi}, {Affer}, {Biazzo}, {Lanza}, {Maggio}, {Gonz{\'a}lez
  Hern{\'a}ndez}, {Perger}, {Pinamonti}, {Scandariato}, {Sozzetti}, {Locci},
  {Di Maio}, {Bignamini}, {Claudi}, {Molinari}, {Rebolo}, {Ribas},
  {Toledo-Padr{\'o}n}, {Covino}, {Desidera}, {Herrero}, {Morales},
  {Su{\'a}rez-Mascare{\~n}o}, {Pagano}, {Petralia}, {Piotto}, \&
  {Poretti}}]{Maldonado:2020}
{Maldonado}, J., {Micela}, G., {Baratella}, M., {et~al.} 2020, \aap, 644, A68,
  \dodoi{10.1051/0004-6361/202039478}

\bibitem[{{Martioli} {et~al.}(2021){Martioli}, {H{\'e}brard}, {Correia},
  {Laskar}, \& {Lecavelier des Etangs}}]{Martioli:2021}
{Martioli}, E., {H{\'e}brard}, G., {Correia}, A.~C.~M., {Laskar}, J., \&
  {Lecavelier des Etangs}, A. 2021, \aap, 649, A177,
  \dodoi{10.1051/0004-6361/202040235}

\bibitem[{{McDonough} \& {Yoshizaki}(2021)}]{McDonough:2021}
{McDonough}, W.~F., \& {Yoshizaki}, T. 2021, Progress in Earth and Planetary
  Science, 8, 39, \dodoi{10.1186/s40645-021-00429-4}

\bibitem[{{Ment} \& {Charbonneau}(2023)}]{Ment:2023}
{Ment}, K., \& {Charbonneau}, D. 2023, arXiv e-prints, arXiv:2302.04242,
  \dodoi{10.48550/arXiv.2302.04242}

\bibitem[{{Millholland} {et~al.}(2017){Millholland}, {Wang}, \&
  {Laughlin}}]{Millholland:2017}
{Millholland}, S., {Wang}, S., \& {Laughlin}, G. 2017, \apjl, 849, L33,
  \dodoi{10.3847/2041-8213/aa9714}

\bibitem[{{Morton} \& {Winn}(2014)}]{Morton:2014}
{Morton}, T.~D., \& {Winn}, J.~N. 2014, \apj, 796, 47,
  \dodoi{10.1088/0004-637X/796/1/47}

\bibitem[{{Mulders} {et~al.}(2015){Mulders}, {Pascucci}, \&
  {Apai}}]{Mulders:2015}
{Mulders}, G.~D., {Pascucci}, I., \& {Apai}, D. 2015, \apj, 814, 130,
  \dodoi{10.1088/0004-637X/814/2/130}

\bibitem[{{Munoz Romero} \& {Kempton}(2018)}]{Munoz-Romero:2018}
{Munoz Romero}, C.~E., \& {Kempton}, E. M.~R. 2018, \aj, 155, 134,
  \dodoi{10.3847/1538-3881/aaab5e}

\bibitem[{{Murgas} {et~al.}(2021){Murgas}, {Astudillo-Defru}, {Bonfils},
  {Crossfield}, {Almenara}, {Livingston}, {Stassun}, {Korth}, {Orell-Miquel},
  {Morello}, {Eastman}, {Lissauer}, {Kane}, {Morales}, {Werner}, {Gorjian},
  {Benneke}, {Dragomir}, {Matthews}, {Howell}, {Ciardi}, {Gonzales}, {Matson},
  {Beichman}, {Schlieder}, {Collins}, {Collins}, {Jensen}, {Evans}, {Pozuelos},
  {Gillon}, {Jehin}, {Barkaoui}, {Artigau}, {Bouchy}, {Charbonneau},
  {Delfosse}, {D{\'\i}az}, {Doyon}, {Figueira}, {Forveille}, {Lovis}, {Melo},
  {Gaisn{\'e}}, {Pepe}, {Santos}, {S{\'e}gransan}, {Udry}, {Goeke}, {Levine},
  {Quintana}, {Guerrero}, {Mireles}, {Caldwell}, {Tenenbaum}, {Brasseur},
  {Ricker}, {Vanderspek}, {Latham}, {Seager}, {Winn}, \&
  {Jenkins}}]{Murgas:2021}
{Murgas}, F., {Astudillo-Defru}, N., {Bonfils}, X., {et~al.} 2021, \aap, 653,
  A60, \dodoi{10.1051/0004-6361/202140718}

\bibitem[{{Owen} \& {Wu}(2017)}]{Owen:2017}
{Owen}, J.~E., \& {Wu}, Y. 2017, \apj, 847, 29,
  \dodoi{10.3847/1538-4357/aa890a}

\bibitem[{{Pecaut} \& {Mamajek}(2013)}]{Pecaut:2013}
{Pecaut}, M.~J., \& {Mamajek}, E.~E. 2013, \apjs, 208, 9,
  \dodoi{10.1088/0067-0049/208/1/9}

\bibitem[{{Piaulet} {et~al.}(2023){Piaulet}, {Benneke}, {Almenara}, {Dragomir},
  {Knutson}, {Thorngren}, {Peterson}, {Crossfield}, {Kempton}, {Kubyshkina},
  {Howard}, {Angus}, {Isaacson}, {Weiss}, {Beichman}, {Fortney}, {Fossati},
  {Lammer}, {McCullough}, {Morley}, \& {Wong}}]{Piaulet:2023}
{Piaulet}, C., {Benneke}, B., {Almenara}, J.~M., {et~al.} 2023, Nature
  Astronomy, 7, 206, \dodoi{10.1038/s41550-022-01835-4}

\bibitem[{{Plotnykov} \& {Valencia}(2020)}]{Plotnykov:2020}
{Plotnykov}, M., \& {Valencia}, D. 2020, \mnras, 499, 932,
  \dodoi{10.1093/mnras/staa2615}

\bibitem[{{Radica} {et~al.}(2022){Radica}, {Artigau}, {Lafreni{\'e}re},
  {Cadieux}, {Cook}, {Doyon}, {Amado}, {Caballero}, {Henning}, {Quirrenbach},
  {Reiners}, \& {Ribas}}]{Radica:2022}
{Radica}, M., {Artigau}, {\'E}., {Lafreni{\'e}re}, D., {et~al.} 2022, \mnras,
  517, 5050, \dodoi{10.1093/mnras/stac3024}

\bibitem[{{Ricker} {et~al.}(2015){Ricker}, {Winn}, {Vanderspek}, {Latham},
  {Bakos}, {Bean}, {Berta-Thompson}, {Brown}, {Buchhave}, {Butler}, {Butler},
  {Chaplin}, {Charbonneau}, {Christensen-Dalsgaard}, {Clampin}, {Deming},
  {Doty}, {De Lee}, {Dressing}, {Dunham}, {Endl}, {Fressin}, {Ge}, {Henning},
  {Holman}, {Howard}, {Ida}, {Jenkins}, {Jernigan}, {Johnson}, {Kaltenegger},
  {Kawai}, {Kjeldsen}, {Laughlin}, {Levine}, {Lin}, {Lissauer}, {MacQueen},
  {Marcy}, {McCullough}, {Morton}, {Narita}, {Paegert}, {Palle}, {Pepe},
  {Pepper}, {Quirrenbach}, {Rinehart}, {Sasselov}, {Sato}, {Seager},
  {Sozzetti}, {Stassun}, {Sullivan}, {Szentgyorgyi}, {Torres}, {Udry}, \&
  {Villasenor}}]{Ricker:2015}
{Ricker}, G.~R., {Winn}, J.~N., {Vanderspek}, R., {et~al.} 2015, Journal of
  Astronomical Telescopes, Instruments, and Systems, 1, 014003,
  \dodoi{10.1117/1.JATIS.1.1.014003}

\bibitem[{{Rodr{\'\i}guez Mart{\'\i}nez} {et~al.}(2023){Rodr{\'\i}guez
  Mart{\'\i}nez}, {Gaudi}, {Schulze}, {Acu{\~n}a}, {Kolecki}, {Johnson},
  {Asnodkar}, {Boley}, {Deleuil}, {Mousis}, {Panero}, \&
  {Wang}}]{RodriguezMartinez:2023}
{Rodr{\'\i}guez Mart{\'\i}nez}, R., {Gaudi}, B.~S., {Schulze}, J.~G., {et~al.}
  2023, \aj, 165, 97, \dodoi{10.3847/1538-3881/acb04b}

\bibitem[{{Rowe} {et~al.}(2014){Rowe}, {Bryson}, {Marcy}, {Lissauer},
  {Jontof-Hutter}, {Mullally}, {Gilliland}, {Issacson}, {Ford}, {Howell},
  {Borucki}, {Haas}, {Huber}, {Steffen}, {Thompson}, {Quintana}, {Barclay},
  {Still}, {Fortney}, {Gautier}, {Hunter}, {Caldwell}, {Ciardi}, {Devore},
  {Cochran}, {Jenkins}, {Agol}, {Carter}, \& {Geary}}]{Rowe:2014}
{Rowe}, J.~F., {Bryson}, S.~T., {Marcy}, G.~W., {et~al.} 2014, \apj, 784, 45,
  \dodoi{10.1088/0004-637X/784/1/45}

\bibitem[{{Sagear} \& {Ballard}(2023)}]{Sagear:2023}
{Sagear}, S., \& {Ballard}, S. 2023, arXiv e-prints, arXiv:2305.17157,
  \dodoi{10.48550/arXiv.2305.17157}

\bibitem[{{Santerne} {et~al.}(2018){Santerne}, {Brugger}, {Armstrong},
  {Adibekyan}, {Lillo-Box}, {Gosselin}, {Aguichine}, {Almenara}, {Barrado},
  {Barros}, {Bayliss}, {Boisse}, {Bonomo}, {Bouchy}, {Brown}, {Deleuil},
  {Delgado Mena}, {Demangeon}, {D{\'\i}az}, {Doyle}, {Dumusque}, {Faedi},
  {Faria}, {Figueira}, {Foxell}, {Giles}, {H{\'e}brard}, {Hojjatpanah},
  {Hobson}, {Jackman}, {King}, {Kirk}, {Lam}, {Ligi}, {Lovis}, {Louden},
  {McCormac}, {Mousis}, {Neal}, {Osborn}, {Pepe}, {Pollacco}, {Santos},
  {Sousa}, {Udry}, \& {Vigan}}]{Santerne:2018}
{Santerne}, A., {Brugger}, B., {Armstrong}, D.~J., {et~al.} 2018, Nature
  Astronomy, 2, 393, \dodoi{10.1038/s41550-018-0420-5}

\bibitem[{{Santos} {et~al.}(2004){Santos}, {Israelian}, \&
  {Mayor}}]{Santos:2004}
{Santos}, N.~C., {Israelian}, G., \& {Mayor}, M. 2004, \aap, 415, 1153,
  \dodoi{10.1051/0004-6361:20034469}

\bibitem[{{Sarkis} {et~al.}(2018){Sarkis}, {Henning}, {K{\"u}rster},
  {Trifonov}, {Zechmeister}, {Tal-Or}, {Anglada-Escud{\'e}}, {Hatzes},
  {Lafarga}, {Dreizler}, {Ribas}, {Caballero}, {Reiners}, {Mallonn}, {Morales},
  {Kaminski}, {Aceituno}, {Amado}, {B{\'e}jar}, {Hagen}, {Jeffers},
  {Quirrenbach}, {Launhardt}, {Marvin}, \& {Montes}}]{sarkis:2018}
{Sarkis}, P., {Henning}, T., {K{\"u}rster}, M., {et~al.} 2018, \aj, 155, 257,
  \dodoi{10.3847/1538-3881/aac108}

\bibitem[{{Schlecker} {et~al.}(2022){Schlecker}, {Burn}, {Sabotta}, {Seifert},
  {Henning}, {Emsenhuber}, {Mordasini}, {Reffert}, {Shan}, \&
  {Klahr}}]{Schlecker:2022}
{Schlecker}, M., {Burn}, R., {Sabotta}, S., {et~al.} 2022, \aap, 664, A180,
  \dodoi{10.1051/0004-6361/202142543}

\bibitem[{{Schulze} {et~al.}(2021){Schulze}, {Wang}, {Johnson}, {Gaudi},
  {Unterborn}, \& {Panero}}]{Schulze:2021}
{Schulze}, J.~G., {Wang}, J., {Johnson}, J.~A., {et~al.} 2021, \psj, 2, 113,
  \dodoi{10.3847/PSJ/abcaa8}

\bibitem[{{Seifahrt} {et~al.}(2018){Seifahrt}, {St{\"u}rmer}, {Bean}, \&
  {Schwab}}]{Seifahrt:2018}
{Seifahrt}, A., {St{\"u}rmer}, J., {Bean}, J.~L., \& {Schwab}, C. 2018, in
  Society of Photo-Optical Instrumentation Engineers (SPIE) Conference Series,
  Vol. 10702, Ground-based and Airborne Instrumentation for Astronomy VII, ed.
  C.~J. {Evans}, L.~{Simard}, \& H.~{Takami}, 107026D,
  \dodoi{10.1117/12.2312936}

\bibitem[{{Shporer} {et~al.}(2020){Shporer}, {Collins}, {Astudillo-Defru},
  {Irwin}, {Bonfils}, {Collins}, {Matthews}, {Winters}, {Anderson},
  {Armstrong}, {Charbonneau}, {Cloutier}, {Daylan}, {Gan}, {G{\"u}nther},
  {Hellier}, {Horne}, {Huang}, {Jensen}, {Kielkopf}, {Palle}, {Sefako},
  {Stassun}, {Tan}, {Vanderburg}, {Ricker}, {Latham}, {Vanderspek}, {Seager},
  {Winn}, {Jenkins}, {Colon}, {Dressing}, {L{\'e}epine}, {Muirhead}, {Rose},
  {Twicken}, \& {Villasenor}}]{Shporer:2020}
{Shporer}, A., {Collins}, K.~A., {Astudillo-Defru}, N., {et~al.} 2020, \apjl,
  890, L7, \dodoi{10.3847/2041-8213/ab7020}

\bibitem[{{Skumanich}(1972)}]{Skumanich:1972}
{Skumanich}, A. 1972, \apj, 171, 565, \dodoi{10.1086/151310}

\bibitem[{{Soto} {et~al.}(2021){Soto}, {Anglada-Escud{\'e}}, {Dreizler},
  {Molaverdikhani}, {Kemmer}, {Rodr{\'\i}guez-L{\'o}pez}, {Lillo-Box},
  {Pall{\'e}}, {Espinoza}, {Caballero}, {Quirrenbach}, {Ribas}, {Reiners},
  {Narita}, {Hirano}, {Amado}, {B{\'e}jar}, {Bluhm}, {Burke}, {Caldwell},
  {Charbonneau}, {Cloutier}, {Collins}, {Cort{\'e}s-Contreras}, {Girardin},
  {Guerra}, {Harakawa}, {Hatzes}, {Irwin}, {Jenkins}, {Jensen}, {Kawauchi},
  {Kotani}, {Kudo}, {Kunimoto}, {Kuzuhara}, {Latham}, {Montes}, {Morales},
  {Mori}, {Nelson}, {Omiya}, {Pedraz}, {Passegger}, {Rackham}, {Rudat},
  {Schlieder}, {Sch{\"o}fer}, {Schweitzer}, {Selezneva}, {Stockdale}, {Tamura},
  {Trifonov}, {Vanderspek}, \& {Watanabe}}]{Soto:2021}
{Soto}, M.~G., {Anglada-Escud{\'e}}, G., {Dreizler}, S., {et~al.} 2021, \aap,
  649, A144, \dodoi{10.1051/0004-6361/202140618}

\bibitem[{{Stefansson} {et~al.}(2020){Stefansson}, {Mahadevan}, {Maney},
  {Ninan}, {Robertson}, {Rajagopal}, {Haase}, {Allen}, {Ford}, {Winn},
  {Wolfgang}, {Dawson}, {Wisniewski}, {Bender}, {Ca{\~n}as}, {Cochran},
  {Diddams}, {Fredrick}, {Halverson}, {Hearty}, {Hebb}, {Kanodia}, {Levi},
  {Metcalf}, {Monson}, {Ramsey}, {Roy}, {Schwab}, {Terrien}, \&
  {Wright}}]{Stefansson:2020}
{Stefansson}, G., {Mahadevan}, S., {Maney}, M., {et~al.} 2020, \aj, 160, 192,
  \dodoi{10.3847/1538-3881/abb13a}

\bibitem[{{Steffen}(2016)}]{Steffen:2016}
{Steffen}, J.~H. 2016, \mnras, 457, 4384, \dodoi{10.1093/mnras/stw241}

\bibitem[{{Steffen} {et~al.}(2012){Steffen}, {Ragozzine}, {Fabrycky}, {Carter},
  {Ford}, {Holman}, {Rowe}, {Welsh}, {Borucki}, {Boss}, {Ciardi}, \&
  {Quinn}}]{Steffen:2012}
{Steffen}, J.~H., {Ragozzine}, D., {Fabrycky}, D.~C., {et~al.} 2012,
  Proceedings of the National Academy of Science, 109, 7982,
  \dodoi{10.1073/pnas.1120970109}

\bibitem[{{Steffen} {et~al.}(2013){Steffen}, {Fabrycky}, {Agol}, {Ford},
  {Morehead}, {Cochran}, {Lissauer}, {Adams}, {Borucki}, {Bryson}, {Caldwell},
  {Dupree}, {Jenkins}, {Robertson}, {Rowe}, {Seader}, {Thompson}, \&
  {Twicken}}]{Steffen:2013}
{Steffen}, J.~H., {Fabrycky}, D.~C., {Agol}, E., {et~al.} 2013, \mnras, 428,
  1077, \dodoi{10.1093/mnras/sts090}

\bibitem[{{Trifonov} {et~al.}(2021){Trifonov}, {Caballero}, {Morales},
  {Seifahrt}, {Ribas}, {Reiners}, {Bean}, {Luque}, {Parviainen}, {Pall{\'e}},
  {Stock}, {Zechmeister}, {Amado}, {Anglada-Escud{\'e}}, {Azzaro}, {Barclay},
  {B{\'e}jar}, {Bluhm}, {Casasayas-Barris}, {Cifuentes}, {Collins}, {Collins},
  {Cort{\'e}s-Contreras}, {de Leon}, {Dreizler}, {Dressing}, {Esparza-Borges},
  {Espinoza}, {Fausnaugh}, {Fukui}, {Hatzes}, {Hellier}, {Henning}, {Henze},
  {Herrero}, {Jeffers}, {Jenkins}, {Jensen}, {Kaminski}, {Kasper},
  {Kossakowski}, {K{\"u}rster}, {Lafarga}, {Latham}, {Mann}, {Molaverdikhani},
  {Montes}, {Montet}, {Murgas}, {Narita}, {Oshagh}, {Passegger}, {Pollacco},
  {Quinn}, {Quirrenbach}, {Ricker}, {Rodr{\'\i}guez L{\'o}pez}, {Sanz-Forcada},
  {Schwarz}, {Schweitzer}, {Seager}, {Shporer}, {Stangret}, {St{\"u}rmer},
  {Tan}, {Tenenbaum}, {Twicken}, {Vanderspek}, \& {Winn}}]{Trifonov:2021}
{Trifonov}, T., {Caballero}, J.~A., {Morales}, J.~C., {et~al.} 2021, Science,
  371, 1038, \dodoi{10.1126/science.abd7645}

\bibitem[{{Unterborn} {et~al.}(2023){Unterborn}, {Desch}, {Haldemann},
  {Lorenzo}, {Schulze}, {Hinkel}, \& {Panero}}]{Unterborn:2023}
{Unterborn}, C.~T., {Desch}, S.~J., {Haldemann}, J., {et~al.} 2023, \apj, 944,
  42, \dodoi{10.3847/1538-4357/acaa3b}

\bibitem[{{Unterborn} {et~al.}(2018){Unterborn}, {Desch}, {Hinkel}, \&
  {Lorenzo}}]{Unterborn:2018}
{Unterborn}, C.~T., {Desch}, S.~J., {Hinkel}, N.~R., \& {Lorenzo}, A. 2018,
  Nature Astronomy, 2, 297, \dodoi{10.1038/s41550-018-0411-6}

\bibitem[{{Unterborn} {et~al.}(2016){Unterborn}, {Dismukes}, \&
  {Panero}}]{Unterborn:2016}
{Unterborn}, C.~T., {Dismukes}, E.~E., \& {Panero}, W.~R. 2016, \apj, 819, 32,
  \dodoi{10.3847/0004-637X/819/1/32}

\bibitem[{{Unterborn} \& {Panero}(2019)}]{Unterborn:2019}
{Unterborn}, C.~T., \& {Panero}, W.~R. 2019, Journal of Geophysical Research
  (Planets), 124, 1704, \dodoi{10.1029/2018JE005844}

\bibitem[{{Van Eylen} {et~al.}(2019){Van Eylen}, {Albrecht}, {Huang},
  {MacDonald}, {Dawson}, {Cai}, {Foreman-Mackey}, {Lundkvist}, {Silva Aguirre},
  {Snellen}, \& {Winn}}]{VanEylen:2019}
{Van Eylen}, V., {Albrecht}, S., {Huang}, X., {et~al.} 2019, \aj, 157, 61,
  \dodoi{10.3847/1538-3881/aaf22f}

\bibitem[{{Van Eylen} {et~al.}(2021){Van Eylen}, {Astudillo-Defru}, {Bonfils},
  {Livingston}, {Hirano}, {Luque}, {Lam}, {Justesen}, {Winn}, {Gandolfi},
  {Nowak}, {Palle}, {Albrecht}, {Dai}, {Campos Estrada}, {Owen},
  {Foreman-Mackey}, {Fridlund}, {Korth}, {Mathur}, {Forveille}, {Mikal-Evans},
  {Osborne}, {Ho}, {Almenara}, {Artigau}, {Barrag{\'a}n}, {Barros}, {Bouchy},
  {Cabrera}, {Caldwell}, {Charbonneau}, {Chaturvedi}, {Cochran}, {Csizmadia},
  {Damasso}, {Delfosse}, {De Medeiros}, {D{\'\i}az}, {Doyon}, {Esposito},
  {F{\H{u}}r{\'e}sz}, {Figueira}, {Georgieva}, {Goffo}, {Grziwa}, {Guenther},
  {Hatzes}, {Jenkins}, {Kabath}, {Knudstrup}, {Latham}, {Lavie}, {Lovis},
  {Mennickent}, {Mullally}, {Murgas}, {Narita}, {Pepe}, {Persson}, {Redfield},
  {Ricker}, {Santos}, {Seager}, {Serrano}, {Smith}, {Su{\'a}rez Mascare{\~n}o},
  {Subjak}, {Twicken}, {Udry}, {Vanderspek}, \& {Zapatero
  Osorio}}]{VanEylen:2021}
{Van Eylen}, V., {Astudillo-Defru}, N., {Bonfils}, X., {et~al.} 2021, \mnras,
  507, 2154, \dodoi{10.1093/mnras/stab2143}

\bibitem[{{Wang} \& {Fischer}(2015)}]{Wang:2015}
{Wang}, J., \& {Fischer}, D.~A. 2015, \aj, 149, 14,
  \dodoi{10.1088/0004-6256/149/1/14}

\bibitem[{{Weiss} {et~al.}(2018{\natexlab{a}}){Weiss}, {Isaacson}, {Marcy},
  {Howard}, {Petigura}, {Fulton}, {Winn}, {Hirsch}, {Sinukoff}, {Rowe}, \&
  {California Kepler Survey}}]{Weiss:2018b}
{Weiss}, L.~M., {Isaacson}, H.~T., {Marcy}, G.~W., {et~al.} 2018{\natexlab{a}},
  \aj, 156, 254, \dodoi{10.3847/1538-3881/aae70a}

\bibitem[{{Weiss} {et~al.}(2018{\natexlab{b}}){Weiss}, {Marcy}, {Petigura},
  {Fulton}, {Howard}, {Winn}, {Isaacson}, {Morton}, {Hirsch}, {Sinukoff},
  {Cumming}, {Hebb}, \& {Cargile}}]{Weiss:2018a}
{Weiss}, L.~M., {Marcy}, G.~W., {Petigura}, E.~A., {et~al.} 2018{\natexlab{b}},
  \aj, 155, 48, \dodoi{10.3847/1538-3881/aa9ff6}

\bibitem[{{Winn} \& {Fabrycky}(2015)}]{Winn:2015}
{Winn}, J.~N., \& {Fabrycky}, D.~C. 2015, \araa, 53, 409,
  \dodoi{10.1146/annurev-astro-082214-122246}

\bibitem[{{Winters} {et~al.}(2022){Winters}, {Cloutier}, {Medina}, {Irwin},
  {Charbonneau}, {Astudillo-Defru}, {Bonfils}, {Howard}, {Isaacson}, {Bean},
  {Seifahrt}, {Teske}, {Eastman}, {Twicken}, {Collins}, {Jensen}, {Quinn},
  {Payne}, {Kristiansen}, {Spencer}, {Vanderburg}, {Zechmeister}, {Weiss},
  {Wang}, {Wang}, {Udry}, {Terentev}, {St{\"u}rmer}, {Stef{\'a}nsson},
  {Shporer}, {Shectman}, {Sefako}, {Schwengeler}, {Schwarz}, {Scarsdale},
  {Rubenzahl}, {Roy}, {Rosenthal}, {Robertson}, {Petigura}, {Pepe},
  {Omohundro}, {Murphy}, {Murgas}, {Mo{\v{c}}nik}, {Montet}, {Mennickent},
  {Mayo}, {Massey}, {Lubin}, {Lovis}, {Lewin}, {Kasper}, {Kane}, {Jenkins},
  {Huber}, {Horne}, {Hill}, {Gorrini}, {Giacalone}, {Fulton}, {Forveille},
  {Figueira}, {Fetherolf}, {Dressing}, {D{\'\i}az}, {Delfosse}, {Dalba}, {Dai},
  {Cort{\'e}s}, {Crossfield}, {Crane}, {Conti}, {Collins}, {Chontos}, {Butler},
  {Brown}, {Brady}, {Behmard}, {Beard}, {Batalha}, \&
  {Almenara}}]{Winters:2022}
{Winters}, J.~G., {Cloutier}, R., {Medina}, A.~A., {et~al.} 2022, \aj, 163,
  168, \dodoi{10.3847/1538-3881/ac50a9}

\bibitem[{{Wolfgang} \& {Lopez}(2015)}]{Wolfgang:2015}
{Wolfgang}, A., \& {Lopez}, E. 2015, \apj, 806, 183,
  \dodoi{10.1088/0004-637X/806/2/183}

\bibitem[{{Wright} {et~al.}(2009){Wright}, {Upadhyay}, {Marcy}, {Fischer},
  {Ford}, \& {Johnson}}]{Wright:2009}
{Wright}, J.~T., {Upadhyay}, S., {Marcy}, G.~W., {et~al.} 2009, \apj, 693,
  1084, \dodoi{10.1088/0004-637X/693/2/1084}

\bibitem[{{Xie} {et~al.}(2016){Xie}, {Dong}, {Zhu}, {Huber}, {Zheng}, {De Cat},
  {Fu}, {Liu}, {Luo}, {Wu}, {Zhang}, {Zhang}, {Zhou}, {Cao}, {Hou}, {Wang}, \&
  {Zhang}}]{Xie:2016}
{Xie}, J.-W., {Dong}, S., {Zhu}, Z., {et~al.} 2016, Proceedings of the National
  Academy of Science, 113, 11431, \dodoi{10.1073/pnas.1604692113}

\bibitem[{{Zeng} {et~al.}(2019){Zeng}, {Jacobsen}, {Sasselov}, {Petaev},
  {Vanderburg}, {Lopez-Morales}, {Perez-Mercader}, {Mattsson}, {Li}, {Heising},
  {Bonomo}, {Damasso}, {Berger}, {Cao}, {Levi}, \& {Wordsworth}}]{Zeng:2019}
{Zeng}, L., {Jacobsen}, S.~B., {Sasselov}, D.~D., {et~al.} 2019, Proceedings of
  the National Academy of Science, 116, 9723, \dodoi{10.1073/pnas.1812905116}

\bibitem[{{Zhang} {et~al.}(2021){Zhang}, {Liu}, {Claytor}, {Best}, {Dupuy}, \&
  {Siverd}}]{Zhang:2021}
{Zhang}, Z., {Liu}, M.~C., {Claytor}, Z.~R., {et~al.} 2021, \apjl, 916, L11,
  \dodoi{10.3847/2041-8213/ac1123}

\end{thebibliography}
\bibliographystyle{aasjournal}

\appendix

\begin{deluxetable*}{lccccccc}[ht]

\tablecaption{Planet Properties \label{table:planet_props}}
\tablehead{
\colhead{Planet Name} & \colhead{Orbital Period} & \colhead{Mass} & \colhead{Radius} & \colhead{Bulk density} &\colhead{$\#$ Planets} & \colhead{CMF or}& \colhead{Reference} \vspace{-2mm} \\
 \colhead{} & \colhead{(days)} & \colhead{($M_{\oplus}$)} & \colhead{($R_{\oplus}$)} & \colhead{(\dens)} & \colhead{} & \colhead{WMF}&  \colhead{}}
\startdata
        TRAPPIST-1 b & 1.510826 & 1.374 $\pm$ 0.069 & 1.11$^{+0.014}_{-0.012}$ & 5.44$^{+0.26}_{-0.27}$ & 7 & CMF = 0.24 $\pm$ 0.07 & \citet{Agol:2021} \\
        TRAPPIST-1c & 2.421937 & 1.308 $\pm$ 0.056 & 1.097$^{+0.014}_{-0.012}$ & 5.46$^{+0.22}_{-0.24}$ & 7 &  CMF = 0.30 $\pm$ 0.07 & \citet{Agol:2021} \\ 
        TRAPPIST-1 d & 4.049219 & 0.388 $\pm$ 0.012 & 0.788$^{+0.011}_{-0.010}$ & 4.37$^{+0.15}_{-0.17}$ & 7 &  CMF = 0.16 $\pm$ 0.04 & \citet{Agol:2021} \\  
        TRAPPIST-1 e & 6.101013 & 0.692 $\pm$ 0.022 & 0.92$^{+0.013}_{-0.012}$ & 4.90$^{+0.17}_{-0.18}$ & 7 &  CMF = 0.22 $\pm$ 0.07 & \citet{Agol:2021} \\  
        TRAPPIST-1 f & 9.20754 & 1.039 $\pm$ 0.031 & 1.045$^{+0.013}_{-0.012}$ & 5.02$^{+0.14}_{-0.16}$ & 7 & CMF = 0.17 $\pm$ 0.06 & \citet{Agol:2021} \\  
        TRAPPIST-1 g & 12.352446 & 1.321 $\pm$ 0.038 & 1.129$^{+0.015}_{-0.013}$ & 5.06$^{+0.14}_{-0.16}$ & 7 & CMF = 0.15 $\pm$ 0.06 & \citet{Agol:2021} \\  
        TRAPPIST-1 h & 18.772866 & 0.326 $\pm$ 0.02 & 0.755 $\pm$ 0.014 & 4.16$^{+0.32}_{-0.30}$ & 7 & CMF = 0.12 $\pm$ 0.06 & \citet{Agol:2021} \\  
        LHS 1140 b & 24.73694 & 6.38$^{+0.46}_{-0.44}$ & 1.635 $\pm$ 0.46 & 8.04$^{+0.84}_{-0.80}$ & 2 & CMF = 0.36 $\pm$ 0.12 & \citet{LilloBox} \\  
        LHS 1140 c & 3.77792 & 1.76$^{+0.17}_{-0.16}$ & 1.169$^{+0.037}_{-0.038}$ & 6.07$^{+0.81}_{-0.74}$ & 2 & CMF = 0.34 $\pm$ 0.15 & \citet{LilloBox} \\  
        K2-25 b & 3.48456408 & 24.5$^{+5.7}_{-5.2}$ & 3.44 $\pm$ 0.12 & 3.31 $\pm$ 0.20$^\dagger$ & 1 &  & \citet{Stefansson:2020}  \\  
        GJ 1214 b & 1.58040433 & 8.17 $\pm$ 0.43 & 2.742$^{+0.050}_{-0.053}$ & 2.2$^{+0.17}_{-0.16}$ & 1 & ~ & \citet{Cloutier:2021} \\  
        GJ 1132 b & 1.628931 & 1.66 $\pm$ 0.23 & 1.13 $\pm$ 0.056 & 6.3 $\pm$ 1.3 & 2 & CMF = 0.40 $\pm$ 0.22 & \citet{Bonfils:2018} \\  
        LTT 3780 b & 0.768448 & 2.62$^{+0.48}_{-0.
        46}$ & 1.332$^{+0.072}_{-0.075}$ & 6.1$^{+1.8}_{-1.5}$ & 2 & CMF = 0.24 $\pm$ 0.26 & \citet{Cloutier:2021} \\  
        LTT 3780 c & 12.2519 & 8.6$^{+1.6}_{-1.3}$ & 2.3$^{+0.16}_{-0.15}$ & 3.9$^{+1.0}_{-0.9}$ & 2 & WMF = 0.90 & \citet{Cloutier:2021} \\  
        GJ 486 b & 1.467119 & 2.82$^{+0.11}_{-0.12}$ & 1.305$^{+0.063}_{-0.067}$ & 7.0$^{+1.2}_{-1.0}$ & 1 & CMF = 0.40 $\pm$ 0.17 & \citet{Trifonov:2021} \\  
        LTT 1445 A b & 5.3587657 & 2.87$^{+0.26}_{-0.25}$ & 1.305$^{+0.66}_{-0.61}$ & 7.1$^{+1.2}_{-1.1}$ & 2 & CMF = 0.41 $\pm$ 0.20 & \citet{Winters:2022} \\  
        LTT 1445 A c & 3.1239035 & 1.54$^{+0.20}_{-0.19}$ & 1.147$^{+0.55}_{-0.54}$ & 5.57$^{+0.68}_{-0.60}$ & 2 & CMF = 0.26 $\pm$ 0.23 & \citet{Winters:2022} \\  
        TOI-2136 b & 7.851928 & 6.37$^{+2.45}_{-2.29}$ & 2.19 $\pm$ 0.17 & 3.34$^{+2.55}_{-1.63}$ & 1 & ~ & \citet{Gan2136:2022} \\  
        GJ 3473 b & 1.1980035 & 1.86 $\pm$ 0.30 & 1.264 $\pm$ 0.05 & 5.03$^{+1.07}_{-0.93}$ & 2 & CMF = 0.07 $\pm$ 0.24 & \citet{Kemmer:2020} \\  
        GJ 3929 b & 2.6162745 & 1.21$^{+0.40}_{-0.42}$ & 1.15$^{+0.040}_{-0.039}$  & 4.4 $\pm$ 1.6 & 2 & WMF = 0.13 & \citet{Kemmer:2022} \\  
        LHS 1478 b & 1.9495378 & 2.33 $\pm$ 0.20 & 1.242$^{+0.051}_{-0.049}$ & 6.67$^{+1.03}_{-0.89}$ & 1 & CMF = 0.39 $\pm$ 0.17 & \citet{Soto:2021} \\  
        K2-146 b & 2.6446 & 5.77 $\pm$ 0.18 & 2.05 $\pm$ 0.06 & 3.69 $\pm$ 0.21 & 2 & WMF = 0.78 & \citet{Hamann:2019} \\  
        K2-146 c & 4.00498 & 7.49 $\pm$ 0.24 & 2.19 $\pm$ 0.07 & 3.92 $\pm$ 0.27 & 2 & WMF = 0.80 & \citet{Hamann:2019} \\  
        HATS-71 b & 3.7955202 & 117.6 $\pm$ 76.3 & 11.478 $\pm$ 0.20& 0.42 $\pm$ 0.28 & 1 & ~ & \citet{Bakos:2020} \\  
        COCONUTS-2 b & 402000000 & 2002.31$^{+476}_{-603}$ & 12.442 $\pm$ 0.36 & 5.71 $\pm$ 1.48$^{\dagger}$ & 1 & ~ & \citet{Zhang:2021} \\ 
        L 98-59 b & 2.2531136 & 0.4$^{+0.16}_{-0.15}$ & 0.85$^{+0.061}_{-0.047}$ & 3.6$^{+1.4}_{-1.5}$ & 4 & WMF = 0.16 & \citet{Demangeon:2021} \\  
        L 98-59 c & 3.6906777 & 2.22$^{+0.26}_{-0.25}$ & 1.385$^{+0.095}_{-0.075}$ & 4.57$^{+0.77}_{-0.85}$ & 4 & WMF = 0.18 & \citet{Demangeon:2021} \\  
        L 98-59 d & 7.4507245 & 1.94 $\pm$ 0.28 & 1.521$^{+0.12}_{-0.09}$ & 2.95$^{+0.79}_{-0.51}$  & 4 & WMF = 0.63 & \citet{Demangeon:2021} \\  
        TOI-1685 b & 0.6691403 & 3.78 $\pm$ 0.63 & 1.7 $\pm$ 0.07 & 4.21$^{+0.95}_{-0.82}$ & 1 & WMF = 0.40 & \citet{Bluhm:2021} \\  
        K2-18 b & 32.939623 & 8.92$^{+1.7}_{-1.6}$ & 2.37 $\pm$ 0.22 & 4.11$^{+1.72}_{-1.18}$ & 2 & ~ & \citet{sarkis:2018} \\  
        GJ 1252 b & 0.5182349 & 2.09 $\pm$ 0.56 & 1.193 $\pm$ 0.074 & 6.76 $\pm$ 0.67$^{\dagger}$ & 1 & CMF = 0.45 $\pm$ 0.35 & \citet{Shporer:2020} \\  
        TOI-1201 b & 2.4919863 & 6.28$^{+0.84}_{-0.88}$ & 2.415$^{+0.091}_{-0.09}$ & 2.45$^{+0.48}_{-0.42}$ & 1 & ~ & \citet{Kossakowski:2021} \\  
        GJ 357 b & 3.93072 & 1.84 $\pm$ 0.31 & 1.217$^{+0.084}_{-0.083}$ & 5.6$^{+1.7}_{-1.3}$ & 3 & CMF = 0.23 $\pm$ 0.29 & \citet{Luque:2019} \\  
        TOI-270 b & 3.3601538 & 1.58 $\pm$ 0.26 & 1.206 $\pm$ 0.039 & 4.97 $\pm$ 0.94 & 3 & CMF = 0.09 $\pm$ 0.21 & \citet{VanEylen:2021} \\  
        TOI-270 c & 5.6605731 & 6.15 $\pm$ 0.37 & 2.355 $\pm$ 0.064 & 2.6 $\pm$ 0.26 & 3 & ~ & \citet{VanEylen:2021} \\  
        TOI-270 d & 11.379573 & 4.78 $\pm$ 0.43 & 2.133 $\pm$ 0.058 & 2.72 $\pm$ 0.33 & 3 & ~ & \citet{VanEylen:2021} \\  
        TOI-269 b & 3.6977104 & 8.8 $\pm 1.4$ & 2.77 $\pm$ 0.12 & 2.28$^{+0.48}_{-0.42}$ & 1 & ~ & \citet{Cointepas:2021} \\  
        TOI-674 b & 1.977143 & 23.6 $\pm$ 3.3 & 5.25 $\pm$ 0.17 & 0.91 $\pm$ 0.15 & 1 & ~ & \citet{Murgas:2021} \\  
        GJ 367 b & 0.321962 & 0.546 $\pm$ 0.078 & 0.718 $\pm$ 0.054 & 8.106 $\pm$ 2.16 & 1 & CMF = 0.87 $\pm$ 0.30 & \citet{Lam:2021} \\ 
        TOI-1634 b & 0.989343 & 4.91$^{+0.68}_{-0.70}$ & 1.79 $\pm$ 0.08 & 4.7$^{+1.0}_{-0.90}$ & 1 & WMF = 0.35 & \citet{Cloutier1634:2021} \\  
        TOI-1231 b & 24.245586 & 15.4 $\pm$ 3.3 & 3.65$^{+0.16}_{-0.15}$ & 1.74$^{+0.47}_{-0.42}$ & 1 & ~ & \citet{Burt:2021} \\  
        GJ 436 b & 2.64388312 & 22.1 $\pm$ 2.3 & 4.17 $\pm$ 0.168 & 1.8 $\pm$ 0.29 & 1 & ~ & \citet{Maciejewski:2014} \\  
        GJ 3470 b & 3.3366496 & 13.9 $\pm$ 1.5 & 4.57 $\pm$ 0.18 & 0.8 $\pm$ 0.13 & 1 & ~ & \citet{Demory:2013} \\  
        TOI-1266 b & 10.894843 & 13.5$^{+11}_{-9}$  & 2.37$^{+0.16}_{-0.12}$ & 5.57$\pm$ 1.14$^{\dagger}$ & 2 & ~ & \citet{Demory:2020} \\  
        TOI-1266 c & 18.80151 & 2.2$^{+2.0}_{-1.5}$ & 1.56$^{+0.15}_{-0.13}$ & 3.18$\pm$ 0.38$^{\dagger}$ & 2 & WMF = 0.60 & \citet{Demory:2020} \\  
        TOI-530 b & 6.387597 & 127.13$^{+28.6}_{-31.8}$ & 9.303 $\pm$ 0.70 & 0.93$^{+0.49}_{-0.35}$  & 1 & ~ & \citet{Gan530:2022} \\  
        TOI-3714 b & 2.15 & 222.48 $\pm$ 10 & 11.321 $\pm$ 0.3 & 0.85 $\pm$ 0.08 & 1 & ~ & \citet{Canas:2022} \\
 \enddata
\tablecomments{The density values marked by a ${\dagger}$ have been calculated in this paper by propagating the errors from the planet's mass and radius. WMF values are calculated assuming a rocky core with a fixed Fe:$\rm MgSiO_{3}$ ratio.}
\end{deluxetable*}

\begin{deluxetable*}{lccccccc}[ht]
\tablecaption{Planet Properties (Continued) 
\label{table:planet_props}}
\tablehead{
\colhead{Planet Name} & \colhead{Orbital Period} & \colhead{Mass} & \colhead{Radius} & \colhead{Bulk density} &\colhead{$\#$ Planets} & \colhead{CMF or}& \colhead{Reference} \vspace{-2mm} \\
 \colhead{} & \colhead{(days)} & \colhead{($M_{\oplus}$)} & \colhead{($R_{\oplus}$)} & \colhead{(\dens)} & \colhead{} & \colhead{WMF}&  \colhead{}}
\startdata
        AU Mic b & 8.4629991 & 20.12$^{+1.57}_{-1.72}$   & 4.07 $\pm$ 0.17 & 1.32$^{+0.19}_{-0.2}$ & 2 & ~ & \citet{Cale:2021} \\   
        ~ & ~ & ~ & ~ & ~ & ~ & ~ & \citet{Gilbert:2022} \\  
        AU Mic c & 18.858991 & 9.6$^{+2.07}_{-2.10}$ & 3.24 $\pm$ 0.17 & 1.22$^{+0.26}_{-0.29}$ & 2 & ~ & \citet{Cale:2021} \\ 
        ~ & ~ & ~ & ~ & ~ & ~ & ~ & \citet{Martioli:2021} \\  
        Kepler-54 b & 8.0109434 & 21.5$^{+5.6}_{-4.9}$ & 2.19 $\pm$ 0.07 & 12.8 $\pm$ 2.0$^{\dagger}$ & 3 & ~ & \citet{Hadden:2014} \\   
        TOI-776 b & 8.24661 & 4.0 $\pm$ 0.9 & 1.85 $\pm$ 0.13 & 3.4$^{+1.1}_{-0.9}$  & 2 & WMF = 0.70 & \citet{Luque:2021} \\   
        TOI-776 c & 15.6653 & 5.3  $\pm$ 1.8 & 2.02 $\pm$ 0.14 & 3.5$^{+1.4}_{-1.3}$  & 2 & WMF = 0.80 & \citet{Luque:2021} \\   
        HATS-6 b & 3.3252725 & 101.38  $\pm$ 22.2 & 11.187 $\pm$ 0.213  & 0.399 $\pm$ 0.089 & 1 & ~ & \citet{Hartman:2015} \\   
        Kepler-231 c & 19.271566 & 24.1$^{+13.5}_{-11.2}$ & 1.93 $\pm$ 0.19 & 18.4 $\pm$ 9.01$^{\dagger}$ & 2 & ~ & \citet{Rowe:2014}, \\   
        ~ & ~ & ~ & ~ & ~ & ~ & ~ & \citet{Hadden:2014} \\   
        HATS-74 A b & 1.73185606 & 464.029 $\pm$ 44 & 11.568 $\pm$ 0.24  & 1.64 $\pm$ 0.19& 1 & ~ & \citet{Jordan:2022} \\ 
        HATS-75 b & 2.7886556 & 156.053 $\pm$ 12 & 9.91 $\pm$ 0.15 & 0.878 $\pm$ 0.013 & 1 & ~ & \citet{Jordan:2022} \\   
        L 168-9 b & 1.4015 & 4.6 $\pm$ 0.56  & 1.39 $\pm$ 0.09 & 9.6$^{+2.4}_{-1.8}$ & 1 & CMF = 0.63 $\pm$ 0.22 & \citet{Astudillo-Defru:2020} \\   
        HD 260655 b & 2.76953 & 2.14 $\pm$ 0.34  & 1.24 $\pm$ 0.023 & 6.2 $\pm$ 1.0 & 2 & CMF = 0.32 $\pm$ 0.19 & \citet{Luque:2022} \\   
        HD 260655 c & 5.70588 & 3.09 $\pm$ 0.48 & 1.533$^{+0.051}_{-0.046}$  & 4.7$^{+0.8}_{-0.9}$  & 2 & WMF = 0.22 & \citet{Luque:2022} \\   
        Kepler-45 b & 2.455239 & 160.49 $\pm$ 28.6  & 10.76 $\pm$ 1.23 & 0.8 $\pm$ 0.5 & 1 & ~ & \citet{Johnson:2012} \\   
        Kepler-138 b & 10.3126 & 0.066$^{+0.059}_{-0.037}$ & 0.522 $\pm$ 0.032 & 2.6$^{+2.4}_{-1.5}$  & 3 & WMF = 0.24 & \citet{Jontof-Hutter:2015} \\   
        Kepler-138 c & 13.7813 & 1.97$^{+1.91}_{-1.12}$ & 1.197 $\pm$ 0.07 & 6.2$^{+5.8}_{-3.4}$  & 3 & CMF = 0.42 $\pm$ 0.49 & \citet{Jontof-Hutter:2015} \\   
        Kepler-138 d & 23.0881 & 0.64$^{+0.674}_{-0.387}$ & 1.212 $\pm$ 0.075  & 2.1$^{+2.2}_{-1.2}$  & 3 & WMF = 0.92 & \citet{Jontof-Hutter:2015} \\   
        TOI-1899 b & 29.02 & 209.76 $\pm$ 22.24 & 12.89 $^{+0.448}_{-0.56}$  & 0.54$^{+0.09}_{-0.1}$  & 1 & ~ & \citet{canas:2020} \\   
        LP 714-47b & 4.052037 & 30.8 $\pm$ 1.5  & 4.7 $\pm$ 0.3 & 1.7 $\pm$ 0.3  & 1 & ~ & \citet{Dreizler:2020} \\ 
         & &  & & &  &  & \citet{Luque:2022} \\
        LHS 1815 b & 3.81433 & 1.58$^{+0.64}_{-0.60}$  & 1.088 $\pm$ 0.064 & 6.74$^{+3.12}_{-2.66}$ & 1 & CMF = 0.46 $\pm$ 0.34 & \citet{Gan:2020} \\ 
        & &  & & &  &  & \citet{Luque:2022} \\
        TOI-1728 b & 3.49151 & 26.8$^{+5.4}_{-5.1}$ & 5.05$^{+0.16}_{-0.17}$  & 1.14$^{+0.26}_{0.24}$ & 1 & ~ & \citet{Kanodia:2020} \\  
        & &  & & &  &  & \citet{Luque:2022} \\
        TOI-1235 b & 3.444717 & 6.69$^{+0.67}_{-0.69}$ & 1.694$^{+0.08}_{-0.077}$ & 7.25$^{+1.3}_{-1.1}$ & 1 & CMF = 0.25 $\pm$ 0.19 & \citet{Bluhm:2020} \\
         & & & & &  &  & \citet{Luque:2022} \\
        K2-3 b & 10.054626 & 6.48$^{+0.99}_{-0.93}$ & 2.103 $\pm$ 0.25 & 3.7$^{+1.67}_{-1.08}$ & 3 & WMF = 0.77 & \citet{Crossfield:2015} \\  
         & &  & & &  &  & \citet{Luque:2022} \\  
        K2-3 c & 24.646582 & 2.14$^{+1.08}_{-1.04}$ & 1.584$^{+0.197}_{-0.195}$ & 2.98$^{+1.96}_{-1.5}$ & 3 & WMF = 0.69 & \citet{Crossfield:2015}\\ 
        & &  & & &  &  & \citet{Luque:2022} \\  
\enddata
\tablecomments{The density values marked by a ${\dagger}$ have been calculated in this paper by propagating the errors from the planet's mass and radius. WMF values are calculated assuming a rocky core with a fixed Fe:$\rm MgSiO_{3}$ ratio.}
\end{deluxetable*}





\end{document}